%% file: main.tex
\documentclass[aps,prb,twocolumn,superscriptaddress,nofootinbib]{revtex4-2}

\usepackage[T1]{fontenc} 
\usepackage{float}
\usepackage{graphicx}
\usepackage{dcolumn}
\usepackage{bm}
\usepackage{color}
\usepackage{amsmath}
\usepackage{amssymb}
\usepackage{xcolor}
\usepackage{braket}
\usepackage{glossaries}
\usepackage{todonotes}
\usepackage[colorlinks=true,urlcolor=red,linkcolor=blue,citecolor=blue]{hyperref}
\usepackage{orcidlink}

\input{glossary.tex}

\newcommand{\bq}{\boldsymbol{q}}

\begin{document}
%%%%%%%%%%%%%%%%%%%%%%%%%%%%%%%%%%%%%%%%%%%%%%%%%%%%%%%%%%%
% TITLE
%%%%%%%%%%%%%%%%%%%%%%%%%%%%%%%%%%%%%%%%%%%%%%%%%%%%%%%%%%%

\title{Theory of electron-phonon interactions in extended correlated systems probed by resonant inelastic x-ray scattering}

%%%%%%%%%%%%%%%%%%%%%%%%%%%%%%%%%%%%%%%%%%%%%%%%%%%%%%%%%%%
% AUTHORS
%%%%%%%%%%%%%%%%%%%%%%%%%%%%%%%%%%%%%%%%%%%%%%%%%%%%%%%%%%%

\author{Jinu Thomas\orcidlink{0000-0003-4818-6660}}
\affiliation{Department of Physics and Astronomy, The University of Tennessee, 
Knoxville, Tennessee 37996, USA}
\affiliation{Institute for Advanced Materials and Manufacturing, University of Tennessee, Knoxville, Tennessee 37996, USA\looseness=-1}
\author{Debshikha Banerjee\orcidlink{0009-0001-2925-9724}}
\affiliation{Department of Physics and Astronomy, The University of Tennessee, Knoxville, Tennessee 37996, USA}
\affiliation{Institute for Advanced Materials and Manufacturing, University of Tennessee, Knoxville, Tennessee 37996, USA\looseness=-1}
\author{Alberto Nocera\orcidlink{0000-0001-9722-6388}}
\affiliation{Stewart Blusson Quantum Matter Institute, University of British Columbia, Vancouver, British Columbia, Canada
V6T 1Z4}
\affiliation{Department of Physics Astronomy, University of British Columbia, Vancouver, British Columbia, Canada V6T 1Z1}
\author{Steven Johnston\orcidlink{0000-0002-2343-0113}}
\affiliation{Department of Physics and Astronomy, The University of Tennessee, Knoxville, Tennessee 37996, USA}
\affiliation{Institute for Advanced Materials and Manufacturing, University of Tennessee, Knoxville, Tennessee 37996, USA\looseness=-1}

\date{\today}

%%%%%%%%%%%%%%%%%%%%%%%%%%%%%%%%%%%%%%%%%%%%%%%%%%%%%%%%%%%
% ABSTRACT
%%%%%%%%%%%%%%%%%%%%%%%%%%%%%%%%%%%%%%%%%%%%%%%%%%%%%%%%%%%

\begin{abstract}
An emerging application of resonant inelastic x-ray scattering (RIXS) is the study of lattice excitations and electron-phonon ($e$-ph) interactions in quantum materials. Despite the growing importance of this area of research, the community lacks a complete understanding of how the RIXS process excites the lattice and how these excitations encode information about the $e$-ph interactions. Here, we present a detailed study of the RIXS spectra of the Hubbard-Holstein model defined on extended one-dimensional lattices. Using the density matrix renormalization group (DMRG) method, we compute the RIXS response while treating the electron mobility, many-body interactions, and core-hole interactions on an equal footing. The predicted spectra exhibit notable differences from those obtained using the commonly adopted Lang-Firsov models, with important implications for analyzing past and future experiments. Our results provide a deeper understanding of how RIXS probes $e$-ph interactions and set the stage for a more realistic analysis of future experiments.
\end{abstract}

\glsresetall

\maketitle

%%%%%%%%%%%%%%%%%%%%%%%%%%%%%%%%%%%%%%%%%%%%%%%%%%%%%%%%%%%
% INTRODUCTION
%%%%%%%%%%%%%%%%%%%%%%%%%%%%%%%%%%%%%%%%%%%%%%%%%%%%%%%%%%%

\section{Introduction}
Studying \gls*{eph} interactions in strongly correlated quantum materials is an exciting and rapidly emerging application of \gls*{rixs}~\cite{Mitriano2024exploring, AmentEPL2011, Gilmore2023quantifying}. This advance has been enabled by significant improvements in instrument resolution, which have facilitated the observation of lattice excitations in diverse families of  materials~\cite{Hancock2010, Yavas2010, LeePRL2013, LeePRB2015, MoserPRL2015, Fatale2016hybridization, JohnstonNatComm2016, Chaix2017, MyersPRL2018, rossi2019experimental, BraicovichPRR2020, LiPNMAS2020, LinPRL2020, Geondzhian2020large, PengPRL2020, Huang2021quantum, dashwood2021probing, PengPRB2022, naamneh2024persistence}. These efforts have been further bolstered by early theoretical models suggesting that the intensity of the lattice excitations provides direct information about the \gls*{eph} vertex $g(\boldsymbol{k}, \boldsymbol{q})$~\cite{AmentEPL2011, DevereauxPRX2016}, as well as its coupling to the system's electronic and magnetic parameters~\cite{LeePRL2013, JohnstonNatComm2016}. However, all theoretical frameworks developed to date to describe these experiments approximate the system in some way, and the validity of each approach has yet to be checked rigorously. For this reason, the mechanics of how the lattice is excited during a \gls*{rixs} experiment are not yet fully understood~\cite{Gilmore2023quantifying, Mitriano2024exploring}. 

%%%%%%%%%%%%%%%%%
%Figure Schematic
%%%%%%%%%%%%%%%%%
\begin{figure*}[t!]
\centering
\includegraphics[width=1\textwidth]{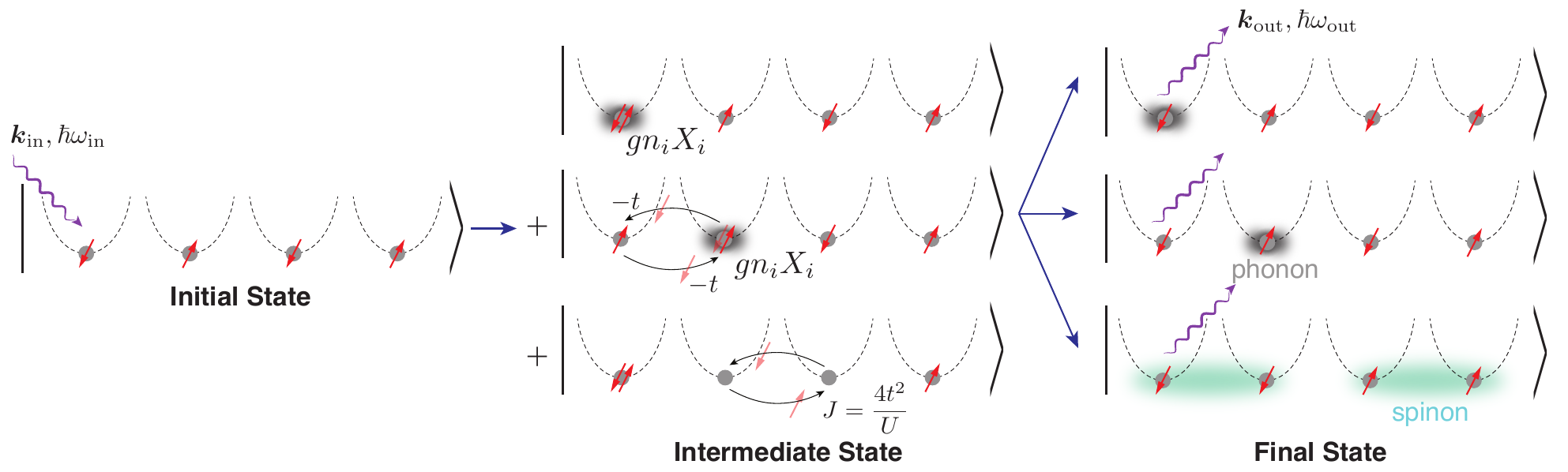}
\caption{Examples of different excitation pathways in \gls*{rixs} measurements on a one-dimensional Mott insulator coupled to the lattice. The system's initial state consists of half-filled sites with strong antiferromagnetic correlations along the chain direction. After photon absorption, a core electron is promoted to the upper Hubbard band, creating a doubly occupied site (a doublon) that interacts with the system via hopping, $e$-ph coupling, and exchange processes. If hopping does not occur (top), the lattice can move in response to the doublon via the $e$-ph interaction, locally generating several phonon quanta before the core hole decays. Electron hopping opens additional excitation pathways that can interfere with this local process. For example, the doublon can delocalize to neighboring sites and create phonons before returning to the core-hole site and decaying (middle). This process produces low-energy phonon excitations on neighbor sites but requires more time than the process shown in the top row. Alternatively, double spin-flip processes can create two-spinon excitations, as sketched in the bottom row. The intermediate stages of these processes all interfere with one another during the \gls*{rixs} process. }
\label{fig:schematic}
\end{figure*}

Many of \gls*{rixs} studies of the \gls*{eph} interaction~\cite{Yavas2010, MyersPRL2018, rossi2019experimental, LinPRL2020, BraicovichPRR2020} have relied on a \gls*{sslf} framework~\cite{AmentEPL2011} to analyze the data. This approach treats the infinite lattice in the atomic limit while coupling the local electron density to a single optical phonon branch. It neglects electron mobility and only allows for a $\boldsymbol{q}$-dependence via a $\boldsymbol{q}$-dependent \gls*{eph} coupling. Nevertheless, one can obtain an exact expression for the \gls*{rixs} cross-section in this limit (see App.~\ref{app:LF}) that captures some aspects of the experimental data. These include the generation of phonon excitations $I_n(\boldsymbol{q}) \equiv I(\boldsymbol{q},\omega = n\Omega)$ at multiples of the phonon energy $\Omega$, whose relative intensities grow in proportion to the \gls*{eph} coupling strength $|g(\boldsymbol{q})|\propto I_{n+1}(\boldsymbol{q})/I_n(\boldsymbol{q})$. 

While the \gls*{sslf}-based models have been widely used to analyze experimental data, subsequent theoretical studies have found that relaxing any of its assumptions can significantly affect the predicted spectra. For example, coupling to multiple modes in the atomic limit qualitatively changes the line shape and relative intensities of the phonon excitations~\cite{GilmorePRB2020}. Similarly, introducing electron mobility~\cite{bieniasz2021} and phonon dispersion \cite{BieniaszPRB2022} in the dilute limit leads to additional $\boldsymbol{q}$-dependence of the lattice excitations, even when the underlying \gls*{eph} coupling is momentum independent. The \gls*{sslf}-based models also fail to accurately capture potential coupling between the lattice and the core hole, which can affect the quantitative analysis of experiments~\cite{Gilmore_vibronic_2018} and lead to nontrivial renormalization of the core-hole potential once electron hopping processes are re-introduced~\cite{bieniasz2021}. These shortcomings highlight an urgent need for detailed modeling of \gls*{eph} coupling in \gls*{rixs} on extended lattices. 

Current approaches in this direction have neglected electron correlations~\cite{DevereauxPRX2016, dashwood2021probing}, focused on small clusters with only a small subset of relevant phonon modes~\cite{LeePRL2013, JohnstonNatComm2016, naamneh2024persistence}, or have been restricted to the dilute limit~\cite{bieniasz2021, BieniaszPRB2022}. In contrast, the majority of \gls*{rixs} studies in this space have been conducted on strongly correlated materials in or close to the half-filled Mott insulating regime. On the one hand, one might hope that the combination of strong Mott correlations and the core hole's attractive potential will suppress the effects of electron mobility. On the other hand, magnetic and charge fluctuations will introduce new excitation pathways that can interfere with the lattice relaxation processes captured at the single-site level, as illustrated in Fig.~\ref{fig:schematic}. It is, therefore, unclear to what extent these effects will alter lattice excitations probed by \gls*{rixs} and influence data analyses. 

We address these open questions in this article by studying the \gls*{rixs} spectra of the \gls*{1d} half-filled \gls*{hh} model. Using the numerically exact \gls*{dmrg} method~\cite{Nocera2018}, we calculate the \gls*{rixs} spectra within the \gls*{kh} formalism on extended chains while treating strong electron correlations, \gls*{eph} interactions, core-hole-lattice interactions, and any electron delocalization on an equal footing. We will demonstrate that each of these aspects can impact the spectra and should be considered in any quantitative analysis of an experiment. However, we also find that the strength of the core-hole potential plays a vital role in determining the relative importance of electron itinerancy. Our results thus contribute to the understanding of lattice excitations in \gls*{rixs} and set the stage for more accurate analysis of experiments. 

%%%%%%%%%%%%%%%%%%%%%%%%%%%%%%%%%%%%%%%%%%%%%%%%%%%%%%%%%%%
% MODEL AND METHODS
%%%%%%%%%%%%%%%%%%%%%%%%%%%%%%%%%%%%%%%%%%%%%%%%%%%%%%%%%%%
\section{Model and Methods}\label{sec:methods}

\subsection{The Hubbard-Holstein model}\label{Subsection: HH}
%%%%%%%%%%%%%%%%%
%1D Hubbard Holstein Model and parameter definitions. 
%%%%%%%%%%%%%%%%%
We consider the \gls*{1d} \gls*{hh} Hamiltonian 
\begin{align}
\label{hamiltonian}
    \begin{split}
         \mathcal{H} =& -t \sum_{j,\sigma}\left( \hat{c}_{j,\sigma}^\dagger \hat{c}^{\phantom\dagger}_{j+1,\sigma} + \  \text{h.c.} \right) + U \sum_{j} \hat{n}_{j,\uparrow}\hat{n}_{j,\downarrow} \\
        &+ \Omega \sum_j \left(\hat{b}_j^\dagger \hat{b}^{\phantom\dagger}_j+\tfrac{1}{2}\right) + g \sum_{j,\sigma} \hat{n}_{j,\sigma}(\hat{b}_j^\dagger+\hat{b}_j^{\phantom\dagger}). 
    \end{split}
\end{align}
Here, $\hat{c}_{j,\sigma}^\dagger$ $(\hat{c}^{\phantom\dagger}_{j,\sigma})$ creates (annihilates) an electron of spin $\sigma$ at site $j$; $\hat{b}_{j}^\dagger $ $(\hat{b}^{\phantom\dagger}_{j})$ creates (annihilates) phonon quanta at site $j$; $\hat{n}_{j,\sigma} = \hat{c}_{j,\sigma}^\dagger \hat{c}^{\phantom\dagger}_{j,\sigma} $ is the number operator; $t=1$ is the nearest-neighbor hopping integral (and our unit of energy); $U$ is the Hubbard interaction strength; $\Omega$ is the phonon energy ($\hbar = 1)$, and $g$ is the momentum-independent \gls*{eph} interaction strength, which we take as positive without loss of generality. 

The physics of the \gls*{eph} interaction is generally governed by two dimensionless parameters. The first is the dimensionless \gls*{eph} coupling $\lambda = g^2 /2\Omega t$, which is equal to the ratio of the lattice deformation energy $g^2/\Omega$ to half the non-interacting bandwidth $W/2 = 2t$. The second is the adiabatic ratio $\Omega/E_\mathrm{F}$, which controls the degree of retardation in the model. Throughout this work, we focus on the half-filled model with $\frac{1}{L}\sum_{i,\sigma}\langle \hat{n}_{i,\sigma}\rangle = 1$ and $\Omega/t = 1$. This choice fixes $\Omega/E_\mathrm{F} = 1/2$ and limits the size of the phonon's local Hilbert space needed for convergence while also retaining some degree of the retardation present in most materials~\cite{Karakuzu2017superconductivity, Costa2020phase}. 

At half filling, the \gls*{hh} model has competing $\boldsymbol{q} = \pi/a$ \gls*{cdw} and Mott insulating phases, with the latter dominating when $\lambda \lessapprox U/W$~\cite{Hardikar2007phase, Tezuka2007phase}. The model also has a \gls*{lel} phase for a region of small $U/W$ and $\lambda$~\cite{Hohenadler_Fehske_2018}. Throughout this work, we focus on $\lambda \le 1/2$, where the model's ground state is a Mott insulator for most values of $U/W$; however, we will also show results in the \gls*{lel} phase, which appears for $U/W < 1$ for our parameters. 

\subsection{Formalism for the \gls*{rixs} Calculations}

%%%%%%%%%%%%%%%%%
%RIXS Kramers-Heisenberg Formalism and Equations
%%%%%%%%%%%%%%%%%
We compute the \gls*{rixs} spectra using the \gls*{kh} formalism, where the intensity is given by 
\begin{equation}\label{rixs_regular}
I(\bq,\omega) \propto\sum_{f}|F_{fi}(\bq)|^{2}\delta(E_{f}-E_{i}-\omega)
\end{equation}
with the scattering amplitude
\begin{equation}\label{scattering_amp}
F_{fi} (\bq) =\sum_{\substack{n,j,\\ \sigma,\sigma^{\prime}}}e^{\mathrm{i}\bq \cdot \boldsymbol{R}_{j}}\frac{\langle f|D_{j,{\sigma^\prime}}^{\dagger}|n\rangle\langle n|D_{j,\sigma}^{\phantom{\dagger}}|i\rangle}{E_{i}-E_{n}+\omega_\mathrm{in}+\mathrm{i}\Gamma/2}.
\end{equation}
Here, $\ket{i}$, $\ket{n}$, and $\ket{f}$ are the initial, intermediate, and final states of the \gls*{rixs} process, with energies $E_i$, $E_n$, and $E_{f}$, respectively; $\omega_{\mathrm{in}}$ is the incident photon energy; $\Gamma/2$ is the inverse core-hole lifetime; $\omega$ is the energy difference between the incoming ($\omega_\mathrm{in}$) and outgoing ($\omega_\mathrm{out}$) photons; $\bq$ is the momentum transfer to the sample; $D_{j,\sigma}^{\dagger}$ and $D_{j,\sigma}^{\phantom{\dagger}}$ are the local transition operators describing the core-valence transition on site $j$ with spin $\sigma$; and $\boldsymbol{R}_j$ is the position of site $j$ in the chain. 

%%%%%%%%%%%%%%%%%
%Focus on the Cu L-edge measurements. 
%%%%%%%%%%%%%%%%%
The single band \gls*{hh} model is most appropriate for modeling materials with a single hole in their active valence states, \textit{e.g.}, Cu-$L$ edge ($2p \rightarrow 3d$) measurements on cuprates. With this in mind, we treat the core-valence transition within the dipole approximation and parameterize $D_{j,\sigma}=M_{\epsilon,\sigma,\alpha}\hat{c}_{j,\sigma}^{\dagger}\hat{p}_{j,\alpha}^{\phantom{\dagger}}$, where $\hat{p}_{j,\alpha}^{\phantom{\dagger}}$ annihilates an electron in the $\alpha$ core level and $M_{\epsilon,\sigma,\alpha} = \bra {3d,\sigma} \hat{\epsilon}\cdot \hat{r}\ket{2p_{\alpha}}$ is a polarization-dependent dipole matrix element.  
Throughout this work, we set $M_{\epsilon,\sigma,\alpha} = 1$ to focus on the lattice excitations rather than the polarization effects and assume that only a single core level is active in the scattering process. If there is a strong \gls*{soc} in the core orbitals, then $\alpha$ corresponds to the total angular momentum, and the electron spin is no longer a good quantum number~\cite{Braicovich_Ament_Bisogni_Forte_Aruta_Balestrino_Brookes2009}. In this case, the cross-section has contributions from \gls*{sc} ($\sigma = \sigma^\prime$) and \gls*{nsc} ($\sigma \ne \sigma^\prime$) channels. We will compute these contributions individually by appropriately restricting the spin summations appearing in Eq.~\eqref{scattering_amp}. These two channels can be separated from one another in quasi-\gls*{1d} systems by analyzing the incoming and outgoing x-ray polarization~\cite{Bisogni2014femtosecond}. 

%%%%%%%%%%%%%%%%%
%Addition of the core-hole piece to the Hamiltonian.
%%%%%%%%%%%%%%%%%
When computing the intermediate states in Eq~\eqref{scattering_amp}, we include the interaction between the core hole and valence electrons and a potential core-hole-lattice coupling. To capture these effects, local interactions of the form  
\begin{align}
\label{eq: h_corehole}
    \mathcal{H}^\mathrm{CH}_j = V^\mathrm{CH}\sum_{\sigma}\hat{n}^{\phantom c}_{j,\sigma}\hat{n}^\mathrm{CH}_j 
    + g_\mathrm{CH} \hat{n}^\mathrm{CH}_j \left(b^\dagger_j + b^{\phantom\dagger}_j\right)
\end{align} 
are added to Eq.~\eqref{hamiltonian}, where $\hat{n}^\mathrm{CH}_j=(1 - \hat{p}^\dagger_{j,\alpha}\hat{p}^{\phantom\dagger}_{j,\alpha})$ is the number operator for the core level. Here, $V^\mathrm{CH}<0$ parameterizes the attractive potential between the core hole and valence electrons, and $g_\mathrm{CH}$ parameterizes the interaction between the core hole and the lattice (with a corresponding dimensionless coupling $\lambda_\mathrm{CH} = g^2_{\mathrm{CH}}/2\Omega t$). The intermediate state Hamiltonian with a core hole at site $j$ is then $\mathcal{H}_j= \mathcal{H}+\mathcal{H}^\mathrm{CH}_j$.

%%%%%%%%%%%%%%%%%
%XAS in the general formalism. 
%%%%%%%%%%%%%%%%%
To obtain the resonant absorption edges for \gls*{rixs}, we also compute the \gls*{xas} using Fermi's golden rule 
\begin{align}\nonumber
I_{\mathrm{XAS}}(\omega_{\mathrm{in}})&\propto  
\sum_n \left\vert\sum_{\sigma}\langle n|D_{j,\sigma}|i\rangle\right\vert^2 
\delta(E_{i}-E_{n}+\omega_{\mathrm{in}}) \\
&\propto - \mathrm{Im} 
\sum_{n}\frac{\left\vert\sum_{\sigma}\langle n|D_{j,\sigma}|i\rangle\right\vert^2}{E_{i}-E_{n}+\omega_{\mathrm{in}}+\mathrm{i}\Gamma/2}.\label{xas_regular}
\end{align}
In the second line, we have followed the common practice of assuming that the inverse core-hole lifetime sets the energy broadening of the \gls*{xas} spectra. 

\subsection{Computational Methods}

%%%%%%%%%%%%%%%%%
%Advantages of using DMRG over current methods.
%%%%%%%%%%%%%%%%%
We use \gls*{dmrg}~\cite{White_1992, White_1993} to compute the \gls*{rixs} and \gls*{xas} spectra, as introduced in Ref.~\cite{Nocera2018}. The approach is based on the Krylov space correction vector algorithm~\cite{Nocera_Alvarez_2016} and is briefly described in App.~\ref{app:dmrgrixs}. It computes the \gls*{rixs} response without relying on any operator \cite{Lu2017nonperturbative}, core-hole-lifetime \cite{Brink2005correlation}, or diagrammatic expansions~\cite{DevereauxPRX2016, Tsvelik2019resonant}. The only approximation that we introduce is the center-site approximation~\cite{PhysRevB.60.335, White2004real} in Eqs.~\eqref{eq:DMRG_2} and \eqref{eq:XAS_CV}, which reduces the computational costs. We also use the Krylov space correction vector algorithm to compute the phonon spectral function $B(q, \omega)$ and dynamical charge $N(q,\omega)$ structure factors (see App.~\ref{app:spectralfunctions}). 

%%%%%%%%%%%%%%%%%
%DMRG++ simulation parameters.
%%%%%%%%%%%%%%%%%
We carried out all \gls*{dmrg} simulations using the \gls*{dmrg}++ code \cite{alvarez2009density}. We typically kept $m=500$ states and restricted the maximum number of phonon modes per site to $M = 10 - 17$ unless otherwise stated. Both values are chosen to ensure convergence of the ground state energy and \gls*{xas} spectra (see also App.~\ref{app:convergence}). We fix the artificial broadening parameter $\eta = 0.2t$ for the \gls*{rixs} calculations, which is consistent with the current experimental resolution assuming $t \approx 300$ meV.

%%%%%%%%%%%%%%%%%%%%%%%%%%%%%%%%%%%%%%%%%%%%%%%%%%%%%%%%%%%
% RESULTS
%%%%%%%%%%%%%%%%%%%%%%%%%%%%%%%%%%%%%%%%%%%%%%%%%%%%%%%%%%%
\section{Results} \label{Section:Results}
We will now begin presenting our results, focusing first on the case without any core-hole-lattice coupling ($\lambda_{\mathrm{CH}}=0$). Sec.~\ref{sec: RIXS Corehole Coupling} will present results with $\lambda_{\mathrm{CH}}\ne 0$. 

\begin{figure}[t!]
\centering
\includegraphics[width=\columnwidth]{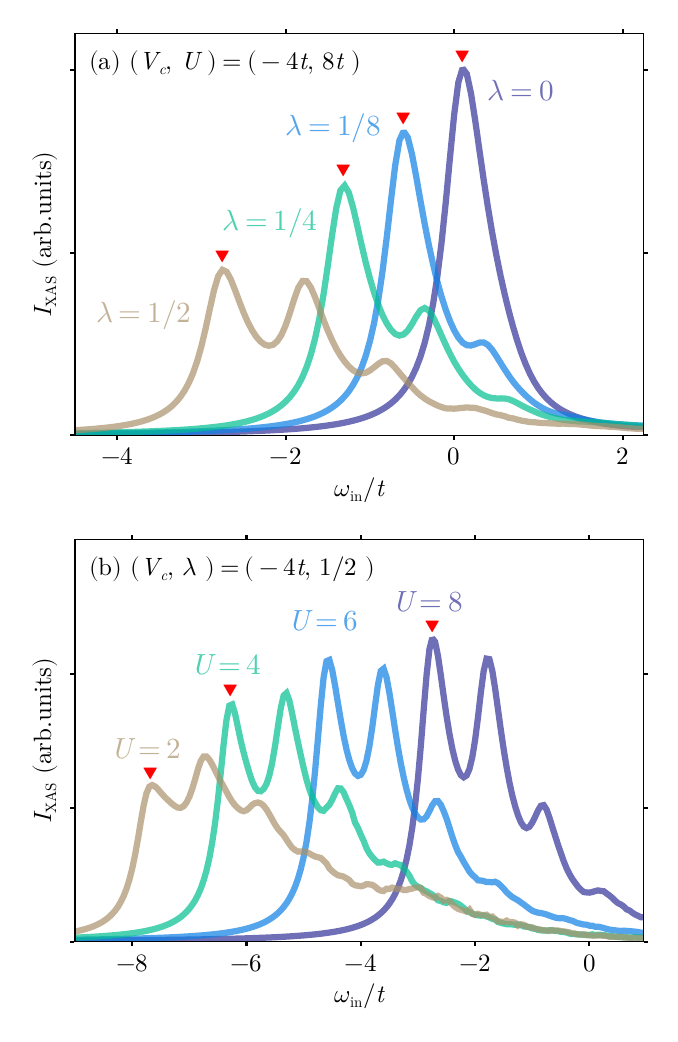}
\vspace{-0.5cm}
\caption{(a) \gls*{xas} as a function of the dimensionless \gls*{eph} coupling $\lambda$ in a half-filled chain with $U=8t$. Additional phonon harmonics are generated as $\lambda$ increases. (b) \gls*{xas} as a function of $U$ for fixed $\lambda=1/2$. Decreasing $U$ shifts the \gls*{xas} to lower energies and redistributes the weight among the phonon harmonics. In both panels, we fixed $L = 24$, $V_c=-4t$ and, $\Gamma/2=t/4$ and neglected the core-hole-lattice coupling ($\lambda_\text{CH} = 0$).  The inverted red triangles indicate the incident energy used in subsequent \gls*{rixs} calculations. }
\label{fig:XAS}
\end{figure}

\subsection{X-ray absorption spectroscopy} \label{sec: XAS results}
Figure~\ref{fig:XAS}a shows the \gls*{xas} spectra as a function of \gls*{eph} coupling for the half-filled $L = 24$ site chains and fixed $U = 8t$, $V_c = -4t$, and $\Gamma/2 = t/4$. The system has a Mott insulating ground state for all values of $\lambda$ shown in this panel. (Similar results for $V_c = -12t$ are shown in App.~\ref{app:xas_vcm12}.) 

\gls*{xas} spectra for the Hubbard model ($\lambda = 0$) have two resonances, which correspond to final states with different electron densities on the site where the core-hole is created~\cite{KourtisRIXStwochannelsHubbard}. The first is the so-called \gls*{ws} resonance, where the core electron is excited into the valence orbital on the same site, creating a double occupation that screens the core-hole potential. The second is the \gls*{ps} resonance, where the core electron is excited into a delocalized state, leaving the valence orbital of the core-hole site half-filled. In the limit $t \rightarrow 0$, these resonances appear at $E_{\mathrm{WS}} \approx U + 2V_c$ and $E_{\mathrm{PS}} \approx U+V_c$, respectively~\cite{Tsutsui_Tohyama_Maekawa_2000, KourtisRIXStwochannelsHubbard}. The peak at $\omega_\mathrm{in} = 0.1t$ in the $\lambda=0$ spectra is the \gls*{ws} resonance.

Introducing the \gls*{eph} coupling shifts both resonances to lower energies and creates one or more phonon satellites at multiples of the phonon energy. The size of the shift can be estimated in the $t \rightarrow 0$ limit as $E_{\mathrm{WS}} \approx U + 2V_c - 3\frac{g^2}{\Omega}$ and $E_{\mathrm{PS}}  \approx U+V_c - 3\frac{g^2}{\Omega}$, see Eq.~\eqref{eq: LF-Single Site}. Here, however, a non-zero hopping further shifts the resonance energies. The phonon satellites are analogous to the Franck-Condon broadening of the transition, which transfers spectral weight from the main resonance to the satellites as $\lambda$ increases, producing an asymmetric line shape. As discussed later, this transfer also reduces the overall intensity of the \gls*{rixs} spectra. 

%%%%%%%%%%%%%%%%%
%XAS response for varying g 
%%%%%%%%%%%%%%%%%
The results in Fig.~\ref{fig:XAS}a are qualitatively similar to those obtained from both the single-site~\cite{AmentEPL2011, GilmorePRB2020} and cluster-based treatments of the problem~\cite{LeePRL2013, JohnstonNatComm2016}; however, we have observed subtle differences in terms of both the width and line shape of the individual phonon peaks (not shown). The importance of treating the extended system becomes much more apparent once we examine the $U$-dependence of the spectra for fixed $\lambda$, as shown in Fig.~\ref{fig:XAS}b. Varying $U$ in the atomic limit only shifts the spectra on the $\omega_\mathrm{in}$ axis. In contrast, decreasing $U$ in the extended model shifts the main \gls*{xas} resonance and changes the relative intensities of the phonon satellites. For example, for $U > 4t$, the intensity of the principal peak is larger than the first phonon harmonic, while this ratio is reversed for $U \le 4t$. If these results are interpreted using a \gls*{sslf} framework, one would incorrectly conclude that the strength of the bare \gls*{eph} coupling increases as $U$ decreases. 

\begin{figure*}[t]
\centering
\includegraphics[width=\textwidth]{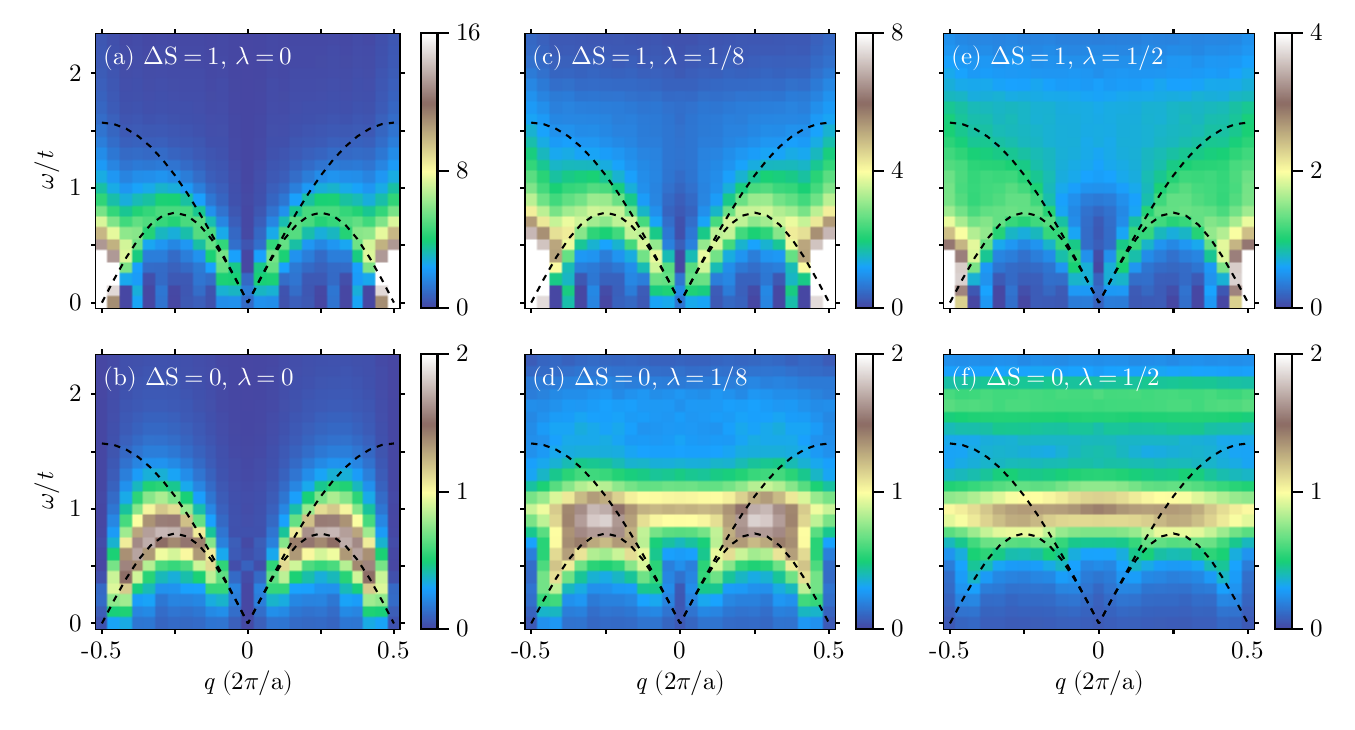}
\vspace{-1cm}
\caption{ The top and bottom rows show 
\gls*{rixs} spectra in the non-spin conserving ($\Delta S = 1$) and spin conserving ($\Delta S = 0$) channels, respectively, for $\lambda=0$, $1/8$, and $1/2$, as indicated. The black dashed lines in each panel are the boundaries of the two-spinon continuum for the Hubbard model ($\lambda=0$), assuming $J=4t^2/U$. The intensity of the two-spinon continuum decreases in the \gls*{nsc} channel 
with increasing $e$-ph coupling without any change to the boundaries. The phonon excitations appear in the spin-conserving channel at multiples of the phonon frequency $\Omega=t$. All results were obtained on an $L = 24$ site chain with fixed $V_c=-4t$, $\Gamma/2=t/4$, and $\lambda_\mathrm{CH} = 0$. The incident photon energies are indicated by the inverted triangles in Fig.~\ref{fig:XAS}a.}
\label{fig: RIXStwochannels}
\end{figure*}

%%%%%%%%%%%%%%%%%
%XAS response for U=2t
%%%%%%%%%%%%%%%%%
The spectra for $U = 2t$, which develops a noisy tail with a secondary peak at $\omega_\mathrm{in} \approx -3t$, is particularly interesting. The model's ground state is a \gls*{lel} phase for this value of $U$, which we verified via the energy analysis described in Ref.~\cite{Hohenadler_Fehske_2018}. In this case, we can resolve the \gls*{ws} and \gls*{ps} states at $\omega_\mathrm{in} \approx -7.65t$ and $\approx -3.6t$, respectively, which are shifted $\approx 1.4t$ higher in energy relative to their locations in the atomic limit. We can also resolve two to three phonon satellites on the \gls*{ws} resonance before their intensity merges with the \gls*{ps} resonance. We have found that the \gls*{ps} resonance does not develop noticeable phonon satellites for all values of $U$ and $V_c$ we have checked (see App. \ref{app:xas_vcm12}), which suggests that the itinerant \gls*{ps} final states are only weakly coupled to the lattice. 

%%%%%%%%%%%%%%%%%
%Figure RIXS response in two channels
%%%%%%%%%%%%%%%%%

\subsection{Resonant inelastic x-ray scattering}\label{sec: NSC and SC RIXS}
%%%%%%%%%%%%%%%%%
%RIXS for the Hubbard Model 
%%%%%%%%%%%%%%%%%
The low-energy \gls*{rixs} spectra provide access to the system's collective excitations~\cite{Mitriano2024exploring}. The elementary magnetic excitations of the \gls*{1d} Hubbard model are spinons, fractionalized spin-density waves that carry spin $S=1/2$ and no charge~\cite{GiamarchiBook}. Spinons must be created in pairs when excited by scattering probes, resulting in a multi-spinon continuum that has been observed in both \gls*{ins} \cite{Tennant1995measurements, Zaliznyak1999anisotropic, Walters2009effects, MourigalNatPhys2013} and \gls*{rixs} experiments~\cite{Schlappa2012spin, Schlappa2018, Bisogni2014femtosecond}. Four spinon excitations are also accessible in \gls*{rixs} measurements, which appear in regions of phase space separated from the two-spinon continuum~\cite{Schlappa2018, Kumar2022unraveling}. \gls*{1d} systems can also host fractionalized charge~\cite{Kim2006distinct, Kumar2018multispinon} and orbital excitations~\cite{Schlappa2012spin}; however, these typically appear at high energies set by the Mott gap and crystal fields of the $3d$ orbitals, respectively~\cite{Schlappa2012spin, Li2021particlehole}. Thus, for large $U/W$, magnetic and lattice excitations dominate the low-energy \gls*{rixs} spectra of systems like the \gls*{1d} cuprates and our model. 

Figures~\ref{fig: RIXStwochannels}a and \ref{fig: RIXStwochannels}b show the momentum-dependent \gls*{rixs} spectra for our model in the \gls*{nsc} and \gls*{sc} channels, respectively. These were calculated on an $L = 24$ site chain with $U=8t$ and $\lambda=\lambda_\mathrm{CH} =0$, $V_c =-4t$, $\Gamma/2 = t/4$, and with an incident energy tuned to the \gls*{ws} resonance, as indicated by the inverted red triangle in Fig.~\ref{fig:XAS}. The \gls*{nsc} magnetic excitations are, by definition, generated by direct spin-flip excitations, which are enabled by the large \gls*{soc} in the core level~\cite{Braicovich_Ament_Bisogni_Forte_Aruta_Balestrino_Brookes2009, AmentPRL2009}. This scattering channel is analogous to spin-flip \gls*{ins} and thus resembles the two-spinon continuum observed in those experiments. The spectral weight in the \gls*{rixs} spectra vanishes at zone-center $q=0$ and is concentrated around the lower bound of the two-spinon continuum, consistent with the expectations for a Hubbard chain in the strong coupling limit~\cite{Bhaseen_Essler_Grage_2005, Klauser_Mossel_Caux_Van_DenBrink_2011, Nocera_Alvarez_2016}. In contrast, the \gls*{sc} magnetic excitations are generated via $\Delta S = 0$ double spin-flip processes like those shown in Fig.~\ref{fig:schematic}~\cite{Klauser_Mossel_Caux_Van_DenBrink_2011, 
Bisogni2014femtosecond, Jia2016_RIXSElement}. The magnetic excitations in this channel thus resemble the dynamical spin-exchange correlation function~\cite{Klauser_Mossel_Caux_Van_DenBrink_2011, forte2011doping}. Notably, magnetic excitations in this channel are confined mainly to the two-spinon phase space and have vanishing spectral weight at $q = 0$ and $\pm \pi/a$~\cite{Klauser_Mossel_Caux_Van_DenBrink_2011, forte2011doping}. 

%%%%%%%%%%%%%%%%%
%RIXS with weak Eph coupling
%%%%%%%%%%%%%%%%%
Introducing the \gls*{eph} coupling generates phonon excitations in \gls*{sc} channel, whose intensity and number depend on the strength of the interaction. For a weak coupling $\lambda=1/8$ (Fig.~\ref{fig: RIXStwochannels}d), the phonon excitation appears as a dispersionless feature that tracks the bare phonon energy across the entire Brillouin zone. The width of this feature in energy loss is set by the broadening parameter $\eta = 0.2t$. A second overtone (two-phonon excitation) appears at twice the phonon energy when the coupling is increased to $\lambda=1/2$ (Fig.~\ref{fig: RIXStwochannels}f). 

\begin{figure}[t!]
\centering
\includegraphics[width=\columnwidth]{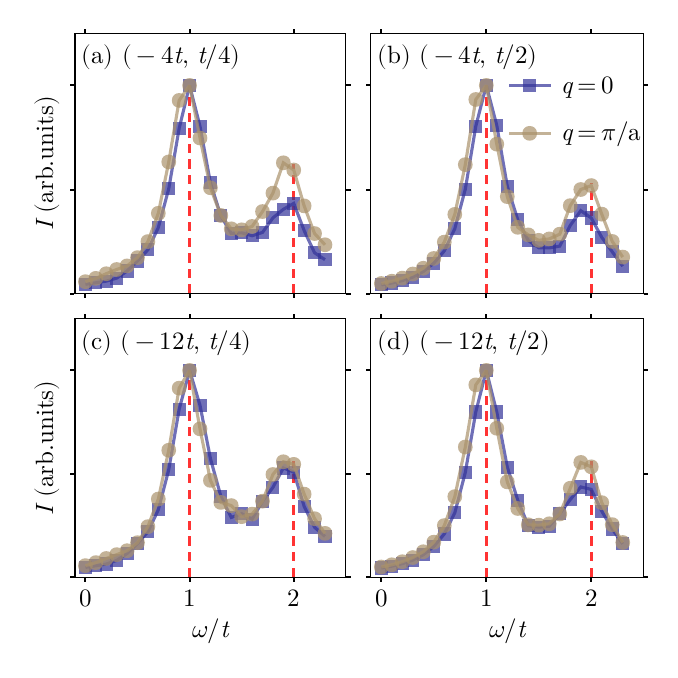}
\vspace{-1cm}
\caption{\gls*{rixs} spectra measured at the center ($q=0$) and boundary ($q=\pi/a$) of the first Brillouin zone. Each panel shows results for different combinations of $(V_c,\Gamma/2)$, as indicated, while fixing $U = 8t$, $\Omega = t$, $\lambda=1/2$, $\lambda_\text{CH} = 0$, and $L = 24$. 
Each spectrum is normalized to the intensity of the first phonon peak. All panels share the legend shown in panel (b). The red dashed lines indicate energy loss values $\omega = \Omega$ and $\omega = 2\Omega$. The \gls*{sslf} predicts no momentum dependence for all cases, while our calculations only observe this in (c).}
\label{fig:MomentumDependence}
\end{figure}

Interestingly, a weak \gls*{eph} interaction does not alter the qualitative structure of the magnetic excitations [Figs.~\ref{fig: RIXStwochannels}(c) \& (d)] but it does change the relative intensities of the \gls*{nsc} and \gls*{sc} scattering channels. (Note the change in the intensity scales in Figs.~\ref{fig: RIXStwochannels}c and \ref{fig: RIXStwochannels}e.) We attribute this reduction to the transfer of spectral weight from the main resonance to the phonon satellites observed in Fig.~\ref{fig:XAS}a, which we confirmed by comparing the ratio of the integrated spectral weight of the \gls*{rixs} spectra to the ratio of the \gls*{xas} peak intensities $I_\mathrm{xas}(\omega_\mathrm{in}+\Omega)/I_\mathrm{xas}(\omega_\mathrm{in})$. This observation has important implications for extracting quantum entanglement measures~\cite{Scheie2021Witnessing, laurell2024witnessing, Mitriano2024exploring, ren2024witnessing} from the \gls*{rixs} measurements of spin-$1/2$ systems like Sr$_2$CuO$_3$, where phonon excitations have been observed~\cite{Schlappa2018} together with the two-spinon continuum. The \gls*{eph} coupling also alters the contrast between the two- and four-spinon excitations~\cite{Kumar2022unraveling}, which is most clearly seen in the $\lambda =1/2$ spectra shown in Fig.~\ref{fig: RIXStwochannels}e. 

Another important observation is how the boundaries of the two-spinon continuum remain unchanged when $g\ne 0$. In general, the \gls*{eph} coupling will mediate an effective, attractive interaction that reduces the effects of the on-site repulsion. In the anti-adiabatic limit ($\Omega \gg t$), the effective Hubbard interaction is $U_\mathrm{eff} = U - \lambda W$. However, the \gls*{eph} interaction also dresses the hopping integrals with $t_\mathrm{eff} = t e^{-g^2/\Omega^2}$ 
in the strong coupling limit ($\lambda \gg 1$). While both effects should alter the effective exchange interaction $J = 4t^2/U$, they act in different directions. The spin excitations in our calculations remain within the two-spinon boundaries determined by the bare exchange interaction, suggesting that the renormalization of $t$ and $U$, together with retardation effects, are conspiring to leave the effective exchange constant unrenormalized. (We have explicitly confirmed this observation also holds for the dynamical spin structure factor and, therefore, is unrelated to the \gls*{rixs} cross-section.)  

At first glance, the results shown in Fig.~\ref{fig: RIXStwochannels} are qualitatively consistent with the predictions of the \gls*{sslf} approach~\cite{AmentEPL2011}. However, there are some important quantitative differences, particularly with respect to the momentum dependence of the phonon intensities. The first phonon excitation overlaps with the magnetic excitations for our choice of parameters. But we can exploit the vanishing weight of the \gls*{sc} magnetic excitations at $q = 0$ and $\pi/a$ (see Fig.~\ref{fig: RIXStwochannels}c) to isolate the contributions from the phonons at these momenta, as shown in Fig.~\ref{fig:MomentumDependence}. The \gls*{sslf} model predicts that the intensity of the phonon excitations is proportional to the momentum-resolved \gls*{eph} coupling $|g(q)|/\Gamma \propto I_2(q)/I_1(q)$, where $I_n(q) = I(q,\omega = n\Omega)$ is the intensity of the $n$th phonon excitation. For the Holstein model, one thus expects the ratio of these intensities to be constant in $q$. Our results, however, do not conform to this expectation. 

Calculations in the dilute limit have shown that electron mobility introduces additional $\boldsymbol{q}$-dependence in the intensity of the lattice excitations~\cite{bieniasz2021}. Our results show that this conclusion persists at finite carrier concentrations and deep into the strongly correlated Mott-insulating regime. For $V_c = -4t$, the ratio $I_2(q)/I_1(q)$ increases for $q = \pi/a$ relative to the value at $q = 0$ (See Fig.~\ref{fig:MomentumDependence}a). The bare \gls*{eph} coupling in our model is momentum-independent, and the renormalization of the phonon dispersion is negligible for these parameters (see also Fig.~\ref{fig: U_dependence}). Therefore, this momentum dependence must arise from the mobility of the electrons. This conclusion is also supported by the spectra's dependence on the strength of core-hole potential $V_c$ and inverse core-hole lifetime $\Gamma/2$, which we discuss next.  

\subsection{Effects of the core-hole potential and lifetime} \label{sec: RIXSVcandGammaResults}

\begin{figure*}[t]
\centering
\includegraphics[width=\textwidth]{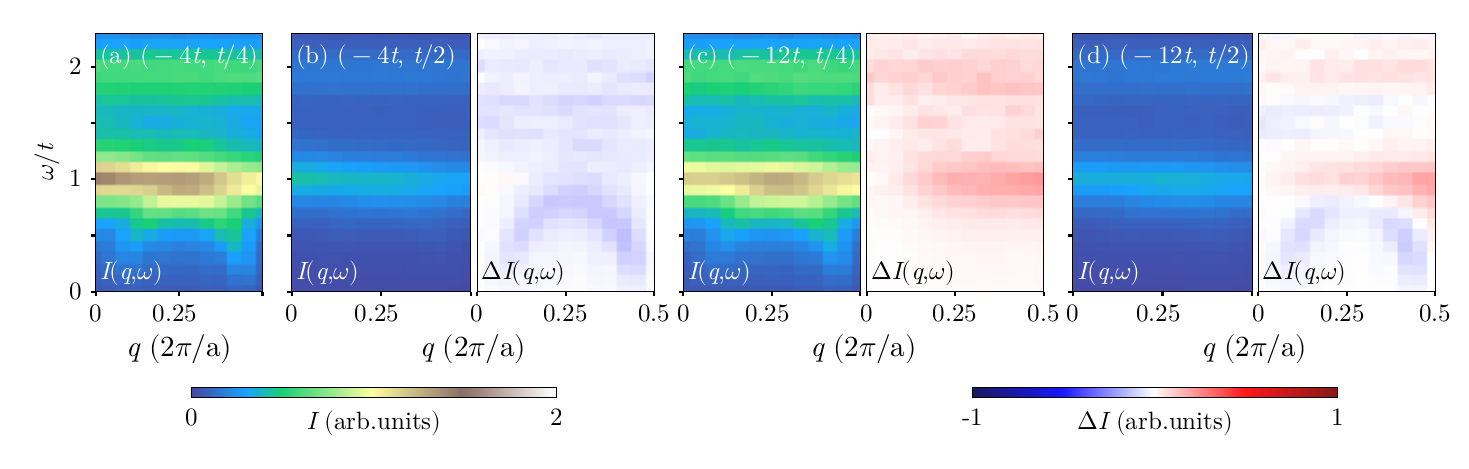}
\vspace{-1cm}
\caption{\gls*{rixs} spectra for $\lambda = 1/2$, $\lambda_\mathrm{CH} = 0$, $U = 8t$, $L = 24$, and different values of $(V_c,\Gamma/2)$, as indicated.~(a) shows the spectrum at $V_c = -4t$ and $\Gamma/2 = t/4$, which serves as our reference $I_\mathrm{ref}(q,\omega)$.~(b)-(d) RIXS spectra for various choices of $V_c$ and $\Gamma/2$. The plot on the left of these panels is the calculated RIXS spectra $I(q,\omega)$ plotted over half of the first Brillouin zone $q=[0,\pi/a]$. The plot on the right of each panel is a difference spectrum $\Delta I(q,\omega)= I(q,\omega)-I_\mathrm{ref}(q,\omega)$, which is obtained from subtracting the reference spectra shown in panel (a). When computing the difference plots, each spectrum was normalized to fix $I(q=0,\omega=\Omega)=1$ before taking the difference to focus on the relative (and sometimes subtle) changes in the magnetic and lattice intensities.}
\label{fig: heatmap_VcandGamma}
\end{figure*}

All transitions to the final states probed by \gls*{rixs} occur via an intermediate state with a core hole and excited core electron. The nature of this intermediate state has two significant consequences for exciting the lattice. First, the inverse core-hole lifetime $\Gamma/2$ limits the time available to the lattice to respond to changes in density created in the intermediate state~\cite{AmentEPL2011}. Second, the core-hole potential $V_c$ controls the degree to which the excited electron can delocalize~\cite{bieniasz2021}. To explore these effects, Fig.~\ref{fig: heatmap_VcandGamma} compares the \gls*{rixs} spectra for fixed $\lambda = 1/2$, $\lambda_\mathrm{CH} = 0$, and $U = 8t$ while varying $V_c$ and $\Gamma/2$. In each case, we tune the incident photon energy to the first peak in the \gls*{xas} spectra. For this discussion, we will use Fig.~\ref{fig: heatmap_VcandGamma}a (identical to Fig.~\ref{fig: RIXStwochannels}f) as a reference spectrum with $V_c = -4t$ and $\Gamma/2 = t/4$. We then examine how changes in $V_c$ and $\Gamma/2$ affect the spectra.

%%%%%%%%%%%%%%%%%
% What happens when core-hole lifetime is halved
%%%%%%%%%%%%%%%%%
Halving the core-hole lifetime (doubling $\Gamma/2$) while keeping $V_c$ fixed  (Fig.~\ref{fig: heatmap_VcandGamma}b) reduces the overall intensity of the spectra. This change is more rapid for the two-spinon excitations (as evident from the difference plot), which almost vanish relative to the phonon excitations. We can rationalize this behavior by considering the relative time scales of the excitations measured against the core-hole lifetime. An intermediate state doublon induces an instantaneous change in the local lattice potential via a displacive mechanism, which will begin generating phonons immediately after it is created. Conversely, the \gls*{sc} magnetic excitations involve ``slow'' double spin-flip processes~\cite{Bisogni2014femtosecond} and will be switched off if the core-hole lifetime is too short. We also find that the intensity of the two-phonon excitations decreases, consistent with the reduced time for lattice relaxation. 
Notably, the differences in the relative phonon intensities $I_2(q)/I_1(q)$ at $q = 0$ and $q = \pi/a$ persist for $\Gamma/2 = t/2$ (see Fig.~\ref{fig:MomentumDependence}b), indicating that the electron itinerancy still contributes to the lattice excitations in this case. 

%%%%%%%%%%%%%%%%%
% V_c is increased
%%%%%%%%%%%%%%%%%
Next, we increase the core-hole potential, which will tend to localize the intermediate state doublon. Setting $V_c=-12t$ while keeping $\Gamma/2 = t/4$, shown in Fig.~\ref{fig: heatmap_VcandGamma}c, redistributes the intensity of the phonons. We don't observe any significant changes in the spinon excitations here, while the intensity of the phonon excitations is enhanced relative to the spin contributions. This behavior is consistent with the picture that a more localized doublon can excite the lattice more effectively than it can induce double spin-flips. The intensity of phonon excitations also become more uniform in momentum space, which is consistent with the predictions of the \gls*{sslf} approach, indicating that $V_c = -12t$ is sufficient to localize the intermediate state doublon. 

Finally, strengthening the core-hole potential and increasing the inverse core-hole lifetime combines these effects (Fig.~\ref{fig: heatmap_VcandGamma}d). 

\subsection{\gls*{rixs} dependence on $U$} \label{sec: RIXSUdependence}

\begin{figure*}[t!]
\centering
\includegraphics[width=\textwidth]{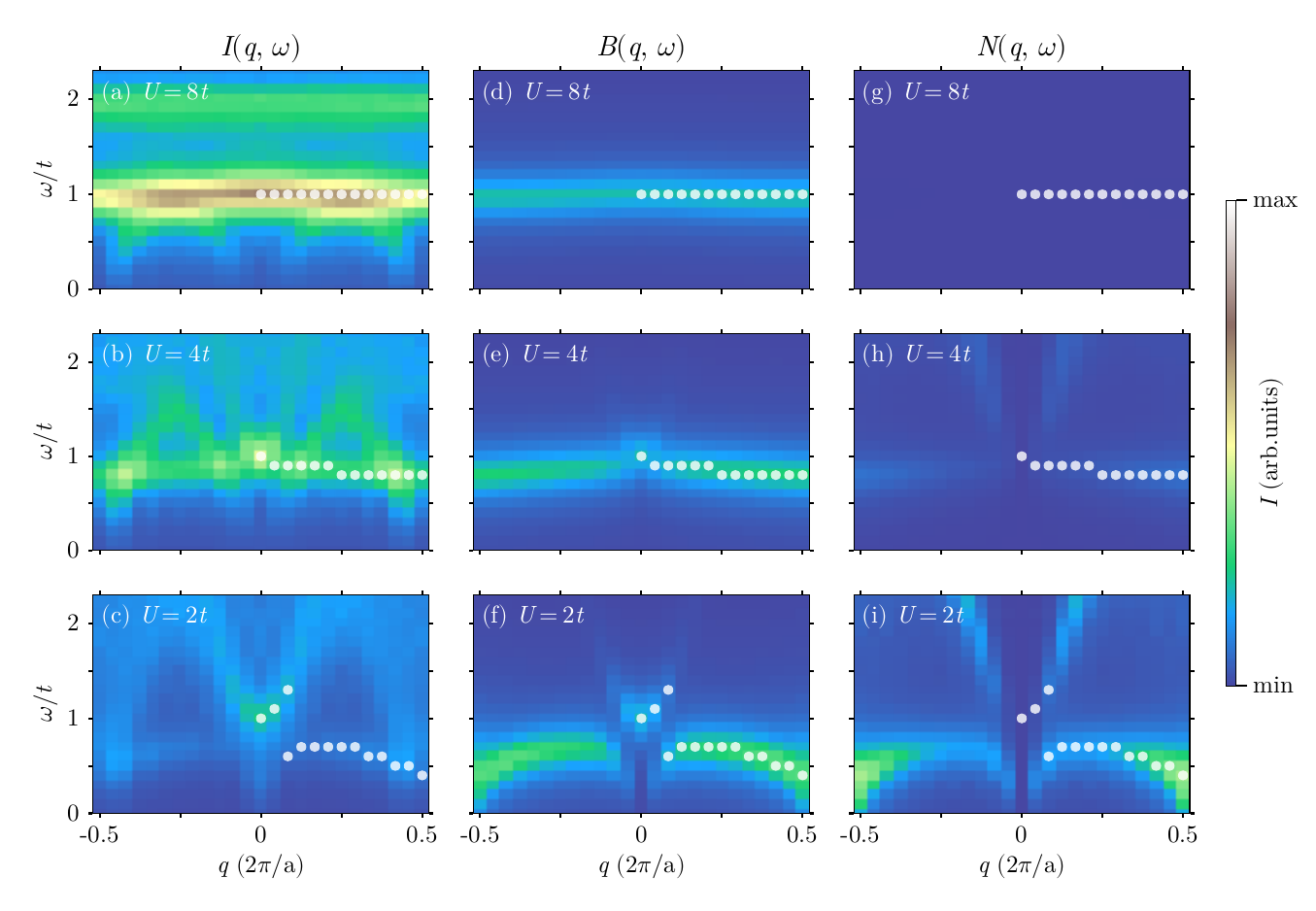}
\vspace{-0.75cm}
\caption{(a)-(c) \gls*{rixs} spectra in the \gls*{sc} channel as a function of $U$ and fixed $\lambda=1/2$, $\Omega = t$, $V_c=-4t$, $\Gamma/2=t/4$, and $\lambda_\mathrm{CH} = 0$. Panels (d)-(f) and (g)-(i) plot the corresponding phonon spectral functions $B(q,\omega)$ and dynamical charge structure factors $N(q,\omega)$, respectively. The top, middle, and bottom rows show results for $U = 8t$, $4t$, and $2t$, respectively. The white dots indicate the renormalized dispersion of the phonon, which was determined from the peak in the phonon spectral function shown in the middle column. We obtained all results on $L = 24$ site chains. The RIXS spectra track the normalized phonon dispersion in all cases, including any hybridization effects with the charge excitations.} 
\label{fig: U_dependence}
\end{figure*}

%%%%%%%%%%%%%%%%%
%Motivate U dependence
%%%%%%%%%%%%%%%%%
We nowxturn toases with weaker electronic correlations, wawhich favorslectron delocalization and allow cs mpeting \gls*{cdw} correlations to re-enter the problem. At half-filling, the charge correlations in the \gls*{1d} \gls*{hh} model are strongest at $q=2k_F=\pi/a$, which will soften the phonon dispersion at this wave vector. The charge fluctuations will also hybridize with the phonon modes near $q = 0$ if they extend low enough in energy~\cite{Weber2015phonon}. To explore these effects, we report the \gls{sc} \gls*{rixs} spectra as a function of $U$ in Fig.~\ref{fig: U_dependence} for the same parameters used in Fig.~\ref{fig: heatmap_VcandGamma}a. For reference, the second and third columns also show the corresponding phonon spectral functions $B(q,\omega)$ and dynamical charge structure factors $N(q,\omega)$, respectively. 

The $U = 8t$ case [Fig.~\ref{fig: U_dependence}(a,d,g)] has a Mott-insulating ground state. In the previous sections, we discussed its \gls*{rixs} spectra. The phonon spectral function closely tracks the bare phonon dispersion (Fig.~\ref{fig: U_dependence}d) and the charge excitations (Fig.~\ref{fig: U_dependence}g) are fully gapped. Reducing $U=4t$ [Fig.~\ref{fig: U_dependence}(b,e,h)] narrows the Mott gap. The charge excitations (Fig.~\ref{fig: U_dependence}h) thus extend to lower energies where they begin to overlap and hybridize with the phonons. The phonons soften slightly near the zone boundary and develop a weak kink-like feature near $q = 0$ due to this hybridization. Similarly, $N(q,\omega)$ acquires a weak dispersionless feature resembling the renormalized phonon branch. The \gls*{rixs} (Fig.~\ref{fig: U_dependence}b) spectra consist of overlapping contributions of the two-spinon, charge, and lattice excitations. In this case, the magnetic excitations extend to higher energies due to the increased exchange coupling $J = 4t^2/U$. The charge excitations also appear in the \gls*{rixs} spectra as an arc of intensities above the magnetic excitations. The first phonon harmonic is easily resolved and tracks the renormalized phonon dispersion (indicated by the white dots). We also see indications of a weak second phonon harmonic; however, its intensity is reduced significantly, and it is much broader in energy—the latter results from the phonon dispersion, which widens the two-phonon phase space~\cite{BieniaszPRB2022}. 

%%%%%%%%%%%%%%%%%
%Results for U=2t
%%%%%%%%%%%%%%%%%
Reducing $U=2t$ drives the system into a \gls*{lel} ground state with significant \gls*{cdw} correlations. The charge excitations (Fig.~\ref{fig: U_dependence}i) extend even lower in energy and strongly hybridize with the phonons, producing a clear and sharp break in phonon dispersion at small $q$ and Kohn anomaly at $q = \pi/a$. The magnetic excitations also extend to much higher energy. The first phonon excitation is extremely weak for this value of $U$, and we cannot resolve the second phonon harmonic.  

%%%%%%%%%%%%%%%%%
%Conclude
%%%%%%%%%%%%%%%%%
The $U$-dependence of our results shows that the dispersion of the phonons observed in \gls*{rixs} essentially tracks the renormalized dispersion, including any hybridization it may have with dispersive charge excitations. This result supports recent interpretations of the dispersive \gls*{cdw} excitations in two-dimensional cuprates~\cite{Chaix2017, LiPNMAS2020, Huang2021quantum}. The intensity of the phonon excitations also depends crucially on the degree of itinerancy in the system. The bare \gls*{eph} coupling used in Fig.~\ref{fig: U_dependence} was fixed to $\lambda = 1/2$. Nevertheless, the intensity of the first and second phonon excitations decreases substantially as $U$ decreases and develops a significant $\boldsymbol{q}$-dependence, particularly for smaller values of $U$. These effects will significantly impact the interpretation of experiments. For example, based on these results, we expect that any mobility introduced by doping an insulating phase will decrease the intensity of the phonon excitations independent of any additional screening effects~\cite{Johnston2010systematic}. Therefore, the strength of the \gls*{eph} coupling extracted from recent \gls*{rixs} measurements on, for example, doped cuprates may be  underestimated~\cite{PengPRB2022, BraicovichPRR2020, rossi2019experimental}. 

%%%%%%%%%%%%%%%%%
%Figure XAS with Core-hole
%%%%%%%%%%%%%%%%%
\begin{figure}[t!]
\centering
\includegraphics[width=\columnwidth]{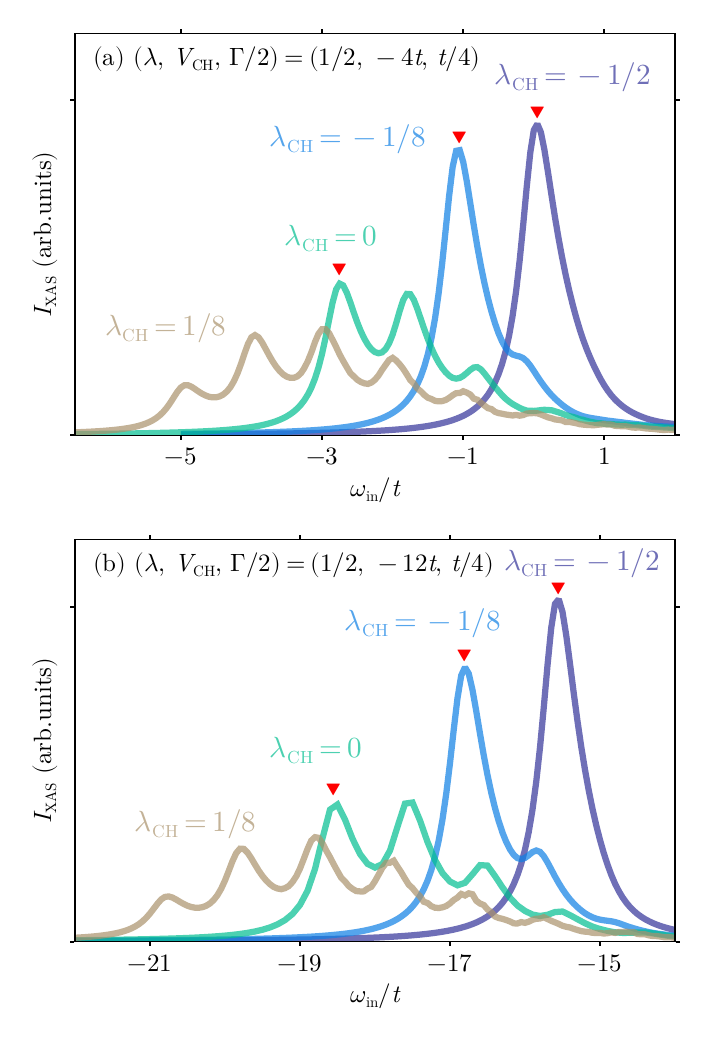}
\vspace{-0.75cm}
\caption{\gls*{xas} spectra for the Hubbard-Holstein model with an additional core-hole-lattice coupling $\lambda_\text{CH}$. 
Results were obtained for (a) $V_\text{CH} = -4t$ and (b) $V_\text{CH} = -12t$ while fixing $L = 24$, $U = 8t$, $\lambda = 1/2$, and $\Gamma/2 = t/4$. The inverted triangles indicate the incident photon energies of the \gls*{rixs} calculations shown in Fig.~\ref{fig: RIXStwochannels_with_corehole}. The \gls*{xas} results qualitatively resemble the spectra for a local coupling of $g_\mathrm{eff} = g+g_\mathrm{CH}$.}
\label{fig:XAS_with_corehole}
\end{figure}

\subsection{Core-hole-lattice coupling effects} \label{sec: RIXS Corehole Coupling}

Geondzhian and Gilmore~\cite{Gilmore_vibronic_2018} have recently argued that the core hole also couples to the lattice, leading to an effective exciton-lattice coupling in \gls*{rixs}'s intermediate state. If true, this would have important implications for interpreting experiments and extracting the magnitude of the valence band \gls*{eph} coupling from the data. In the \gls*{sslf} description, any core-hole-lattice interaction enters as an effective coupling $g_\mathrm{eff} = g+g_\mathrm{CH}$, which will enhance or suppress the phonon excitations depending on the relative sign of two coupling constants. However, the situation can be more complicated in an extended system. A core-hole-lattice coupling can promote or frustrate polaron formation at the core-hole site in the dilute limit, renormalizing the effective core-hole potential~\cite{bieniasz2021}. The core-hole-lattice interaction can thus enhance or suppress mobility effects in the intermediate state, depending on its sign. However, whether this result holds at finite densities or in correlated systems like those considered here is unclear. 

\begin{figure*}[t] 
\centering
\includegraphics[width=\textwidth]{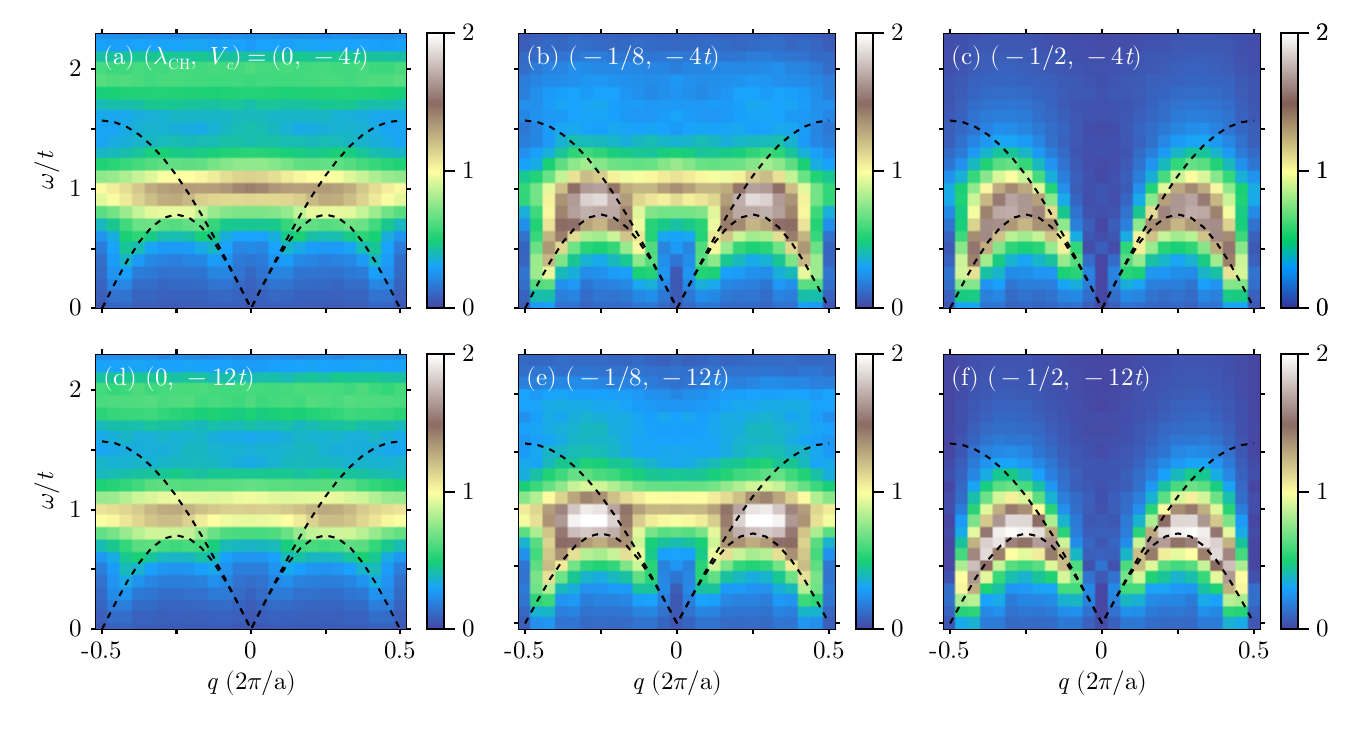} 
\vspace{-1cm}
\caption{\gls*{rixs} spectra in the spin conserving channel
as a function of the core-hole-lattice coupling and 
core-hole potential. The top row [panels (a) - (c)] shows results for $\lambda_{\mathrm{CH}}=0$, $-1/8$, and $-1/2$ and weak core-hole potential $V_c = -4t$. The bottom row [panels (d) - (f)] shows similar results for $V_c=-12t$. The black dashed lines in all panels are the boundaries of the two-spinon continuum expected for the Hubbard model ($\lambda=0$), assuming  $J=4t^2/U$. In all cases, we tuned the incident photon energies to the first peak in the \gls*{xas} spectra, as indicated by the inverted triangles in Fig.~\ref{fig:XAS_with_corehole}. The \gls*{rixs} results mostly follow the \gls*{xas} resembling a local coupling $g_\mathrm{eff} = g+g_\mathrm{CH}$. However, the phonon intensities show highly nontrivial effects for $V_c=-4t$, especially in (c), where they are completely suppressed at the zone center but still visible at the zone boundary. All results were obtained on $L = 24$ site chains with fixed $\lambda=1/2$, $U = 8t$, and $\Gamma/2=t/4$. }\label{fig: RIXStwochannels_with_corehole}
\end{figure*}

Motivated by this, we now examine these effects of setting $g_\mathrm{CH} \ne 0$ [see Eq.~\eqref{eq: h_corehole}]. Our model assumes that the core-hole-lattice coupling is local and is only active in the intermediate state. First, we consider scenarios where the core-hole coupling either augments ($g_\mathrm{CH}/g > 0$) or competes ($g_\mathrm{CH}/g < 0$) with the local valence band coupling; however, the latter is expected if \gls*{eph} coupling arises from an electrostatic interaction between the charge density and atomic positions. [If an electron in the valence band decreases the lattice displacement $(g > 0)$,  then a core hole should increase it $(g_\mathrm{CH} < 0)$.] As the dimensionless coupling constant $\lambda_\mathrm{CH}$ is positive by definition, we will append the sign of $g_\mathrm{CH}$ to distinguish the two scenarios [i.e., $\pm \lambda_\mathrm{CH}$ for $g_\mathrm{CH}/g > 0$ ($g_\mathrm{CH}/g < 0$)]. 

Figure~\ref{fig:XAS_with_corehole} shows \gls*{xas} results as a function of $\lambda_\mathrm{CH}$ for fixed $\lambda = 1/2$, $U = 8t$, and $\Gamma/2=t/4$. The additional core-hole coupling shifts the \gls*{ws} resonance. It redistributes the spectral weight of the phonon satellites in a way that is qualitatively consistent with the $g_\mathrm{eff} \approx g + g_\mathrm{CH}$ picture. However, we have observed quantitative and subtle qualitative differences when we compared the results shown here to spectra with an equivalent value of $g=g_\mathrm{eff}$ and $g_\mathrm{CH} = 0$ (not shown). These differences occur because $g_\mathrm{CH}$ only acts at the core-hole site and thus only affects local phonon creation while $t \ne 0$ opens up pathways for creating phonons at neighboring sites that do not couple to the core-hole. 

Figure~\ref{fig: RIXStwochannels_with_corehole} shows the corresponding \gls*{rixs} spectra. As mentioned, we expect $g_\mathrm{CH}/g < 0$ for any coupling arising from electrostatic interactions, so we focus on this regime. For a shallow potential $V_c=-4t$ and a weak core-hole-lattice coupling $\lambda_\mathrm{CH}=-1/8$ (Fig.~\ref{fig: RIXStwochannels_with_corehole}b), the spectra resemble the $\lambda=1/8$ case in Fig~\ref{fig: RIXStwochannels}d. The second phonon harmonic is suppressed as the local effective coupling $g_\mathrm{eff}$ is reduced; however, the momentum-dependence of this feature becomes much more pronounced with $I(q=\pi/a,\omega=\Omega)<I(q=0,\omega=\Omega)$. This result suggests that mobility effects are \textit{enhanced} by the core-hole-lattice coupling. Setting $\lambda_\mathrm{CH}=-1/2$, shown in Fig.~\ref{fig: RIXStwochannels_with_corehole}c, results in more substantial changes; the phonon excitations nearly vanish because the local effective coupling $g_\mathrm{eff}=0$. However, there is a remnant at $q = \pi/a$, resulting from phonon excitation on neighboring sites. 

When we increase the strength of the core-hole potential to $V_c=-12t$, the momentum dependence in the phonon excitation intensities is reduced, and spectra are more consistent with the predictions of the \gls*{sslf} model. For example, for $\lambda_\mathrm{CH}=-1/2$, shown in Fig.~\ref{fig: RIXStwochannels_with_corehole}f, the phonon excitations completely disappear, suggesting that the local effective coupling is zero and phonons are not being created on neighboring sites. 

%%%%%%%%%%%%%%%%%%%%%%%%%%%%%%%%%%%%%%%%%%%%%%%%%%%%%%%%%%%
% DISCUSSION
%%%%%%%%%%%%%%%%%%%%%%%%%%%%%%%%%%%%%%%%%%%%%%%%%%%%%%%%%%%

\section{Summary \& Conclusions}
We have presented a detailed study of the \gls*{xas} and \gls*{rixs} spectra of a strongly correlated \gls*{eph} coupled system defined on an extended lattice. Using the \gls*{dmrg} method, we obtained numerically exact results while treating all aspects of the problem (electron kinetic energy, many-body interactions, and core-hole interactions) on an equal footing. We find that including electron mobility leads to additional $\boldsymbol{q}$-modulations in the phonon excitations, which is not present in the microscopic \gls*{eph} coupling constant. While the strength of these modulations appears to decrease with increasing $U$, they are still present deep in the Mott insulating regime. 

Working on extended clusters has also allowed us to understand how many-body renormalizations and additional core-hole-lattice interactions shape the resulting spectra. For example, \gls*{cdw} correlations appearing for smaller values of $U$ significantly renormalized the phonon dispersion. The \gls*{rixs} spectra reflect this renormalized dispersion, which lends support to the interpretation of several \gls*{rixs} experiments where anomalous softening has been observed in proximity to \gls*{cdw} wave order~\cite{Chaix2017, LiPNMAS2020, Huang2021quantum}. Coupling between the lattice and the core hole was also found to modify the spectra in nontrivial ways, which will need to be accounted for. 

Notably, the strength of many of these effects strongly depended on the magnitude of the core-hole potential; deep potentials tend to produce more localized intermediate states and bring our predictions more in line with the \gls*{sslf} model. Therefore, we expect these effects to be more pronounced at, for example, ligand $K$-edges than TM $L$-edges. Based on these observations, it may be possible to design an experimental protocol to contrast results obtained at different edges to unravel contributions from the intrinsic \gls*{eph} coupling, potential core-hole-lattice coupling, and electron delocalization. 

Our results demonstrate that quantitative information about the \gls*{eph} coupling in correlated quantum materials cannot be directly obtained from data as implied by the \gls*{sslf} framework. For example, we believe that recent estimates~\cite{PengPRB2022, rossi2019experimental, BraicovichPRR2020} for the \gls*{eph} coupling extracted from measurements of doped high-$T_\mathrm{c}$ cuprates are likely underestimated. Instead, this analysis will require detailed modeling, particularly when the core-hole potential is weak. Nevertheless, our \gls*{dmrg} framework sets the stage for a more detailed analysis of \gls*{eph} interactions in quasi-\gls*{1d} systems like the cuprate ladders and spin chains. Several experiments have already reported the observation of the Cu-O bond stretching phonons in these materials~\cite{LeePRL2013, JohnstonNatComm2016, Schlappa2018, Tseng22}. A recent ARPES study of the doped cuprate Ba$_{2-x}$Sr$_x$CuO$_{3+\delta}$~\cite{ChenScience2021} has inferred a sizable next-nearest-neighbor attraction that has been attributed to long-range \gls*{eph} coupling~\cite{Wang2021phonon}. If true, this interaction should produce one or more phonon harmonics in the \gls*{rixs} spectra, which can be simulated directly with our methods. These systems thus offer a unique opportunity for quantitative comparisons between theory and experiment. Finally, our work will stimulate future theoretical investigations of the \gls*{eph} coupling in time-resolved \gls*{rixs} experiments, either using the Keldysh~\cite{chen2019theory} or time-dependent scattering~\cite{zawadzki2023time} approaches. 

%%%%%%%%%%%%%%%%%%%%%%%%%%%%%%%%%%%%%%%%%%%%%%%%%%%%%%%%%%%
% ACKNOWLEDGEMENTS & BIBLIOGRAPHY
%%%%%%%%%%%%%%%%%%%%%%%%%%%%%%%%%%%%%%%%%%%%%%%%%%%%%%%%%%%
\section*{Acknowledgments} 
We thank M. P. M. Dean and A. F. Kemper for valuable discussions. This work is supported by the National Science Foundation under Grant No.~DMR-1842056. A. N. acknowledges the support of the Canada First Research Excellence Fund. This work used computational resources and services provided by Advanced Research Computing at the University of British Columbia.

%%%%%%%%%%%%%%%%%%%%%%%%%%%%%%%%%%%%%%%%%%%%%%%%%%%%%%%%%%%
%%%%%%%%%%%%%%%%%%%%%%%%%%%%%%%%%%%%%%%%%%%%%%%%%%%%%%%%%%%
% APPENDICES
%%%%%%%%%%%%%%%%%%%%%%%%%%%%%%%%%%%%%%%%%%%%%%%%%%%%%%%%%%%
%%%%%%%%%%%%%%%%%%%%%%%%%%%%%%%%%%%%%%%%%%%%%%%%%%%%%%%%%%%
\appendix
\section{Lang-Firsov Limit}\label{app:LF}
%%%%%%%%%%%%%%%%%
%The Hamiltonian after LF transformation
%%%%%%%%%%%%%%%%%
This section briefly reviews the single-site framework for analyzing \gls*{rixs} data~\cite{AmentEPL2011}. The starting point is the local Hamiltonian
\begin{align}
    \label{Single-Site}
    \mathcal{H}_{\mathrm{local}} = U \hat{n}_{\uparrow}\hat{n}_{\downarrow} + \Omega  \hat{b}^\dagger \hat{b}^{\phantom\dagger} + g  \hat{n} (\hat{b}^\dagger+\hat{b}). 
\end{align}
We have assumed that the localized electrons couple to dispersionless Einstein oscillators via a local coupling to the total density. In contrast, the derivation in Ref.~\cite{AmentEPL2011} assumes a coupling to the change in electron density in the intermediate state, i.e. $H_{e-\mathrm{ph}}=g\Delta n(b^\dagger+b)$, where $\Delta n = (n-\langle n \rangle)$ and $\langle n \rangle$ is the electron density in the ground state. The two forms for the \gls*{eph} interaction can be mapped to one another via a simple shift of the lattice equilibrium position in the atomic limit. 

One can diagonalize Eq.~\eqref{Single-Site} by applying the canonical \gls*{lf} transformation $\mathcal{H}_{\mathrm{LF}}=e^S \mathcal{H}_{\mathrm{local}} e^{-S}$, where $S =  \frac{g}{\Omega} \hat{n} (\hat{b}^\dagger-\hat{b}^{\phantom\dagger})$, to obtain  
\begin{align}
      \label{eq: LF-Single Site}
      \mathcal{H}_{\mathrm{LF}} = U \hat{n}_{\uparrow}\hat{n}_{\downarrow} + \Omega  \hat{b}^\dagger \hat{b}- \frac{g^2}{\Omega}  \hat{n}^2 .
\end{align}
Eq.~\eqref{eq: LF-Single Site} must be supplemented with an additional core-hole term $\mathcal{H}^\mathrm{CH}$
when computing the intermediate states of the \gls*{rixs} process, which results in a trivial shift of the eigenvalues $E_n\rightarrow E_n + 2V_c$. 

%%%%%%%%%%%%%%%%%
%RIXS cross-section using Franck-Condon Factors. 
%%%%%%%%%%%%%%%%%
The corresponding scattering amplitude to transition from the initial state $i$ with zero phonons 
to the final state $f$ with $M$ phonons is given by~\cite{AmentEPL2011}
\begin{align}
     F_{fi} = \sum_{M^\prime=0}^\infty \frac{B_{\text{max}(M,M^\prime),\text{min}(M,M^\prime)}(\frac{g}{\Omega})B_{M^\prime,0}(\frac{g}{\Omega})}{ \frac{4g^2}{\Omega} - M^\prime \Omega -U - 2V_c + \omega_\text{in }+ \mathrm{i}\Gamma/2}.
\end{align} 
Here, $M^\prime$ is the number of phonons in the intermediate state and 
\begin{align*}
   B_{M,M^\prime} (x) = {e^{-\tfrac{x^2}{2}}\sqrt{M!M^\prime!}} \sum_{l=0}^{M^\prime} \frac{{(-1)^{M+l}} \ x^{(2l + M - M^\prime)}}{(M^\prime-l)!  \ l!  \ (M -M^\prime+1)!}
\end{align*} 
is the Franck-Condon overlap factor. We obtain the total \gls*{rixs} intensity by summing over final states using~Eq.~\eqref{rixs_regular}.

\section{DMRG formalism for RIXS}\label{app:dmrgrixs}
%%%%%%%%%%%%%%%%%
%RIXS and XAS Formulation in DMRG
%%%%%%%%%%%%%%%%%
Our approach is based on the Krylov space correction vector algorithm~\cite{Nocera_Alvarez_2016}, which we briefly outline here. The first step is to expand the square of the matrix elements in Eq.~\eqref{scattering_amp} to reformulate 
the problem using correction vectors~\cite{PhysRevB.60.335, PhysRevB.66.045114} 
\begin{align}
\centering
\label{rixsdmrg}
I(\bq,\omega) &\propto -\mathrm{Im} \left[\sum_{j=1}^{N}\sum e^{\mathrm{i}\bq \cdot(\boldsymbol{R}_{j}-\boldsymbol{R}_c)}\mathcal{I}_{j,\sigma, \sigma^\prime,\gamma, \gamma^\prime}\right], 
\end{align} 
where 
\begin{align}\label{eq:DMRG_2}
\mathcal{I}_{j,\sigma, \sigma^\prime,\gamma, \gamma^\prime} &= \bra{\Psi_{j,\gamma}} D_{j,\gamma^{\prime}}^{\phantom\dagger} G D_{c,\sigma^{\prime}}^{\dagger} \ket{\Psi_{c,\sigma}}  
\end{align}
and
\begin{align}\label{eq:DMRG_3}
\ket{\Psi_{j,\sigma}} = \mathcal{G}_{j}^{\phantom\dagger} D_{j,\sigma}^{\phantom\dagger}  \ket{i}. 
\end{align}
Here, $\ket{i}$ is the ground state, $\sigma, \sigma^\prime, \gamma,$ and $\gamma^\prime$ are spin indices, $c$ denotes a site in the center of the cluster, and 
\begin{align}\label{eq:Gintermediate}    
\mathcal{G}_{j}^{\phantom\dagger} &=  \frac{1}{E_{i}-\hat{\mathcal{H}}_{j}+\omega_{\mathrm{in}}+\mathrm{i}\Gamma/2},
\end{align}
and
\begin{align}\label{eq:Gfinal}
G &=  \frac{1}{E_{i}-\hat{\mathcal{H}}+\omega+\mathrm{i}\eta}
\end{align} 
are the electron Green's function with and without a core-hole, respectively. The intermediate state Hamiltonian $\mathcal{H}_j$ in Eq.~\eqref{eq:Gintermediate} includes the interaction with the core-hole created at site $j$. The broadening parameter $\eta$ in Eq.~\eqref{eq:Gfinal} sets the resolution in energy loss $\omega$ and is typically set to the instrument broadening.

The \gls*{xas} spectra are similarly formulated in the correction-vector method as
\begin{equation}\label{eq:XAS_CV}
I_{\mathrm{XAS}}(\omega_{\mathrm{in}}) \propto -\mathrm{{Im}}\langle i|D_{c,\sigma}^{\dagger}\mathcal{G}_{c}D_{c,\sigma}|i\rangle.
\end{equation}

\section{Dynamical correlation functions} \label{app:spectralfunctions}
In the main text, we show results for the dynamical charge structure factor and phonon spectral function, which are calculated using the correction vector method. The real-space charge correlation function is given by
\begin{equation}\label{eq:Nrw}
    N_{j,c}(\omega) = -\frac{1}{\pi} \operatorname{Im} [\bra{i}\Tilde{n}_j G \Tilde{n}_c\ket{i}].
\end{equation}
The phonon spectral function is given by
\begin{equation}\label{eq:Brw}
    B_{j,c}(\omega) = -\frac{1}{\pi} \operatorname{Im} [\bra{i}\Tilde{X}_j G\Tilde{X}_c \ket{i}].
\end{equation}
Here, $\ket{i}$ is the ground state and $\Tilde{n}_c = (\hat{n}_c-\langle \hat{n}_c \rangle)$ and $\Tilde{X}_c = (\hat{X}_c-\langle \hat{X}_c \rangle)$, where $c$ is the center site of the \gls*{1d} chain. We then compute the momentum space dynamical correlation functions by Fourier transform~\cite{Nocera_Alvarez_2016}. 

\section{Convergence tests}\label{app:convergence}

\begin{figure}[t]
\centering
\includegraphics[width=\columnwidth]{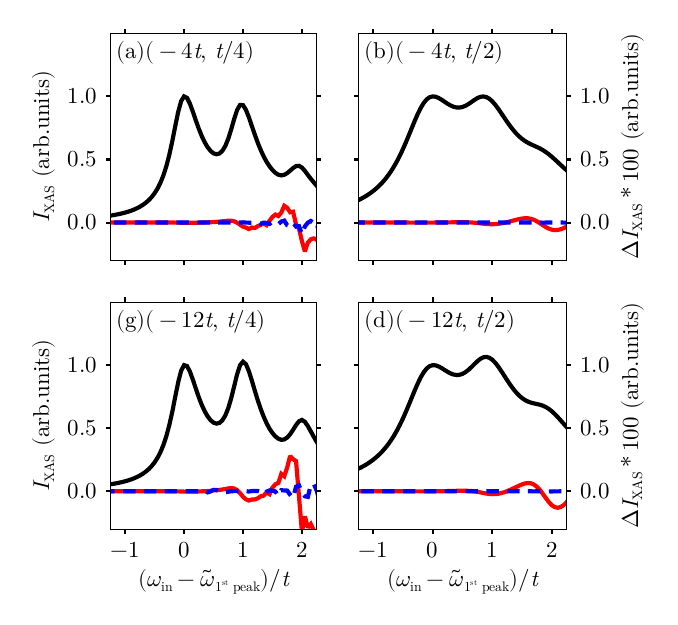}
\vspace{-1cm}
\caption{Convergence tests of the \gls*{xas} spectra obtained on an $L = 24$ site chain with $U = 8t$, $\lambda = 0.5$, $\lambda_\mathrm{CH} = 0$. The solid black line is a reference \gls*{xas} spectrum computed 
for different $(V_c,\Gamma/2)$ values as indicated in each panel. We kept a local phonon Hilbert space of $M = 16$ and $m = 500$ \gls*{dmrg} states for the reference case. The red (blue) lines show the difference of the \gls*{xas} spectra calculated with $M = 17$, $m = 500$ ($M = 16$, $m=600$), and the reference spectrum is shown in each panel.}
\label{fig:xas_convergence}
\end{figure}

We have performed extensive convergence tests for our calculations. Here, we show representative results for a select few parameter regimes. 

Figure~\ref{fig:xas_convergence} shows \gls*{xas} spectra calculated on $L = 24$ site chains with $U = 8t$, $\lambda = 0.5$, $ \lambda_\mathrm{CH} = 0$ and different values of $(V_c,\Gamma/2)$, as indicated in each panel. Each panel shows a reference \gls*{xas} spectrum computed after truncating the local phonon Hilbert space to $M = 16$ phonons per site and keeping $m = 500$ \gls*{dmrg} states. The red and blue lines show the difference between the reference spectrum and analogous \gls*{xas} spectra calculated with ($M = 17$, $m = 500$) and ($M = 16$, $m=600$), respectively. (Note that we have multiplied the difference curves by $100$ to place them on a scale similar to the \gls*{xas} spectra.) These results indicate that our spectra are well converged with respect to the size of the local phonon Hilbert space and the number of \gls*{dmrg} states. 

\begin{figure}[b]
\centering
\includegraphics[width=\columnwidth]{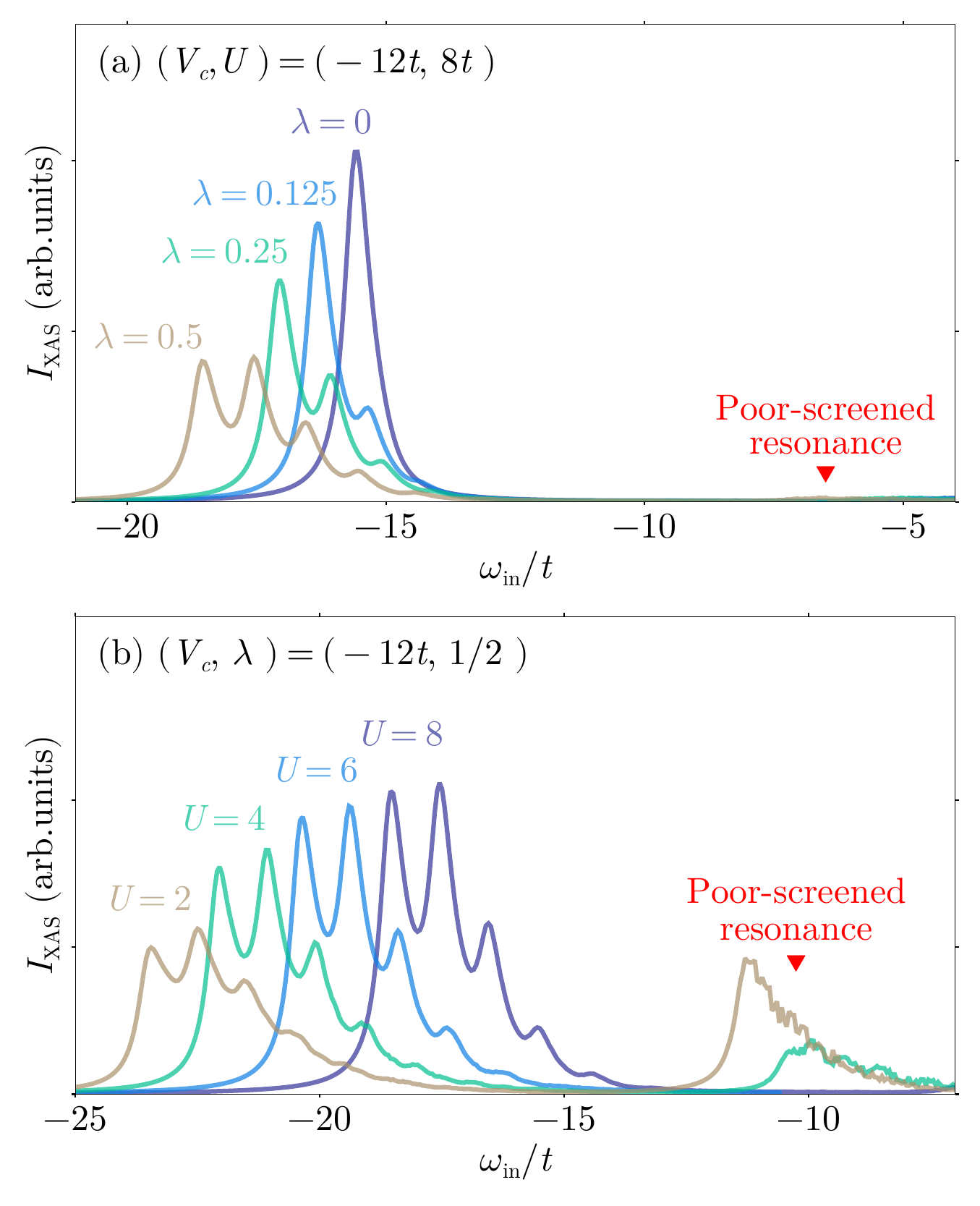}
\vspace{-0.75cm}
\caption{Additional \gls*{xas} results obtained for a strong core-hole potential $V_c = -12t$ and inverse core-hole lifetime $\Gamma/2 = t/4$. (a) The dependence of the \gls*{xas} spectra as a function of the dimensionless \gls*{eph} coupling $\lambda$ for a fixed $U=8t$ and $\lambda_\mathrm{CH} = 0$. (b) The dependence of the \gls*{xas} spectra as a function of $U$ for fixed $\lambda=0.5$ and $\lambda_\mathrm{CH} = 0$. The weight of the \gls*{ps} resonance is extremely small in the \gls*{mi} limit and is enhanced for smaller $U$ values without generating any harmonics.}
\label{fig:XAS_Vcminus12}
\end{figure}

\section{Additional XAS results}\label{app:xas_vcm12}
Figure~\ref{fig:XAS_Vcminus12} plots additional \gls*{xas} results for a strong core-hole potential $V_c=-12t$. We have fixed the core-hole lifetime to $\Gamma/2=t/4$ and plotted results over a wider range of incident photon energies covering the \gls*{ps} resonance. 
The results indicate that only the \gls*{ws} resonance develops strong phonon satellites once the \gls*{eph} coupling is introduced. 

\bibliography{references}

\end{document}

%% file: glossary.tex
\newacronym{xas}{XAS}{x-ray absorption spectra}
\newacronym{rixs}{RIXS}{resonant inelastic x-ray scattering}
\newacronym{ins}{INS}{inelastic neutron scattering}
\newacronym{dmrg}{DMRG}{density matrix renormalization group}
\newacronym{ed}{ED}{exact diagonalization}
\newacronym{cdw}{CDW}{charge density wave}
\newacronym{ucl}{UCL}{ultrashort core-hole lifetime}
\newacronym{sc}{SC}{spin conserving}
\newacronym{nsc}{NSC}{non-spin conserving}
\newacronym{sqw}{$S(\vec{q},\omega$)}{dynamical spin structure factor}
\newacronym{nqw}{$\tilde{N}(\vec{q},\omega$)}{dynamical charge structure factor}
\newacronym{bqw}{$B(\vec{q},\omega$)}{phonon spectral function}
\newacronym{gs}{GS}{ground state}
\newacronym{hh}{HH}{Hubbard-Holstein}
\newacronym{soc}{SOC}{spin-orbit coupling}
\newacronym{eph}{$e$-ph}{electron-phonon}
\newacronym{afm}{AFM}{antiferromagnetic}
\newacronym{kh}{KH}{Kramers-Heisenberg}
\newacronym{mi}{MI}{Mott insulating limit}
\newacronym{ee}{$e$-$e$}{electron-electron}
\newacronym{oqw}{$O(\vec{q},\omega$)}{spin-exchange dynamical structure factor}
\newacronym{ws}{WS}{``well-screened''}
\newacronym{ps}{PS}{``poor-screened''}
\newacronym{lel}{LEL}{Luther-Emery liquid}
\newacronym{lf}{LF}{Lang-Firsov}
\newacronym{sslf}{SS-LF}{single-site Lang-Firsov}
\newacronym{2d}{2D}{two-dimensional}
\newacronym{1d}{1D}{one-dimensional}
\newacronym{ssh}{SSH}{Su–Schrieffer–Heeger}

%% file: main.bbl
\begin{thebibliography}{73}%
\makeatletter
\providecommand \@ifxundefined [1]{%
 \@ifx{#1\undefined}
}%
\providecommand \@ifnum [1]{%
 \ifnum #1\expandafter \@firstoftwo
 \else \expandafter \@secondoftwo
 \fi
}%
\providecommand \@ifx [1]{%
 \ifx #1\expandafter \@firstoftwo
 \else \expandafter \@secondoftwo
 \fi
}%
\providecommand \natexlab [1]{#1}%
\providecommand \enquote  [1]{``#1''}%
\providecommand \bibnamefont  [1]{#1}%
\providecommand \bibfnamefont [1]{#1}%
\providecommand \citenamefont [1]{#1}%
\providecommand \href@noop [0]{\@secondoftwo}%
\providecommand \href [0]{\begingroup \@sanitize@url \@href}%
\providecommand \@href[1]{\@@startlink{#1}\@@href}%
\providecommand \@@href[1]{\endgroup#1\@@endlink}%
\providecommand \@sanitize@url [0]{\catcode `\\12\catcode `\$12\catcode `\&12\catcode `\#12\catcode `\^12\catcode `\_12\catcode `\%12\relax}%
\providecommand \@@startlink[1]{}%
\providecommand \@@endlink[0]{}%
\providecommand \url  [0]{\begingroup\@sanitize@url \@url }%
\providecommand \@url [1]{\endgroup\@href {#1}{\urlprefix }}%
\providecommand \urlprefix  [0]{URL }%
\providecommand \Eprint [0]{\href }%
\providecommand \doibase [0]{https://doi.org/}%
\providecommand \selectlanguage [0]{\@gobble}%
\providecommand \bibinfo  [0]{\@secondoftwo}%
\providecommand \bibfield  [0]{\@secondoftwo}%
\providecommand \translation [1]{[#1]}%
\providecommand \BibitemOpen [0]{}%
\providecommand \bibitemStop [0]{}%
\providecommand \bibitemNoStop [0]{.\EOS\space}%
\providecommand \EOS [0]{\spacefactor3000\relax}%
\providecommand \BibitemShut  [1]{\csname bibitem#1\endcsname}%
\let\auto@bib@innerbib\@empty
%</preamble>
\bibitem [{\citenamefont {Mitrano}\ \emph {et~al.}(2024)\citenamefont {Mitrano}, \citenamefont {Johnston}, \citenamefont {Kim},\ and\ \citenamefont {Dean}}]{Mitriano2024exploring}%
  \BibitemOpen
  \bibfield  {author} {\bibinfo {author} {\bibfnamefont {M.}~\bibnamefont {Mitrano}}, \bibinfo {author} {\bibfnamefont {S.}~\bibnamefont {Johnston}}, \bibinfo {author} {\bibfnamefont {Y.-J.}\ \bibnamefont {Kim}},\ and\ \bibinfo {author} {\bibfnamefont {M.~P.~M.}\ \bibnamefont {Dean}},\ }\bibfield  {title} {\bibinfo {title} {Exploring quantum materials with resonant inelastic x-ray scattering},\ }\href {https://doi.org/10.1103/PhysRevX.14.040501} {\bibfield  {journal} {\bibinfo  {journal} {Phys. Rev. X}\ }\textbf {\bibinfo {volume} {14}},\ \bibinfo {pages} {040501} (\bibinfo {year} {2024})}\BibitemShut {NoStop}%
\bibitem [{\citenamefont {Ament}\ \emph {et~al.}(2011)\citenamefont {Ament}, \citenamefont {van Veenendaal},\ and\ \citenamefont {van~den Brink}}]{AmentEPL2011}%
  \BibitemOpen
  \bibfield  {author} {\bibinfo {author} {\bibfnamefont {L.~J.~P.}\ \bibnamefont {Ament}}, \bibinfo {author} {\bibfnamefont {M.}~\bibnamefont {van Veenendaal}},\ and\ \bibinfo {author} {\bibfnamefont {J.}~\bibnamefont {van~den Brink}},\ }\bibfield  {title} {\bibinfo {title} {Determining the electron-phonon coupling strength from resonant inelastic x-ray scattering at transition metal {L}-edges},\ }\href {https://doi.org/10.1209/0295-5075/95/27008} {\bibfield  {journal} {\bibinfo  {journal} {{EPL} (Europhysics Letters)}\ }\textbf {\bibinfo {volume} {95}},\ \bibinfo {pages} {27008} (\bibinfo {year} {2011})}\BibitemShut {NoStop}%
\bibitem [{\citenamefont {Gilmore}(2023)}]{Gilmore2023quantifying}%
  \BibitemOpen
  \bibfield  {author} {\bibinfo {author} {\bibfnamefont {K.}~\bibnamefont {Gilmore}},\ }\bibfield  {title} {\bibinfo {title} {Quantifying vibronic coupling with resonant inelastic x-ray scattering},\ }\href {https://doi.org/10.1039/D2CP00968D} {\bibfield  {journal} {\bibinfo  {journal} {Phys. Chem. Chem. Phys.}\ }\textbf {\bibinfo {volume} {25}},\ \bibinfo {pages} {217} (\bibinfo {year} {2023})}\BibitemShut {NoStop}%
\bibitem [{\citenamefont {Hancock}\ \emph {et~al.}(2010)\citenamefont {Hancock}, \citenamefont {Chabot-Couture},\ and\ \citenamefont {Greven}}]{Hancock2010}%
  \BibitemOpen
  \bibfield  {author} {\bibinfo {author} {\bibfnamefont {J.~N.}\ \bibnamefont {Hancock}}, \bibinfo {author} {\bibfnamefont {G.}~\bibnamefont {Chabot-Couture}},\ and\ \bibinfo {author} {\bibfnamefont {M.}~\bibnamefont {Greven}},\ }\bibfield  {title} {\bibinfo {title} {Lattice coupling and {F}ranck{\textendash}{C}ondon effects in {K}-edge resonant inelastic x-ray scattering},\ }\href {https://doi.org/10.1088/1367-2630/12/3/033001} {\bibfield  {journal} {\bibinfo  {journal} {New Journal of Physics}\ }\textbf {\bibinfo {volume} {12}},\ \bibinfo {pages} {033001} (\bibinfo {year} {2010})}\BibitemShut {NoStop}%
\bibitem [{\citenamefont {Yava{\c{s}}}\ \emph {et~al.}(2010)\citenamefont {Yava{\c{s}}}, \citenamefont {van Veenendaal}, \citenamefont {van~den Brink}, \citenamefont {Ament}, \citenamefont {Alatas}, \citenamefont {Leu}, \citenamefont {Apostu}, \citenamefont {Wizent}, \citenamefont {Behr}, \citenamefont {Sturhahn}, \citenamefont {Sinn},\ and\ \citenamefont {Alp}}]{Yavas2010}%
  \BibitemOpen
  \bibfield  {author} {\bibinfo {author} {\bibfnamefont {H.}~\bibnamefont {Yava{\c{s}}}}, \bibinfo {author} {\bibfnamefont {M.}~\bibnamefont {van Veenendaal}}, \bibinfo {author} {\bibfnamefont {J.}~\bibnamefont {van~den Brink}}, \bibinfo {author} {\bibfnamefont {L.~J.~P.}\ \bibnamefont {Ament}}, \bibinfo {author} {\bibfnamefont {A.}~\bibnamefont {Alatas}}, \bibinfo {author} {\bibfnamefont {B.~M.}\ \bibnamefont {Leu}}, \bibinfo {author} {\bibfnamefont {M.-O.}\ \bibnamefont {Apostu}}, \bibinfo {author} {\bibfnamefont {N.}~\bibnamefont {Wizent}}, \bibinfo {author} {\bibfnamefont {G.}~\bibnamefont {Behr}}, \bibinfo {author} {\bibfnamefont {W.}~\bibnamefont {Sturhahn}}, \bibinfo {author} {\bibfnamefont {H.}~\bibnamefont {Sinn}},\ and\ \bibinfo {author} {\bibfnamefont {E.~E.}\ \bibnamefont {Alp}},\ }\bibfield  {title} {\bibinfo {title} {Observation of phonons with resonant inelastic x-ray scattering},\ }\href {https://doi.org/10.1088/0953-8984/22/48/485601} {\bibfield  {journal} {\bibinfo  {journal} {Journal of
  Physics: Condensed Matter}\ }\textbf {\bibinfo {volume} {22}},\ \bibinfo {pages} {485601} (\bibinfo {year} {2010})}\BibitemShut {NoStop}%
\bibitem [{\citenamefont {Lee}\ \emph {et~al.}(2013)\citenamefont {Lee}, \citenamefont {Johnston}, \citenamefont {Moritz}, \citenamefont {Lee}, \citenamefont {Yi}, \citenamefont {Zhou}, \citenamefont {Schmitt}, \citenamefont {Patthey}, \citenamefont {Strocov}, \citenamefont {Kudo}, \citenamefont {Koike}, \citenamefont {van~den Brink}, \citenamefont {Devereaux},\ and\ \citenamefont {Shen}}]{LeePRL2013}%
  \BibitemOpen
  \bibfield  {author} {\bibinfo {author} {\bibfnamefont {W.~S.}\ \bibnamefont {Lee}}, \bibinfo {author} {\bibfnamefont {S.}~\bibnamefont {Johnston}}, \bibinfo {author} {\bibfnamefont {B.}~\bibnamefont {Moritz}}, \bibinfo {author} {\bibfnamefont {J.}~\bibnamefont {Lee}}, \bibinfo {author} {\bibfnamefont {M.}~\bibnamefont {Yi}}, \bibinfo {author} {\bibfnamefont {K.~J.}\ \bibnamefont {Zhou}}, \bibinfo {author} {\bibfnamefont {T.}~\bibnamefont {Schmitt}}, \bibinfo {author} {\bibfnamefont {L.}~\bibnamefont {Patthey}}, \bibinfo {author} {\bibfnamefont {V.}~\bibnamefont {Strocov}}, \bibinfo {author} {\bibfnamefont {K.}~\bibnamefont {Kudo}}, \bibinfo {author} {\bibfnamefont {Y.}~\bibnamefont {Koike}}, \bibinfo {author} {\bibfnamefont {J.}~\bibnamefont {van~den Brink}}, \bibinfo {author} {\bibfnamefont {T.~P.}\ \bibnamefont {Devereaux}},\ and\ \bibinfo {author} {\bibfnamefont {Z.~X.}\ \bibnamefont {Shen}},\ }\bibfield  {title} {\bibinfo {title} {Role of lattice coupling in establishing electronic and magnetic
  properties in quasi-one-dimensional cuprates},\ }\href {https://doi.org/10.1103/PhysRevLett.110.265502} {\bibfield  {journal} {\bibinfo  {journal} {Phys. Rev. Lett.}\ }\textbf {\bibinfo {volume} {110}},\ \bibinfo {pages} {265502} (\bibinfo {year} {2013})}\BibitemShut {NoStop}%
\bibitem [{\citenamefont {Lee}\ \emph {et~al.}(2014)\citenamefont {Lee}, \citenamefont {Moritz}, \citenamefont {Lee}, \citenamefont {Yi}, \citenamefont {Jia}, \citenamefont {Sorini}, \citenamefont {Kudo}, \citenamefont {Koike}, \citenamefont {Zhou}, \citenamefont {Monney}, \citenamefont {Strocov}, \citenamefont {Patthey}, \citenamefont {Schmitt}, \citenamefont {Devereaux},\ and\ \citenamefont {Shen}}]{LeePRB2015}%
  \BibitemOpen
  \bibfield  {author} {\bibinfo {author} {\bibfnamefont {J.~J.}\ \bibnamefont {Lee}}, \bibinfo {author} {\bibfnamefont {B.}~\bibnamefont {Moritz}}, \bibinfo {author} {\bibfnamefont {W.~S.}\ \bibnamefont {Lee}}, \bibinfo {author} {\bibfnamefont {M.}~\bibnamefont {Yi}}, \bibinfo {author} {\bibfnamefont {C.~J.}\ \bibnamefont {Jia}}, \bibinfo {author} {\bibfnamefont {A.~P.}\ \bibnamefont {Sorini}}, \bibinfo {author} {\bibfnamefont {K.}~\bibnamefont {Kudo}}, \bibinfo {author} {\bibfnamefont {Y.}~\bibnamefont {Koike}}, \bibinfo {author} {\bibfnamefont {K.~J.}\ \bibnamefont {Zhou}}, \bibinfo {author} {\bibfnamefont {C.}~\bibnamefont {Monney}}, \bibinfo {author} {\bibfnamefont {V.}~\bibnamefont {Strocov}}, \bibinfo {author} {\bibfnamefont {L.}~\bibnamefont {Patthey}}, \bibinfo {author} {\bibfnamefont {T.}~\bibnamefont {Schmitt}}, \bibinfo {author} {\bibfnamefont {T.~P.}\ \bibnamefont {Devereaux}},\ and\ \bibinfo {author} {\bibfnamefont {Z.~X.}\ \bibnamefont {Shen}},\ }\bibfield  {title} {\bibinfo {title}
  {Charge-orbital-lattice coupling effects in the $dd$ excitation profile of one-dimensional cuprates},\ }\href {https://doi.org/10.1103/PhysRevB.89.041104} {\bibfield  {journal} {\bibinfo  {journal} {Phys. Rev. B}\ }\textbf {\bibinfo {volume} {89}},\ \bibinfo {pages} {041104} (\bibinfo {year} {2014})}\BibitemShut {NoStop}%
\bibitem [{\citenamefont {Moser}\ \emph {et~al.}(2015)\citenamefont {Moser}, \citenamefont {Fatale}, \citenamefont {Kr\"uger}, \citenamefont {Berger}, \citenamefont {Bugnon}, \citenamefont {Magrez}, \citenamefont {Niwa}, \citenamefont {Miyawaki}, \citenamefont {Harada},\ and\ \citenamefont {Grioni}}]{MoserPRL2015}%
  \BibitemOpen
  \bibfield  {author} {\bibinfo {author} {\bibfnamefont {S.}~\bibnamefont {Moser}}, \bibinfo {author} {\bibfnamefont {S.}~\bibnamefont {Fatale}}, \bibinfo {author} {\bibfnamefont {P.}~\bibnamefont {Kr\"uger}}, \bibinfo {author} {\bibfnamefont {H.}~\bibnamefont {Berger}}, \bibinfo {author} {\bibfnamefont {P.}~\bibnamefont {Bugnon}}, \bibinfo {author} {\bibfnamefont {A.}~\bibnamefont {Magrez}}, \bibinfo {author} {\bibfnamefont {H.}~\bibnamefont {Niwa}}, \bibinfo {author} {\bibfnamefont {J.}~\bibnamefont {Miyawaki}}, \bibinfo {author} {\bibfnamefont {Y.}~\bibnamefont {Harada}},\ and\ \bibinfo {author} {\bibfnamefont {M.}~\bibnamefont {Grioni}},\ }\bibfield  {title} {\bibinfo {title} {Electron-phonon coupling in the bulk of {A}natase {${\mathrm{TiO}}_{2}$} measured by resonant inelastic x-ray spectroscopy},\ }\href {https://doi.org/10.1103/PhysRevLett.115.096404} {\bibfield  {journal} {\bibinfo  {journal} {Phys. Rev. Lett.}\ }\textbf {\bibinfo {volume} {115}},\ \bibinfo {pages} {096404} (\bibinfo {year}
  {2015})}\BibitemShut {NoStop}%
\bibitem [{\citenamefont {Fatale}\ \emph {et~al.}(2016)\citenamefont {Fatale}, \citenamefont {Moser}, \citenamefont {Miyawaki}, \citenamefont {Harada},\ and\ \citenamefont {Grioni}}]{Fatale2016hybridization}%
  \BibitemOpen
  \bibfield  {author} {\bibinfo {author} {\bibfnamefont {S.}~\bibnamefont {Fatale}}, \bibinfo {author} {\bibfnamefont {S.}~\bibnamefont {Moser}}, \bibinfo {author} {\bibfnamefont {J.}~\bibnamefont {Miyawaki}}, \bibinfo {author} {\bibfnamefont {Y.}~\bibnamefont {Harada}},\ and\ \bibinfo {author} {\bibfnamefont {M.}~\bibnamefont {Grioni}},\ }\bibfield  {title} {\bibinfo {title} {Hybridization and electron-phonon coupling in ferroelectric {$\mathrm{BaTiO}_{3}$} probed by resonant inelastic x-ray scattering},\ }\href {https://doi.org/10.1103/PhysRevB.94.195131} {\bibfield  {journal} {\bibinfo  {journal} {Phys. Rev. B}\ }\textbf {\bibinfo {volume} {94}},\ \bibinfo {pages} {195131} (\bibinfo {year} {2016})}\BibitemShut {NoStop}%
\bibitem [{\citenamefont {Johnston}\ \emph {et~al.}(2016)\citenamefont {Johnston}, \citenamefont {Monney}, \citenamefont {Bisogni}, \citenamefont {Zhou}, \citenamefont {Kraus}, \citenamefont {Behr}, \citenamefont {Strocov}, \citenamefont {M{\'a}lek}, \citenamefont {Drechsler}, \citenamefont {Geck}, \citenamefont {Schmitt},\ and\ \citenamefont {van~den Brink}}]{JohnstonNatComm2016}%
  \BibitemOpen
  \bibfield  {author} {\bibinfo {author} {\bibfnamefont {S.}~\bibnamefont {Johnston}}, \bibinfo {author} {\bibfnamefont {C.}~\bibnamefont {Monney}}, \bibinfo {author} {\bibfnamefont {V.}~\bibnamefont {Bisogni}}, \bibinfo {author} {\bibfnamefont {K.-J.}\ \bibnamefont {Zhou}}, \bibinfo {author} {\bibfnamefont {R.}~\bibnamefont {Kraus}}, \bibinfo {author} {\bibfnamefont {G.}~\bibnamefont {Behr}}, \bibinfo {author} {\bibfnamefont {V.~N.}\ \bibnamefont {Strocov}}, \bibinfo {author} {\bibfnamefont {J.}~\bibnamefont {M{\'a}lek}}, \bibinfo {author} {\bibfnamefont {S.-L.}\ \bibnamefont {Drechsler}}, \bibinfo {author} {\bibfnamefont {J.}~\bibnamefont {Geck}}, \bibinfo {author} {\bibfnamefont {T.}~\bibnamefont {Schmitt}},\ and\ \bibinfo {author} {\bibfnamefont {J.}~\bibnamefont {van~den Brink}},\ }\bibfield  {title} {\bibinfo {title} {Electron-lattice interactions strongly renormalize the charge-transfer energy in the spin-chain cuprate \protect{Li$_2$CuO$_2$}},\ }\href {https://doi.org/10.1038/ncomms10563} {\bibfield
  {journal} {\bibinfo  {journal} {Nature Communications}\ }\textbf {\bibinfo {volume} {7}},\ \bibinfo {pages} {10563} (\bibinfo {year} {2016})}\BibitemShut {NoStop}%
\bibitem [{\citenamefont {Chaix}\ \emph {et~al.}(2017)\citenamefont {Chaix}, \citenamefont {Ghiringhelli}, \citenamefont {Peng}, \citenamefont {Hashimoto}, \citenamefont {Moritz}, \citenamefont {Kummer}, \citenamefont {Brookes}, \citenamefont {He}, \citenamefont {Chen}, \citenamefont {Ishida}, \citenamefont {Yoshida}, \citenamefont {Eisaki}, \citenamefont {Salluzzo}, \citenamefont {Braicovich}, \citenamefont {Shen}, \citenamefont {Devereaux},\ and\ \citenamefont {Lee}}]{Chaix2017}%
  \BibitemOpen
  \bibfield  {author} {\bibinfo {author} {\bibfnamefont {L.}~\bibnamefont {Chaix}}, \bibinfo {author} {\bibfnamefont {G.}~\bibnamefont {Ghiringhelli}}, \bibinfo {author} {\bibfnamefont {Y.~Y.}\ \bibnamefont {Peng}}, \bibinfo {author} {\bibfnamefont {M.}~\bibnamefont {Hashimoto}}, \bibinfo {author} {\bibfnamefont {B.}~\bibnamefont {Moritz}}, \bibinfo {author} {\bibfnamefont {K.}~\bibnamefont {Kummer}}, \bibinfo {author} {\bibfnamefont {N.~B.}\ \bibnamefont {Brookes}}, \bibinfo {author} {\bibfnamefont {Y.}~\bibnamefont {He}}, \bibinfo {author} {\bibfnamefont {S.}~\bibnamefont {Chen}}, \bibinfo {author} {\bibfnamefont {S.}~\bibnamefont {Ishida}}, \bibinfo {author} {\bibfnamefont {Y.}~\bibnamefont {Yoshida}}, \bibinfo {author} {\bibfnamefont {H.}~\bibnamefont {Eisaki}}, \bibinfo {author} {\bibfnamefont {M.}~\bibnamefont {Salluzzo}}, \bibinfo {author} {\bibfnamefont {L.}~\bibnamefont {Braicovich}}, \bibinfo {author} {\bibfnamefont {Z.~X.}\ \bibnamefont {Shen}}, \bibinfo {author} {\bibfnamefont {T.~P.}\
  \bibnamefont {Devereaux}},\ and\ \bibinfo {author} {\bibfnamefont {W.~S.}\ \bibnamefont {Lee}},\ }\bibfield  {title} {\bibinfo {title} {Dispersive charge density wave excitations in {Bi$_2$Sr$_2$CaCu$_2$O$_{8+\delta}$}},\ }\href {https://doi.org/10.1038/nphys4157} {\bibfield  {journal} {\bibinfo  {journal} {Nature Physics}\ }\textbf {\bibinfo {volume} {13}},\ \bibinfo {pages} {952} (\bibinfo {year} {2017})}\BibitemShut {NoStop}%
\bibitem [{\citenamefont {Meyers}\ \emph {et~al.}(2018)\citenamefont {Meyers}, \citenamefont {Nakatsukasa}, \citenamefont {Mu}, \citenamefont {Hao}, \citenamefont {Yang}, \citenamefont {Cao}, \citenamefont {Fabbris}, \citenamefont {Miao}, \citenamefont {Pelliciari}, \citenamefont {McNally}, \citenamefont {Dantz}, \citenamefont {Paris}, \citenamefont {Karapetrova}, \citenamefont {Choi}, \citenamefont {Haskel}, \citenamefont {Shafer}, \citenamefont {Arenholz}, \citenamefont {Schmitt}, \citenamefont {Berlijn}, \citenamefont {Johnston}, \citenamefont {Liu},\ and\ \citenamefont {Dean}}]{MyersPRL2018}%
  \BibitemOpen
  \bibfield  {author} {\bibinfo {author} {\bibfnamefont {D.}~\bibnamefont {Meyers}}, \bibinfo {author} {\bibfnamefont {K.}~\bibnamefont {Nakatsukasa}}, \bibinfo {author} {\bibfnamefont {S.}~\bibnamefont {Mu}}, \bibinfo {author} {\bibfnamefont {L.}~\bibnamefont {Hao}}, \bibinfo {author} {\bibfnamefont {J.}~\bibnamefont {Yang}}, \bibinfo {author} {\bibfnamefont {Y.}~\bibnamefont {Cao}}, \bibinfo {author} {\bibfnamefont {G.}~\bibnamefont {Fabbris}}, \bibinfo {author} {\bibfnamefont {H.}~\bibnamefont {Miao}}, \bibinfo {author} {\bibfnamefont {J.}~\bibnamefont {Pelliciari}}, \bibinfo {author} {\bibfnamefont {D.}~\bibnamefont {McNally}}, \bibinfo {author} {\bibfnamefont {M.}~\bibnamefont {Dantz}}, \bibinfo {author} {\bibfnamefont {E.}~\bibnamefont {Paris}}, \bibinfo {author} {\bibfnamefont {E.}~\bibnamefont {Karapetrova}}, \bibinfo {author} {\bibfnamefont {Y.}~\bibnamefont {Choi}}, \bibinfo {author} {\bibfnamefont {D.}~\bibnamefont {Haskel}}, \bibinfo {author} {\bibfnamefont {P.}~\bibnamefont {Shafer}}, \bibinfo
  {author} {\bibfnamefont {E.}~\bibnamefont {Arenholz}}, \bibinfo {author} {\bibfnamefont {T.}~\bibnamefont {Schmitt}}, \bibinfo {author} {\bibfnamefont {T.}~\bibnamefont {Berlijn}}, \bibinfo {author} {\bibfnamefont {S.}~\bibnamefont {Johnston}}, \bibinfo {author} {\bibfnamefont {J.}~\bibnamefont {Liu}},\ and\ \bibinfo {author} {\bibfnamefont {M.~P.~M.}\ \bibnamefont {Dean}},\ }\bibfield  {title} {\bibinfo {title} {Decoupling carrier concentration and electron-phonon coupling in oxide heterostructures observed with resonant inelastic x-ray scattering},\ }\href {https://doi.org/10.1103/PhysRevLett.121.236802} {\bibfield  {journal} {\bibinfo  {journal} {Phys. Rev. Lett.}\ }\textbf {\bibinfo {volume} {121}},\ \bibinfo {pages} {236802} (\bibinfo {year} {2018})}\BibitemShut {NoStop}%
\bibitem [{\citenamefont {Rossi}\ \emph {et~al.}(2019)\citenamefont {Rossi}, \citenamefont {Arpaia}, \citenamefont {Fumagalli}, \citenamefont {Moretti~Sala}, \citenamefont {Betto}, \citenamefont {Kummer}, \citenamefont {De~Luca}, \citenamefont {van~den Brink}, \citenamefont {Salluzzo}, \citenamefont {Brookes}, \citenamefont {Braicovich},\ and\ \citenamefont {Ghiringhelli}}]{rossi2019experimental}%
  \BibitemOpen
  \bibfield  {author} {\bibinfo {author} {\bibfnamefont {M.}~\bibnamefont {Rossi}}, \bibinfo {author} {\bibfnamefont {R.}~\bibnamefont {Arpaia}}, \bibinfo {author} {\bibfnamefont {R.}~\bibnamefont {Fumagalli}}, \bibinfo {author} {\bibfnamefont {M.}~\bibnamefont {Moretti~Sala}}, \bibinfo {author} {\bibfnamefont {D.}~\bibnamefont {Betto}}, \bibinfo {author} {\bibfnamefont {K.}~\bibnamefont {Kummer}}, \bibinfo {author} {\bibfnamefont {G.~M.}\ \bibnamefont {De~Luca}}, \bibinfo {author} {\bibfnamefont {J.}~\bibnamefont {van~den Brink}}, \bibinfo {author} {\bibfnamefont {M.}~\bibnamefont {Salluzzo}}, \bibinfo {author} {\bibfnamefont {N.~B.}\ \bibnamefont {Brookes}}, \bibinfo {author} {\bibfnamefont {L.}~\bibnamefont {Braicovich}},\ and\ \bibinfo {author} {\bibfnamefont {G.}~\bibnamefont {Ghiringhelli}},\ }\bibfield  {title} {\bibinfo {title} {Experimental determination of momentum-resolved electron-phonon coupling},\ }\href {https://doi.org/10.1103/PhysRevLett.123.027001} {\bibfield  {journal} {\bibinfo  {journal}
  {Phys. Rev. Lett.}\ }\textbf {\bibinfo {volume} {123}},\ \bibinfo {pages} {027001} (\bibinfo {year} {2019})}\BibitemShut {NoStop}%
\bibitem [{\citenamefont {Braicovich}\ \emph {et~al.}(2020)\citenamefont {Braicovich}, \citenamefont {Rossi}, \citenamefont {Fumagalli}, \citenamefont {Peng}, \citenamefont {Wang}, \citenamefont {Arpaia}, \citenamefont {Betto}, \citenamefont {De~Luca}, \citenamefont {Di~Castro}, \citenamefont {Kummer}, \citenamefont {Moretti~Sala}, \citenamefont {Pagetti}, \citenamefont {Balestrino}, \citenamefont {Brookes}, \citenamefont {Salluzzo}, \citenamefont {Johnston}, \citenamefont {van~den Brink},\ and\ \citenamefont {Ghiringhelli}}]{BraicovichPRR2020}%
  \BibitemOpen
  \bibfield  {author} {\bibinfo {author} {\bibfnamefont {L.}~\bibnamefont {Braicovich}}, \bibinfo {author} {\bibfnamefont {M.}~\bibnamefont {Rossi}}, \bibinfo {author} {\bibfnamefont {R.}~\bibnamefont {Fumagalli}}, \bibinfo {author} {\bibfnamefont {Y.}~\bibnamefont {Peng}}, \bibinfo {author} {\bibfnamefont {Y.}~\bibnamefont {Wang}}, \bibinfo {author} {\bibfnamefont {R.}~\bibnamefont {Arpaia}}, \bibinfo {author} {\bibfnamefont {D.}~\bibnamefont {Betto}}, \bibinfo {author} {\bibfnamefont {G.~M.}\ \bibnamefont {De~Luca}}, \bibinfo {author} {\bibfnamefont {D.}~\bibnamefont {Di~Castro}}, \bibinfo {author} {\bibfnamefont {K.}~\bibnamefont {Kummer}}, \bibinfo {author} {\bibfnamefont {M.}~\bibnamefont {Moretti~Sala}}, \bibinfo {author} {\bibfnamefont {M.}~\bibnamefont {Pagetti}}, \bibinfo {author} {\bibfnamefont {G.}~\bibnamefont {Balestrino}}, \bibinfo {author} {\bibfnamefont {N.~B.}\ \bibnamefont {Brookes}}, \bibinfo {author} {\bibfnamefont {M.}~\bibnamefont {Salluzzo}}, \bibinfo {author} {\bibfnamefont
  {S.}~\bibnamefont {Johnston}}, \bibinfo {author} {\bibfnamefont {J.}~\bibnamefont {van~den Brink}},\ and\ \bibinfo {author} {\bibfnamefont {G.}~\bibnamefont {Ghiringhelli}},\ }\bibfield  {title} {\bibinfo {title} {Determining the electron-phonon coupling in superconducting cuprates by resonant inelastic x-ray scattering: methods and results on {${\mathrm{Nd}}_{1+x}{\mathrm{Ba}}_{2\ensuremath{-}x}{\mathrm{Cu}}_{3}{\mathrm{O}}_{7\ensuremath{-}\ensuremath{\delta}}$}},\ }\href {https://doi.org/10.1103/PhysRevResearch.2.023231} {\bibfield  {journal} {\bibinfo  {journal} {Phys. Rev. Research}\ }\textbf {\bibinfo {volume} {2}},\ \bibinfo {pages} {023231} (\bibinfo {year} {2020})}\BibitemShut {NoStop}%
\bibitem [{\citenamefont {Li}\ \emph {et~al.}(2020)\citenamefont {Li}, \citenamefont {Nag}, \citenamefont {Pelliciari}, \citenamefont {Robarts}, \citenamefont {Walters}, \citenamefont {Garcia-Fernandez}, \citenamefont {Eisaki}, \citenamefont {Song}, \citenamefont {Ding}, \citenamefont {Johnston}, \citenamefont {Comin},\ and\ \citenamefont {Zhou}}]{LiPNMAS2020}%
  \BibitemOpen
  \bibfield  {author} {\bibinfo {author} {\bibfnamefont {J.}~\bibnamefont {Li}}, \bibinfo {author} {\bibfnamefont {A.}~\bibnamefont {Nag}}, \bibinfo {author} {\bibfnamefont {J.}~\bibnamefont {Pelliciari}}, \bibinfo {author} {\bibfnamefont {H.}~\bibnamefont {Robarts}}, \bibinfo {author} {\bibfnamefont {A.}~\bibnamefont {Walters}}, \bibinfo {author} {\bibfnamefont {M.}~\bibnamefont {Garcia-Fernandez}}, \bibinfo {author} {\bibfnamefont {H.}~\bibnamefont {Eisaki}}, \bibinfo {author} {\bibfnamefont {D.}~\bibnamefont {Song}}, \bibinfo {author} {\bibfnamefont {H.}~\bibnamefont {Ding}}, \bibinfo {author} {\bibfnamefont {S.}~\bibnamefont {Johnston}}, \bibinfo {author} {\bibfnamefont {R.}~\bibnamefont {Comin}},\ and\ \bibinfo {author} {\bibfnamefont {K.-J.}\ \bibnamefont {Zhou}},\ }\bibfield  {title} {\bibinfo {title} {Multiorbital charge-density wave excitations and concomitant phonon anomalies in {Bi$_2$Sr$_2$LaCuO$_{6+\delta}$}},\ }\href {https://doi.org/10.1073/pnas.2001755117} {\bibfield  {journal} {\bibinfo
  {journal} {Proceedings of the National Academy of Sciences}\ }\textbf {\bibinfo {volume} {117}},\ \bibinfo {pages} {16219} (\bibinfo {year} {2020})}\BibitemShut {NoStop}%
\bibitem [{\citenamefont {Lin}\ \emph {et~al.}(2020)\citenamefont {Lin}, \citenamefont {Miao}, \citenamefont {Mazzone}, \citenamefont {Gu}, \citenamefont {Nag}, \citenamefont {Walters}, \citenamefont {Garc\'{\i}a-Fern\'andez}, \citenamefont {Barbour}, \citenamefont {Pelliciari}, \citenamefont {Jarrige}, \citenamefont {Oda}, \citenamefont {Kurosawa}, \citenamefont {Momono}, \citenamefont {Zhou}, \citenamefont {Bisogni}, \citenamefont {Liu},\ and\ \citenamefont {Dean}}]{LinPRL2020}%
  \BibitemOpen
  \bibfield  {author} {\bibinfo {author} {\bibfnamefont {J.~Q.}\ \bibnamefont {Lin}}, \bibinfo {author} {\bibfnamefont {H.}~\bibnamefont {Miao}}, \bibinfo {author} {\bibfnamefont {D.~G.}\ \bibnamefont {Mazzone}}, \bibinfo {author} {\bibfnamefont {G.~D.}\ \bibnamefont {Gu}}, \bibinfo {author} {\bibfnamefont {A.}~\bibnamefont {Nag}}, \bibinfo {author} {\bibfnamefont {A.~C.}\ \bibnamefont {Walters}}, \bibinfo {author} {\bibfnamefont {M.}~\bibnamefont {Garc\'{\i}a-Fern\'andez}}, \bibinfo {author} {\bibfnamefont {A.}~\bibnamefont {Barbour}}, \bibinfo {author} {\bibfnamefont {J.}~\bibnamefont {Pelliciari}}, \bibinfo {author} {\bibfnamefont {I.}~\bibnamefont {Jarrige}}, \bibinfo {author} {\bibfnamefont {M.}~\bibnamefont {Oda}}, \bibinfo {author} {\bibfnamefont {K.}~\bibnamefont {Kurosawa}}, \bibinfo {author} {\bibfnamefont {N.}~\bibnamefont {Momono}}, \bibinfo {author} {\bibfnamefont {K.-J.}\ \bibnamefont {Zhou}}, \bibinfo {author} {\bibfnamefont {V.}~\bibnamefont {Bisogni}}, \bibinfo {author} {\bibfnamefont
  {X.}~\bibnamefont {Liu}},\ and\ \bibinfo {author} {\bibfnamefont {M.~P.~M.}\ \bibnamefont {Dean}},\ }\bibfield  {title} {\bibinfo {title} {Strongly correlated charge density wave in {${\mathrm{La}}_{2\ensuremath{-}x}{\mathrm{Sr}}_{x}{\mathrm{CuO}}_{4}$} evidenced by doping-dependent phonon anomaly},\ }\href {https://doi.org/10.1103/PhysRevLett.124.207005} {\bibfield  {journal} {\bibinfo  {journal} {Phys. Rev. Lett.}\ }\textbf {\bibinfo {volume} {124}},\ \bibinfo {pages} {207005} (\bibinfo {year} {2020})}\BibitemShut {NoStop}%
\bibitem [{\citenamefont {Geondzhian}\ \emph {et~al.}(2020)\citenamefont {Geondzhian}, \citenamefont {Sambri}, \citenamefont {De~Luca}, \citenamefont {Di~Capua}, \citenamefont {Di~Gennaro}, \citenamefont {Betto}, \citenamefont {Rossi}, \citenamefont {Peng}, \citenamefont {Fumagalli}, \citenamefont {Brookes}, \citenamefont {Braicovich}, \citenamefont {Gilmore}, \citenamefont {Ghiringhelli},\ and\ \citenamefont {Salluzzo}}]{Geondzhian2020large}%
  \BibitemOpen
  \bibfield  {author} {\bibinfo {author} {\bibfnamefont {A.}~\bibnamefont {Geondzhian}}, \bibinfo {author} {\bibfnamefont {A.}~\bibnamefont {Sambri}}, \bibinfo {author} {\bibfnamefont {G.~M.}\ \bibnamefont {De~Luca}}, \bibinfo {author} {\bibfnamefont {R.}~\bibnamefont {Di~Capua}}, \bibinfo {author} {\bibfnamefont {E.}~\bibnamefont {Di~Gennaro}}, \bibinfo {author} {\bibfnamefont {D.}~\bibnamefont {Betto}}, \bibinfo {author} {\bibfnamefont {M.}~\bibnamefont {Rossi}}, \bibinfo {author} {\bibfnamefont {Y.~Y.}\ \bibnamefont {Peng}}, \bibinfo {author} {\bibfnamefont {R.}~\bibnamefont {Fumagalli}}, \bibinfo {author} {\bibfnamefont {N.~B.}\ \bibnamefont {Brookes}}, \bibinfo {author} {\bibfnamefont {L.}~\bibnamefont {Braicovich}}, \bibinfo {author} {\bibfnamefont {K.}~\bibnamefont {Gilmore}}, \bibinfo {author} {\bibfnamefont {G.}~\bibnamefont {Ghiringhelli}},\ and\ \bibinfo {author} {\bibfnamefont {M.}~\bibnamefont {Salluzzo}},\ }\bibfield  {title} {\bibinfo {title} {Large polarons as key quasiparticles in
  {$\mathrm{SrTiO}_{3}$} and {${\mathrm{SrTiO}}_{3}$}-based heterostructures},\ }\href {https://doi.org/10.1103/PhysRevLett.125.126401} {\bibfield  {journal} {\bibinfo  {journal} {Phys. Rev. Lett.}\ }\textbf {\bibinfo {volume} {125}},\ \bibinfo {pages} {126401} (\bibinfo {year} {2020})}\BibitemShut {NoStop}%
\bibitem [{\citenamefont {Peng}\ \emph {et~al.}(2020)\citenamefont {Peng}, \citenamefont {Husain}, \citenamefont {Mitrano}, \citenamefont {Sun}, \citenamefont {Johnson}, \citenamefont {Zakrzewski}, \citenamefont {MacDougall}, \citenamefont {Barbour}, \citenamefont {Jarrige}, \citenamefont {Bisogni},\ and\ \citenamefont {Abbamonte}}]{PengPRL2020}%
  \BibitemOpen
  \bibfield  {author} {\bibinfo {author} {\bibfnamefont {Y.~Y.}\ \bibnamefont {Peng}}, \bibinfo {author} {\bibfnamefont {A.~A.}\ \bibnamefont {Husain}}, \bibinfo {author} {\bibfnamefont {M.}~\bibnamefont {Mitrano}}, \bibinfo {author} {\bibfnamefont {S.~X.-L.}\ \bibnamefont {Sun}}, \bibinfo {author} {\bibfnamefont {T.~A.}\ \bibnamefont {Johnson}}, \bibinfo {author} {\bibfnamefont {A.~V.}\ \bibnamefont {Zakrzewski}}, \bibinfo {author} {\bibfnamefont {G.~J.}\ \bibnamefont {MacDougall}}, \bibinfo {author} {\bibfnamefont {A.}~\bibnamefont {Barbour}}, \bibinfo {author} {\bibfnamefont {I.}~\bibnamefont {Jarrige}}, \bibinfo {author} {\bibfnamefont {V.}~\bibnamefont {Bisogni}},\ and\ \bibinfo {author} {\bibfnamefont {P.}~\bibnamefont {Abbamonte}},\ }\bibfield  {title} {\bibinfo {title} {Enhanced electron-phonon coupling for charge-density-wave formation in {${\mathrm{La}}_{1.8\ensuremath{-}x}{\mathrm{Eu}}_{0.2}{\mathrm{Sr}}_{x}{\mathrm{CuO}}_{4+\ensuremath{\delta}}$}},\ }\href
  {https://doi.org/10.1103/PhysRevLett.125.097002} {\bibfield  {journal} {\bibinfo  {journal} {Phys. Rev. Lett.}\ }\textbf {\bibinfo {volume} {125}},\ \bibinfo {pages} {097002} (\bibinfo {year} {2020})}\BibitemShut {NoStop}%
\bibitem [{\citenamefont {Huang}\ \emph {et~al.}(2021)\citenamefont {Huang}, \citenamefont {Singh}, \citenamefont {Mou}, \citenamefont {Johnston}, \citenamefont {Kemper}, \citenamefont {van~den Brink}, \citenamefont {Chen}, \citenamefont {Lee}, \citenamefont {Okamoto}, \citenamefont {Chu}, \citenamefont {Li}, \citenamefont {Komiya}, \citenamefont {Komarek}, \citenamefont {Fujimori}, \citenamefont {Chen},\ and\ \citenamefont {Huang}}]{Huang2021quantum}%
  \BibitemOpen
  \bibfield  {author} {\bibinfo {author} {\bibfnamefont {H.~Y.}\ \bibnamefont {Huang}}, \bibinfo {author} {\bibfnamefont {A.}~\bibnamefont {Singh}}, \bibinfo {author} {\bibfnamefont {C.~Y.}\ \bibnamefont {Mou}}, \bibinfo {author} {\bibfnamefont {S.}~\bibnamefont {Johnston}}, \bibinfo {author} {\bibfnamefont {A.~F.}\ \bibnamefont {Kemper}}, \bibinfo {author} {\bibfnamefont {J.}~\bibnamefont {van~den Brink}}, \bibinfo {author} {\bibfnamefont {P.~J.}\ \bibnamefont {Chen}}, \bibinfo {author} {\bibfnamefont {T.~K.}\ \bibnamefont {Lee}}, \bibinfo {author} {\bibfnamefont {J.}~\bibnamefont {Okamoto}}, \bibinfo {author} {\bibfnamefont {Y.~Y.}\ \bibnamefont {Chu}}, \bibinfo {author} {\bibfnamefont {J.~H.}\ \bibnamefont {Li}}, \bibinfo {author} {\bibfnamefont {S.}~\bibnamefont {Komiya}}, \bibinfo {author} {\bibfnamefont {A.~C.}\ \bibnamefont {Komarek}}, \bibinfo {author} {\bibfnamefont {A.}~\bibnamefont {Fujimori}}, \bibinfo {author} {\bibfnamefont {C.~T.}\ \bibnamefont {Chen}},\ and\ \bibinfo {author} {\bibfnamefont
  {D.~J.}\ \bibnamefont {Huang}},\ }\bibfield  {title} {\bibinfo {title} {Quantum fluctuations of charge order induce phonon softening in a superconducting cuprate},\ }\href {https://doi.org/10.1103/PhysRevX.11.041038} {\bibfield  {journal} {\bibinfo  {journal} {Phys. Rev. X}\ }\textbf {\bibinfo {volume} {11}},\ \bibinfo {pages} {041038} (\bibinfo {year} {2021})}\BibitemShut {NoStop}%
\bibitem [{\citenamefont {Dashwood}\ \emph {et~al.}(2021)\citenamefont {Dashwood}, \citenamefont {Geondzhian}, \citenamefont {Vale}, \citenamefont {Pakpour-Tabrizi}, \citenamefont {Howard}, \citenamefont {Faure}, \citenamefont {Veiga}, \citenamefont {Meyers}, \citenamefont {Chiuzb\ifmmode~\u{a}\else \u{a}\fi{}ian}, \citenamefont {Nicolaou}, \citenamefont {Jaouen}, \citenamefont {Jackman}, \citenamefont {Nag}, \citenamefont {Garc\'{\i}a-Fern\'andez}, \citenamefont {Zhou}, \citenamefont {Walters}, \citenamefont {Gilmore}, \citenamefont {McMorrow},\ and\ \citenamefont {Dean}}]{dashwood2021probing}%
  \BibitemOpen
  \bibfield  {author} {\bibinfo {author} {\bibfnamefont {C.~D.}\ \bibnamefont {Dashwood}}, \bibinfo {author} {\bibfnamefont {A.}~\bibnamefont {Geondzhian}}, \bibinfo {author} {\bibfnamefont {J.~G.}\ \bibnamefont {Vale}}, \bibinfo {author} {\bibfnamefont {A.~C.}\ \bibnamefont {Pakpour-Tabrizi}}, \bibinfo {author} {\bibfnamefont {C.~A.}\ \bibnamefont {Howard}}, \bibinfo {author} {\bibfnamefont {Q.}~\bibnamefont {Faure}}, \bibinfo {author} {\bibfnamefont {L.~S.~I.}\ \bibnamefont {Veiga}}, \bibinfo {author} {\bibfnamefont {D.}~\bibnamefont {Meyers}}, \bibinfo {author} {\bibfnamefont {S.~G.}\ \bibnamefont {Chiuzb\ifmmode~\u{a}\else \u{a}\fi{}ian}}, \bibinfo {author} {\bibfnamefont {A.}~\bibnamefont {Nicolaou}}, \bibinfo {author} {\bibfnamefont {N.}~\bibnamefont {Jaouen}}, \bibinfo {author} {\bibfnamefont {R.~B.}\ \bibnamefont {Jackman}}, \bibinfo {author} {\bibfnamefont {A.}~\bibnamefont {Nag}}, \bibinfo {author} {\bibfnamefont {M.}~\bibnamefont {Garc\'{\i}a-Fern\'andez}}, \bibinfo {author} {\bibfnamefont {K.-J.}\
  \bibnamefont {Zhou}}, \bibinfo {author} {\bibfnamefont {A.~C.}\ \bibnamefont {Walters}}, \bibinfo {author} {\bibfnamefont {K.}~\bibnamefont {Gilmore}}, \bibinfo {author} {\bibfnamefont {D.~F.}\ \bibnamefont {McMorrow}},\ and\ \bibinfo {author} {\bibfnamefont {M.~P.~M.}\ \bibnamefont {Dean}},\ }\bibfield  {title} {\bibinfo {title} {Probing electron-phonon interactions away from the {F}ermi level with resonant inelastic x-ray scattering},\ }\href {https://doi.org/10.1103/PhysRevX.11.041052} {\bibfield  {journal} {\bibinfo  {journal} {Phys. Rev. X}\ }\textbf {\bibinfo {volume} {11}},\ \bibinfo {pages} {041052} (\bibinfo {year} {2021})}\BibitemShut {NoStop}%
\bibitem [{\citenamefont {Peng}\ \emph {et~al.}(2022)\citenamefont {Peng}, \citenamefont {Martinelli}, \citenamefont {Li}, \citenamefont {Rossi}, \citenamefont {Mitrano}, \citenamefont {Arpaia}, \citenamefont {Sala}, \citenamefont {Gao}, \citenamefont {Guo}, \citenamefont {De~Luca}, \citenamefont {Walters}, \citenamefont {Nag}, \citenamefont {Barbour}, \citenamefont {Gu}, \citenamefont {Pelliciari}, \citenamefont {Brookes}, \citenamefont {Abbamonte}, \citenamefont {Salluzzo}, \citenamefont {Zhou}, \citenamefont {Zhou}, \citenamefont {Bisogni}, \citenamefont {Braicovich}, \citenamefont {Johnston},\ and\ \citenamefont {Ghiringhelli}}]{PengPRB2022}%
  \BibitemOpen
  \bibfield  {author} {\bibinfo {author} {\bibfnamefont {Y.}~\bibnamefont {Peng}}, \bibinfo {author} {\bibfnamefont {L.}~\bibnamefont {Martinelli}}, \bibinfo {author} {\bibfnamefont {Q.}~\bibnamefont {Li}}, \bibinfo {author} {\bibfnamefont {M.}~\bibnamefont {Rossi}}, \bibinfo {author} {\bibfnamefont {M.}~\bibnamefont {Mitrano}}, \bibinfo {author} {\bibfnamefont {R.}~\bibnamefont {Arpaia}}, \bibinfo {author} {\bibfnamefont {M.~M.}\ \bibnamefont {Sala}}, \bibinfo {author} {\bibfnamefont {Q.}~\bibnamefont {Gao}}, \bibinfo {author} {\bibfnamefont {X.}~\bibnamefont {Guo}}, \bibinfo {author} {\bibfnamefont {G.~M.}\ \bibnamefont {De~Luca}}, \bibinfo {author} {\bibfnamefont {A.}~\bibnamefont {Walters}}, \bibinfo {author} {\bibfnamefont {A.}~\bibnamefont {Nag}}, \bibinfo {author} {\bibfnamefont {A.}~\bibnamefont {Barbour}}, \bibinfo {author} {\bibfnamefont {G.}~\bibnamefont {Gu}}, \bibinfo {author} {\bibfnamefont {J.}~\bibnamefont {Pelliciari}}, \bibinfo {author} {\bibfnamefont {N.~B.}\ \bibnamefont {Brookes}},
  \bibinfo {author} {\bibfnamefont {P.}~\bibnamefont {Abbamonte}}, \bibinfo {author} {\bibfnamefont {M.}~\bibnamefont {Salluzzo}}, \bibinfo {author} {\bibfnamefont {X.}~\bibnamefont {Zhou}}, \bibinfo {author} {\bibfnamefont {K.-J.}\ \bibnamefont {Zhou}}, \bibinfo {author} {\bibfnamefont {V.}~\bibnamefont {Bisogni}}, \bibinfo {author} {\bibfnamefont {L.}~\bibnamefont {Braicovich}}, \bibinfo {author} {\bibfnamefont {S.}~\bibnamefont {Johnston}},\ and\ \bibinfo {author} {\bibfnamefont {G.}~\bibnamefont {Ghiringhelli}},\ }\bibfield  {title} {\bibinfo {title} {Doping dependence of the electron-phonon coupling in two families of bilayer superconducting cuprates},\ }\href {https://doi.org/10.1103/PhysRevB.105.115105} {\bibfield  {journal} {\bibinfo  {journal} {Phys. Rev. B}\ }\textbf {\bibinfo {volume} {105}},\ \bibinfo {pages} {115105} (\bibinfo {year} {2022})}\BibitemShut {NoStop}%
\bibitem [{\citenamefont {Naamneh}\ \emph {et~al.}(2024)\citenamefont {Naamneh}, \citenamefont {Paris}, \citenamefont {McNally}, \citenamefont {Tseng}, \citenamefont {Pudelko}, \citenamefont {Gawryluk}, \citenamefont {Shamblin}, \citenamefont {OQuinn}, \citenamefont {Cohen-Stead}, \citenamefont {Shi}, \citenamefont {Radovic}, \citenamefont {Lang}, \citenamefont {Schmitt}, \citenamefont {Johnston},\ and\ \citenamefont {Plumb}}]{naamneh2024persistence}%
  \BibitemOpen
  \bibfield  {author} {\bibinfo {author} {\bibfnamefont {M.}~\bibnamefont {Naamneh}}, \bibinfo {author} {\bibfnamefont {E.}~\bibnamefont {Paris}}, \bibinfo {author} {\bibfnamefont {D.}~\bibnamefont {McNally}}, \bibinfo {author} {\bibfnamefont {Y.}~\bibnamefont {Tseng}}, \bibinfo {author} {\bibfnamefont {W.~R.}\ \bibnamefont {Pudelko}}, \bibinfo {author} {\bibfnamefont {D.~J.}\ \bibnamefont {Gawryluk}}, \bibinfo {author} {\bibfnamefont {J.}~\bibnamefont {Shamblin}}, \bibinfo {author} {\bibfnamefont {E.}~\bibnamefont {OQuinn}}, \bibinfo {author} {\bibfnamefont {B.}~\bibnamefont {Cohen-Stead}}, \bibinfo {author} {\bibfnamefont {M.}~\bibnamefont {Shi}}, \bibinfo {author} {\bibfnamefont {M.}~\bibnamefont {Radovic}}, \bibinfo {author} {\bibfnamefont {M.}~\bibnamefont {Lang}}, \bibinfo {author} {\bibfnamefont {T.}~\bibnamefont {Schmitt}}, \bibinfo {author} {\bibfnamefont {S.}~\bibnamefont {Johnston}},\ and\ \bibinfo {author} {\bibfnamefont {N.~C.}\ \bibnamefont {Plumb}},\ }\bibfield  {title} {\bibinfo {title}
  {Persistence of small polarons into the superconducting phase of {Ba$_{1-x}$K$_x$BiO$_3$}},\ }\href {https://arxiv.org/abs/2408.00401} {\bibfield  {journal} {\bibinfo  {journal} {arXiv:2408.00401}\ } (\bibinfo {year} {2024})}\BibitemShut {NoStop}%
\bibitem [{\citenamefont {Devereaux}\ \emph {et~al.}(2016)\citenamefont {Devereaux}, \citenamefont {Shvaika}, \citenamefont {Wu}, \citenamefont {Wohlfeld}, \citenamefont {Jia}, \citenamefont {Wang}, \citenamefont {Moritz}, \citenamefont {Chaix}, \citenamefont {Lee}, \citenamefont {Shen}, \citenamefont {Ghiringhelli},\ and\ \citenamefont {Braicovich}}]{DevereauxPRX2016}%
  \BibitemOpen
  \bibfield  {author} {\bibinfo {author} {\bibfnamefont {T.~P.}\ \bibnamefont {Devereaux}}, \bibinfo {author} {\bibfnamefont {A.~M.}\ \bibnamefont {Shvaika}}, \bibinfo {author} {\bibfnamefont {K.}~\bibnamefont {Wu}}, \bibinfo {author} {\bibfnamefont {K.}~\bibnamefont {Wohlfeld}}, \bibinfo {author} {\bibfnamefont {C.~J.}\ \bibnamefont {Jia}}, \bibinfo {author} {\bibfnamefont {Y.}~\bibnamefont {Wang}}, \bibinfo {author} {\bibfnamefont {B.}~\bibnamefont {Moritz}}, \bibinfo {author} {\bibfnamefont {L.}~\bibnamefont {Chaix}}, \bibinfo {author} {\bibfnamefont {W.-S.}\ \bibnamefont {Lee}}, \bibinfo {author} {\bibfnamefont {Z.-X.}\ \bibnamefont {Shen}}, \bibinfo {author} {\bibfnamefont {G.}~\bibnamefont {Ghiringhelli}},\ and\ \bibinfo {author} {\bibfnamefont {L.}~\bibnamefont {Braicovich}},\ }\bibfield  {title} {\bibinfo {title} {Directly characterizing the relative strength and momentum dependence of electron-phonon coupling using resonant inelastic x-ray scattering},\ }\href
  {https://doi.org/10.1103/PhysRevX.6.041019} {\bibfield  {journal} {\bibinfo  {journal} {Phys. Rev. X}\ }\textbf {\bibinfo {volume} {6}},\ \bibinfo {pages} {041019} (\bibinfo {year} {2016})}\BibitemShut {NoStop}%
\bibitem [{\citenamefont {Geondzhian}\ and\ \citenamefont {Gilmore}(2020)}]{GilmorePRB2020}%
  \BibitemOpen
  \bibfield  {author} {\bibinfo {author} {\bibfnamefont {A.}~\bibnamefont {Geondzhian}}\ and\ \bibinfo {author} {\bibfnamefont {K.}~\bibnamefont {Gilmore}},\ }\bibfield  {title} {\bibinfo {title} {Generalization of the {F}ranck-{C}ondon model for phonon excitations by resonant inelastic x-ray scattering},\ }\href {https://doi.org/10.1103/PhysRevB.101.214307} {\bibfield  {journal} {\bibinfo  {journal} {Phys. Rev. B}\ }\textbf {\bibinfo {volume} {101}},\ \bibinfo {pages} {214307} (\bibinfo {year} {2020})}\BibitemShut {NoStop}%
\bibitem [{\citenamefont {Bieniasz}\ \emph {et~al.}(2021)\citenamefont {Bieniasz}, \citenamefont {Johnston},\ and\ \citenamefont {Berciu}}]{bieniasz2021}%
  \BibitemOpen
  \bibfield  {author} {\bibinfo {author} {\bibfnamefont {K.}~\bibnamefont {Bieniasz}}, \bibinfo {author} {\bibfnamefont {S.}~\bibnamefont {Johnston}},\ and\ \bibinfo {author} {\bibfnamefont {M.}~\bibnamefont {Berciu}},\ }\bibfield  {title} {\bibinfo {title} {{Beyond the single-site approximation modeling of electron-phonon coupling effects on resonant inelastic X-ray scattering spectra}},\ }\href {https://doi.org/10.21468/SciPostPhys.11.3.062} {\bibfield  {journal} {\bibinfo  {journal} {SciPost Phys.}\ }\textbf {\bibinfo {volume} {11}},\ \bibinfo {pages} {62} (\bibinfo {year} {2021})}\BibitemShut {NoStop}%
\bibitem [{\citenamefont {Bieniasz}\ \emph {et~al.}(2022)\citenamefont {Bieniasz}, \citenamefont {Johnston},\ and\ \citenamefont {Berciu}}]{BieniaszPRB2022}%
  \BibitemOpen
  \bibfield  {author} {\bibinfo {author} {\bibfnamefont {K.}~\bibnamefont {Bieniasz}}, \bibinfo {author} {\bibfnamefont {S.}~\bibnamefont {Johnston}},\ and\ \bibinfo {author} {\bibfnamefont {M.}~\bibnamefont {Berciu}},\ }\bibfield  {title} {\bibinfo {title} {Theory of dispersive optical phonons in resonant inelastic x-ray scattering experiments},\ }\href {https://doi.org/10.1103/PhysRevB.105.L180302} {\bibfield  {journal} {\bibinfo  {journal} {Phys. Rev. B}\ }\textbf {\bibinfo {volume} {105}},\ \bibinfo {pages} {L180302} (\bibinfo {year} {2022})}\BibitemShut {NoStop}%
\bibitem [{\citenamefont {Geondzhian}\ and\ \citenamefont {Gilmore}(2018)}]{Gilmore_vibronic_2018}%
  \BibitemOpen
  \bibfield  {author} {\bibinfo {author} {\bibfnamefont {A.}~\bibnamefont {Geondzhian}}\ and\ \bibinfo {author} {\bibfnamefont {K.}~\bibnamefont {Gilmore}},\ }\bibfield  {title} {\bibinfo {title} {Demonstration of resonant inelastic x-ray scattering as a probe of exciton-phonon coupling},\ }\href {https://doi.org/10.1103/PhysRevB.98.214305} {\bibfield  {journal} {\bibinfo  {journal} {Phys. Rev. B}\ }\textbf {\bibinfo {volume} {98}},\ \bibinfo {pages} {214305} (\bibinfo {year} {2018})}\BibitemShut {NoStop}%
\bibitem [{\citenamefont {Nocera}\ \emph {et~al.}(2018)\citenamefont {Nocera}, \citenamefont {Kumar}, \citenamefont {Kaushal}, \citenamefont {Alvarez}, \citenamefont {Dagotto},\ and\ \citenamefont {Johnston}}]{Nocera2018}%
  \BibitemOpen
  \bibfield  {author} {\bibinfo {author} {\bibfnamefont {A.}~\bibnamefont {Nocera}}, \bibinfo {author} {\bibfnamefont {U.}~\bibnamefont {Kumar}}, \bibinfo {author} {\bibfnamefont {N.}~\bibnamefont {Kaushal}}, \bibinfo {author} {\bibfnamefont {G.}~\bibnamefont {Alvarez}}, \bibinfo {author} {\bibfnamefont {E.}~\bibnamefont {Dagotto}},\ and\ \bibinfo {author} {\bibfnamefont {S.}~\bibnamefont {Johnston}},\ }\bibfield  {title} {\bibinfo {title} {Computing resonant inelastic x-ray scattering spectra using the density matrix renormalization group method},\ }\href {https://doi.org/10.1038/s41598-018-29218-8} {\bibfield  {journal} {\bibinfo  {journal} {Scientific Reports}\ }\textbf {\bibinfo {volume} {8}},\ \bibinfo {pages} {11080} (\bibinfo {year} {2018})}\BibitemShut {NoStop}%
\bibitem [{\citenamefont {Karakuzu}\ \emph {et~al.}(2017)\citenamefont {Karakuzu}, \citenamefont {Tocchio}, \citenamefont {Sorella},\ and\ \citenamefont {Becca}}]{Karakuzu2017superconductivity}%
  \BibitemOpen
  \bibfield  {author} {\bibinfo {author} {\bibfnamefont {S.}~\bibnamefont {Karakuzu}}, \bibinfo {author} {\bibfnamefont {L.~F.}\ \bibnamefont {Tocchio}}, \bibinfo {author} {\bibfnamefont {S.}~\bibnamefont {Sorella}},\ and\ \bibinfo {author} {\bibfnamefont {F.}~\bibnamefont {Becca}},\ }\bibfield  {title} {\bibinfo {title} {Superconductivity, charge-density waves, antiferromagnetism, and phase separation in the {H}ubbard-{H}olstein model},\ }\href {https://doi.org/10.1103/PhysRevB.96.205145} {\bibfield  {journal} {\bibinfo  {journal} {Phys. Rev. B}\ }\textbf {\bibinfo {volume} {96}},\ \bibinfo {pages} {205145} (\bibinfo {year} {2017})}\BibitemShut {NoStop}%
\bibitem [{\citenamefont {Costa}\ \emph {et~al.}(2020)\citenamefont {Costa}, \citenamefont {Seki}, \citenamefont {Yunoki},\ and\ \citenamefont {Sorella}}]{Costa2020phase}%
  \BibitemOpen
  \bibfield  {author} {\bibinfo {author} {\bibfnamefont {N.~C.}\ \bibnamefont {Costa}}, \bibinfo {author} {\bibfnamefont {K.}~\bibnamefont {Seki}}, \bibinfo {author} {\bibfnamefont {S.}~\bibnamefont {Yunoki}},\ and\ \bibinfo {author} {\bibfnamefont {S.}~\bibnamefont {Sorella}},\ }\bibfield  {title} {\bibinfo {title} {Phase diagram of the two-dimensional {H}ubbard-{H}olstein model},\ }\href {https://doi.org/10.1038/s42005-020-0342-2} {\bibfield  {journal} {\bibinfo  {journal} {Communications Physics}\ }\textbf {\bibinfo {volume} {3}},\ \bibinfo {pages} {80} (\bibinfo {year} {2020})}\BibitemShut {NoStop}%
\bibitem [{\citenamefont {Hardikar}\ and\ \citenamefont {Clay}(2007)}]{Hardikar2007phase}%
  \BibitemOpen
  \bibfield  {author} {\bibinfo {author} {\bibfnamefont {R.~P.}\ \bibnamefont {Hardikar}}\ and\ \bibinfo {author} {\bibfnamefont {R.~T.}\ \bibnamefont {Clay}},\ }\bibfield  {title} {\bibinfo {title} {Phase diagram of the one-dimensional {H}ubbard-{H}olstein model at half and quarter filling},\ }\href {https://doi.org/10.1103/PhysRevB.75.245103} {\bibfield  {journal} {\bibinfo  {journal} {Phys. Rev. B}\ }\textbf {\bibinfo {volume} {75}},\ \bibinfo {pages} {245103} (\bibinfo {year} {2007})}\BibitemShut {NoStop}%
\bibitem [{\citenamefont {Tezuka}\ \emph {et~al.}(2007)\citenamefont {Tezuka}, \citenamefont {Arita},\ and\ \citenamefont {Aoki}}]{Tezuka2007phase}%
  \BibitemOpen
  \bibfield  {author} {\bibinfo {author} {\bibfnamefont {M.}~\bibnamefont {Tezuka}}, \bibinfo {author} {\bibfnamefont {R.}~\bibnamefont {Arita}},\ and\ \bibinfo {author} {\bibfnamefont {H.}~\bibnamefont {Aoki}},\ }\bibfield  {title} {\bibinfo {title} {Phase diagram for the one-dimensional {H}ubbard-{H}olstein model: a density-matrix renormalization group study},\ }\href {https://doi.org/10.1103/PhysRevB.76.155114} {\bibfield  {journal} {\bibinfo  {journal} {Phys. Rev. B}\ }\textbf {\bibinfo {volume} {76}},\ \bibinfo {pages} {155114} (\bibinfo {year} {2007})}\BibitemShut {NoStop}%
\bibitem [{\citenamefont {Hohenadler}\ and\ \citenamefont {Fehske}(2018)}]{Hohenadler_Fehske_2018}%
  \BibitemOpen
  \bibfield  {author} {\bibinfo {author} {\bibfnamefont {M.}~\bibnamefont {Hohenadler}}\ and\ \bibinfo {author} {\bibfnamefont {H.}~\bibnamefont {Fehske}},\ }\bibfield  {title} {\bibinfo {title} {Density waves in strongly correlated quantum chains},\ }\href {https://doi.org/10.1140/epjb/e2018-90354-7} {\bibfield  {journal} {\bibinfo  {journal} {The European Physical Journal B}\ }\textbf {\bibinfo {volume} {91}},\ \bibinfo {pages} {204} (\bibinfo {year} {2018})}\BibitemShut {NoStop}%
\bibitem [{\citenamefont {Braicovich}\ \emph {et~al.}(2009)\citenamefont {Braicovich}, \citenamefont {Ament}, \citenamefont {Bisogni}, \citenamefont {Forte}, \citenamefont {Aruta}, \citenamefont {Balestrino}, \citenamefont {Brookes}, \citenamefont {De~Luca}, \citenamefont {Medaglia}, \citenamefont {Granozio}, \citenamefont {Radovic}, \citenamefont {Salluzzo}, \citenamefont {van~den Brink},\ and\ \citenamefont {Ghiringhelli}}]{Braicovich_Ament_Bisogni_Forte_Aruta_Balestrino_Brookes2009}%
  \BibitemOpen
  \bibfield  {author} {\bibinfo {author} {\bibfnamefont {L.}~\bibnamefont {Braicovich}}, \bibinfo {author} {\bibfnamefont {L.~J.~P.}\ \bibnamefont {Ament}}, \bibinfo {author} {\bibfnamefont {V.}~\bibnamefont {Bisogni}}, \bibinfo {author} {\bibfnamefont {F.}~\bibnamefont {Forte}}, \bibinfo {author} {\bibfnamefont {C.}~\bibnamefont {Aruta}}, \bibinfo {author} {\bibfnamefont {G.}~\bibnamefont {Balestrino}}, \bibinfo {author} {\bibfnamefont {N.~B.}\ \bibnamefont {Brookes}}, \bibinfo {author} {\bibfnamefont {G.~M.}\ \bibnamefont {De~Luca}}, \bibinfo {author} {\bibfnamefont {P.~G.}\ \bibnamefont {Medaglia}}, \bibinfo {author} {\bibfnamefont {F.~M.}\ \bibnamefont {Granozio}}, \bibinfo {author} {\bibfnamefont {M.}~\bibnamefont {Radovic}}, \bibinfo {author} {\bibfnamefont {M.}~\bibnamefont {Salluzzo}}, \bibinfo {author} {\bibfnamefont {J.}~\bibnamefont {van~den Brink}},\ and\ \bibinfo {author} {\bibfnamefont {G.}~\bibnamefont {Ghiringhelli}},\ }\bibfield  {title} {\bibinfo {title} {Dispersion of magnetic excitations
  in the cuprate {${\mathrm{La}}_{2}{\mathrm{CuO}}_{4}$} and {${\mathrm{CaCuO}}_{2}$} compounds measured using resonant x-ray scattering},\ }\href {https://doi.org/10.1103/PhysRevLett.102.167401} {\bibfield  {journal} {\bibinfo  {journal} {Phys. Rev. Lett.}\ }\textbf {\bibinfo {volume} {102}},\ \bibinfo {pages} {167401} (\bibinfo {year} {2009})}\BibitemShut {NoStop}%
\bibitem [{\citenamefont {Bisogni}\ \emph {et~al.}(2014)\citenamefont {Bisogni}, \citenamefont {Kourtis}, \citenamefont {Monney}, \citenamefont {Zhou}, \citenamefont {Kraus}, \citenamefont {Sekar}, \citenamefont {Strocov}, \citenamefont {B\"uchner}, \citenamefont {van~den Brink}, \citenamefont {Braicovich}, \citenamefont {Schmitt}, \citenamefont {Daghofer},\ and\ \citenamefont {Geck}}]{Bisogni2014femtosecond}%
  \BibitemOpen
  \bibfield  {author} {\bibinfo {author} {\bibfnamefont {V.}~\bibnamefont {Bisogni}}, \bibinfo {author} {\bibfnamefont {S.}~\bibnamefont {Kourtis}}, \bibinfo {author} {\bibfnamefont {C.}~\bibnamefont {Monney}}, \bibinfo {author} {\bibfnamefont {K.}~\bibnamefont {Zhou}}, \bibinfo {author} {\bibfnamefont {R.}~\bibnamefont {Kraus}}, \bibinfo {author} {\bibfnamefont {C.}~\bibnamefont {Sekar}}, \bibinfo {author} {\bibfnamefont {V.}~\bibnamefont {Strocov}}, \bibinfo {author} {\bibfnamefont {B.}~\bibnamefont {B\"uchner}}, \bibinfo {author} {\bibfnamefont {J.}~\bibnamefont {van~den Brink}}, \bibinfo {author} {\bibfnamefont {L.}~\bibnamefont {Braicovich}}, \bibinfo {author} {\bibfnamefont {T.}~\bibnamefont {Schmitt}}, \bibinfo {author} {\bibfnamefont {M.}~\bibnamefont {Daghofer}},\ and\ \bibinfo {author} {\bibfnamefont {J.}~\bibnamefont {Geck}},\ }\bibfield  {title} {\bibinfo {title} {Femtosecond dynamics of momentum-dependent magnetic excitations from resonant inelastic x-ray scattering in
  {${\mathrm{CaCu}}_{2}{\mathrm{O}}_{3}$}},\ }\href {https://doi.org/10.1103/PhysRevLett.112.147401} {\bibfield  {journal} {\bibinfo  {journal} {Phys. Rev. Lett.}\ }\textbf {\bibinfo {volume} {112}},\ \bibinfo {pages} {147401} (\bibinfo {year} {2014})}\BibitemShut {NoStop}%
\bibitem [{\citenamefont {White}(1992)}]{White_1992}%
  \BibitemOpen
  \bibfield  {author} {\bibinfo {author} {\bibfnamefont {S.~R.}\ \bibnamefont {White}},\ }\bibfield  {title} {\bibinfo {title} {Density matrix formulation for quantum renormalization groups},\ }\href {https://doi.org/10.1103/PhysRevLett.69.2863} {\bibfield  {journal} {\bibinfo  {journal} {Physical Review Letters}\ }\textbf {\bibinfo {volume} {69}},\ \bibinfo {pages} {2863–2866} (\bibinfo {year} {1992})}\BibitemShut {NoStop}%
\bibitem [{\citenamefont {White}(1993)}]{White_1993}%
  \BibitemOpen
  \bibfield  {author} {\bibinfo {author} {\bibfnamefont {S.~R.}\ \bibnamefont {White}},\ }\bibfield  {title} {\bibinfo {title} {Density-matrix algorithms for quantum renormalization groups},\ }\href {https://doi.org/10.1103/PhysRevB.48.10345} {\bibfield  {journal} {\bibinfo  {journal} {Physical Review B}\ }\textbf {\bibinfo {volume} {48}},\ \bibinfo {pages} {10345–10356} (\bibinfo {year} {1993})}\BibitemShut {NoStop}%
\bibitem [{\citenamefont {Nocera}\ and\ \citenamefont {Alvarez}(2016)}]{Nocera_Alvarez_2016}%
  \BibitemOpen
  \bibfield  {author} {\bibinfo {author} {\bibfnamefont {A.}~\bibnamefont {Nocera}}\ and\ \bibinfo {author} {\bibfnamefont {G.}~\bibnamefont {Alvarez}},\ }\bibfield  {title} {\bibinfo {title} {Spectral functions with the density matrix renormalization group: Krylov-space approach for correction vectors},\ }\href {https://doi.org/10.1103/PhysRevE.94.053308} {\bibfield  {journal} {\bibinfo  {journal} {Physical Review E}\ }\textbf {\bibinfo {volume} {94}},\ \bibinfo {pages} {053308} (\bibinfo {year} {2016})}\BibitemShut {NoStop}%
\bibitem [{\citenamefont {Lu}\ and\ \citenamefont {Haverkort}(2017)}]{Lu2017nonperturbative}%
  \BibitemOpen
  \bibfield  {author} {\bibinfo {author} {\bibfnamefont {Y.}~\bibnamefont {Lu}}\ and\ \bibinfo {author} {\bibfnamefont {M.~W.}\ \bibnamefont {Haverkort}},\ }\bibfield  {title} {\bibinfo {title} {Nonperturbative series expansion of {Green's} functions: the anatomy of resonant inelastic x-ray scattering in the doped {H}ubbard model},\ }\href {https://doi.org/10.1103/PhysRevLett.119.256401} {\bibfield  {journal} {\bibinfo  {journal} {Phys. Rev. Lett.}\ }\textbf {\bibinfo {volume} {119}},\ \bibinfo {pages} {256401} (\bibinfo {year} {2017})}\BibitemShut {NoStop}%
\bibitem [{\citenamefont {van~den Brink}\ and\ \citenamefont {van Veenendaal}(2005)}]{Brink2005correlation}%
  \BibitemOpen
  \bibfield  {author} {\bibinfo {author} {\bibfnamefont {J.}~\bibnamefont {van~den Brink}}\ and\ \bibinfo {author} {\bibfnamefont {M.}~\bibnamefont {van Veenendaal}},\ }\bibfield  {title} {\bibinfo {title} {Correlation functions measured by indirect resonant inelastic x-ray scattering},\ }\href {https://doi.org/10.1209/epl/i2005-10366-9} {\bibfield  {journal} {\bibinfo  {journal} {Europhys. Lett.}\ }\textbf {\bibinfo {volume} {73}},\ \bibinfo {pages} {121} (\bibinfo {year} {2005})}\BibitemShut {NoStop}%
\bibitem [{\citenamefont {Tsvelik}\ \emph {et~al.}(2019)\citenamefont {Tsvelik}, \citenamefont {Konik}, \citenamefont {Prokof'ev},\ and\ \citenamefont {Tupitsyn}}]{Tsvelik2019resonant}%
  \BibitemOpen
  \bibfield  {author} {\bibinfo {author} {\bibfnamefont {A.~M.}\ \bibnamefont {Tsvelik}}, \bibinfo {author} {\bibfnamefont {R.~M.}\ \bibnamefont {Konik}}, \bibinfo {author} {\bibfnamefont {N.~V.}\ \bibnamefont {Prokof'ev}},\ and\ \bibinfo {author} {\bibfnamefont {I.~S.}\ \bibnamefont {Tupitsyn}},\ }\bibfield  {title} {\bibinfo {title} {Resonant inelastic x-ray scattering in metals: a diagrammatic approach},\ }\href {https://doi.org/10.1103/PhysRevResearch.1.033093} {\bibfield  {journal} {\bibinfo  {journal} {Phys. Rev. Research}\ }\textbf {\bibinfo {volume} {1}},\ \bibinfo {pages} {033093} (\bibinfo {year} {2019})}\BibitemShut {NoStop}%
\bibitem [{\citenamefont {K\"uhner}\ and\ \citenamefont {White}(1999)}]{PhysRevB.60.335}%
  \BibitemOpen
  \bibfield  {author} {\bibinfo {author} {\bibfnamefont {T.~D.}\ \bibnamefont {K\"uhner}}\ and\ \bibinfo {author} {\bibfnamefont {S.~R.}\ \bibnamefont {White}},\ }\bibfield  {title} {\bibinfo {title} {Dynamical correlation functions using the density matrix renormalization group},\ }\href {https://doi.org/10.1103/PhysRevB.60.335} {\bibfield  {journal} {\bibinfo  {journal} {Phys. Rev. B}\ }\textbf {\bibinfo {volume} {60}},\ \bibinfo {pages} {335} (\bibinfo {year} {1999})}\BibitemShut {NoStop}%
\bibitem [{\citenamefont {White}\ and\ \citenamefont {Feiguin}(2004)}]{White2004real}%
  \BibitemOpen
  \bibfield  {author} {\bibinfo {author} {\bibfnamefont {S.~R.}\ \bibnamefont {White}}\ and\ \bibinfo {author} {\bibfnamefont {A.~E.}\ \bibnamefont {Feiguin}},\ }\bibfield  {title} {\bibinfo {title} {Real-time evolution using the density matrix renormalization group},\ }\href {https://doi.org/10.1103/PhysRevLett.93.076401} {\bibfield  {journal} {\bibinfo  {journal} {Phys. Rev. Lett.}\ }\textbf {\bibinfo {volume} {93}},\ \bibinfo {pages} {076401} (\bibinfo {year} {2004})}\BibitemShut {NoStop}%
\bibitem [{\citenamefont {Alvarez}(2009)}]{alvarez2009density}%
  \BibitemOpen
  \bibfield  {author} {\bibinfo {author} {\bibfnamefont {G.}~\bibnamefont {Alvarez}},\ }\bibfield  {title} {\bibinfo {title} {The density matrix renormalization group for strongly correlated electron systems: a generic implementation},\ }\href {https://doi.org/10.1016/j.cpc.2009.02.016} {\bibfield  {journal} {\bibinfo  {journal} {Computer Physics Communications}\ }\textbf {\bibinfo {volume} {180}},\ \bibinfo {pages} {1572} (\bibinfo {year} {2009})}\BibitemShut {NoStop}%
\bibitem [{\citenamefont {Kourtis}\ \emph {et~al.}(2012)\citenamefont {Kourtis}, \citenamefont {van~den Brink},\ and\ \citenamefont {Daghofer}}]{KourtisRIXStwochannelsHubbard}%
  \BibitemOpen
  \bibfield  {author} {\bibinfo {author} {\bibfnamefont {S.}~\bibnamefont {Kourtis}}, \bibinfo {author} {\bibfnamefont {J.}~\bibnamefont {van~den Brink}},\ and\ \bibinfo {author} {\bibfnamefont {M.}~\bibnamefont {Daghofer}},\ }\bibfield  {title} {\bibinfo {title} {Exact diagonalization results for resonant inelastic x-ray scattering spectra of one-dimensional mott insulators},\ }\href {https://doi.org/10.1103/PhysRevB.85.064423} {\bibfield  {journal} {\bibinfo  {journal} {Phys. Rev. B}\ }\textbf {\bibinfo {volume} {85}},\ \bibinfo {pages} {064423} (\bibinfo {year} {2012})}\BibitemShut {NoStop}%
\bibitem [{\citenamefont {Tsutsui}\ \emph {et~al.}(2000)\citenamefont {Tsutsui}, \citenamefont {Tohyama},\ and\ \citenamefont {Maekawa}}]{Tsutsui_Tohyama_Maekawa_2000}%
  \BibitemOpen
  \bibfield  {author} {\bibinfo {author} {\bibfnamefont {K.}~\bibnamefont {Tsutsui}}, \bibinfo {author} {\bibfnamefont {T.}~\bibnamefont {Tohyama}},\ and\ \bibinfo {author} {\bibfnamefont {S.}~\bibnamefont {Maekawa}},\ }\bibfield  {title} {\bibinfo {title} {Resonant inelastic x-ray scattering in one-dimensional copper oxides},\ }\href {https://doi.org/10.1103/PhysRevB.61.7180} {\bibfield  {journal} {\bibinfo  {journal} {Physical Review B}\ }\textbf {\bibinfo {volume} {61}},\ \bibinfo {pages} {7180–7182} (\bibinfo {year} {2000})}\BibitemShut {NoStop}%
\bibitem [{\citenamefont {Giamarchi}(2003)}]{GiamarchiBook}%
  \BibitemOpen
  \bibfield  {author} {\bibinfo {author} {\bibfnamefont {T.}~\bibnamefont {Giamarchi}},\ }\href {https://doi.org/10.1093/acprof:oso/9780198525004.001.0001} {\emph {\bibinfo {title} {{Quantum Physics in One Dimension}}}}\ (\bibinfo  {publisher} {Oxford University Press},\ \bibinfo {year} {2003})\BibitemShut {NoStop}%
\bibitem [{\citenamefont {Tennant}\ \emph {et~al.}(1995)\citenamefont {Tennant}, \citenamefont {Cowley}, \citenamefont {Nagler},\ and\ \citenamefont {Tsvelik}}]{Tennant1995measurements}%
  \BibitemOpen
  \bibfield  {author} {\bibinfo {author} {\bibfnamefont {D.~A.}\ \bibnamefont {Tennant}}, \bibinfo {author} {\bibfnamefont {R.~A.}\ \bibnamefont {Cowley}}, \bibinfo {author} {\bibfnamefont {S.~E.}\ \bibnamefont {Nagler}},\ and\ \bibinfo {author} {\bibfnamefont {A.~M.}\ \bibnamefont {Tsvelik}},\ }\bibfield  {title} {\bibinfo {title} {Measurement of the spin-excitation continuum in one-dimensional {${\mathrm{KCuF}}_{3}$} using neutron scattering},\ }\href {https://doi.org/10.1103/PhysRevB.52.13368} {\bibfield  {journal} {\bibinfo  {journal} {Phys. Rev. B}\ }\textbf {\bibinfo {volume} {52}},\ \bibinfo {pages} {13368} (\bibinfo {year} {1995})}\BibitemShut {NoStop}%
\bibitem [{\citenamefont {Zaliznyak}\ \emph {et~al.}(1999)\citenamefont {Zaliznyak}, \citenamefont {Broholm}, \citenamefont {Kibune}, \citenamefont {Nohara},\ and\ \citenamefont {Takagi}}]{Zaliznyak1999anisotropic}%
  \BibitemOpen
  \bibfield  {author} {\bibinfo {author} {\bibfnamefont {I.~A.}\ \bibnamefont {Zaliznyak}}, \bibinfo {author} {\bibfnamefont {C.}~\bibnamefont {Broholm}}, \bibinfo {author} {\bibfnamefont {M.}~\bibnamefont {Kibune}}, \bibinfo {author} {\bibfnamefont {M.}~\bibnamefont {Nohara}},\ and\ \bibinfo {author} {\bibfnamefont {H.}~\bibnamefont {Takagi}},\ }\bibfield  {title} {\bibinfo {title} {Anisotropic spin freezing in the {$\mathit{S}\phantom{\rule{0ex}{0ex}}=\phantom{\rule{0ex}{0ex}}1/2$} zigzag chain compound {${\mathrm{SrCuO}}_{2}$}},\ }\href {https://doi.org/10.1103/PhysRevLett.83.5370} {\bibfield  {journal} {\bibinfo  {journal} {Phys. Rev. Lett.}\ }\textbf {\bibinfo {volume} {83}},\ \bibinfo {pages} {5370} (\bibinfo {year} {1999})}\BibitemShut {NoStop}%
\bibitem [{\citenamefont {Walters}\ \emph {et~al.}(2009)\citenamefont {Walters}, \citenamefont {Perring}, \citenamefont {Caux}, \citenamefont {Savici}, \citenamefont {Gu}, \citenamefont {Lee}, \citenamefont {Ku},\ and\ \citenamefont {Zaliznyak}}]{Walters2009effects}%
  \BibitemOpen
  \bibfield  {author} {\bibinfo {author} {\bibfnamefont {A.~C.}\ \bibnamefont {Walters}}, \bibinfo {author} {\bibfnamefont {T.~G.}\ \bibnamefont {Perring}}, \bibinfo {author} {\bibfnamefont {J.-S.}\ \bibnamefont {Caux}}, \bibinfo {author} {\bibfnamefont {A.~T.}\ \bibnamefont {Savici}}, \bibinfo {author} {\bibfnamefont {G.~D.}\ \bibnamefont {Gu}}, \bibinfo {author} {\bibfnamefont {C.-C.}\ \bibnamefont {Lee}}, \bibinfo {author} {\bibfnamefont {W.}~\bibnamefont {Ku}},\ and\ \bibinfo {author} {\bibfnamefont {I.~A.}\ \bibnamefont {Zaliznyak}},\ }\bibfield  {title} {\bibinfo {title} {Effect of covalent bonding on magnetism and the missing neutron intensity in copper oxide compounds},\ }\href {https://doi.org/10.1038/nphys1405} {\bibfield  {journal} {\bibinfo  {journal} {Nature Physics}\ }\textbf {\bibinfo {volume} {5}},\ \bibinfo {pages} {867} (\bibinfo {year} {2009})}\BibitemShut {NoStop}%
\bibitem [{\citenamefont {Mourigal}\ \emph {et~al.}(2013)\citenamefont {Mourigal}, \citenamefont {Enderle}, \citenamefont {Kl{\"o}pperpieper}, \citenamefont {Caux}, \citenamefont {Stunault},\ and\ \citenamefont {R{\o}nnow}}]{MourigalNatPhys2013}%
  \BibitemOpen
  \bibfield  {author} {\bibinfo {author} {\bibfnamefont {M.}~\bibnamefont {Mourigal}}, \bibinfo {author} {\bibfnamefont {M.}~\bibnamefont {Enderle}}, \bibinfo {author} {\bibfnamefont {A.}~\bibnamefont {Kl{\"o}pperpieper}}, \bibinfo {author} {\bibfnamefont {J.-S.}\ \bibnamefont {Caux}}, \bibinfo {author} {\bibfnamefont {A.}~\bibnamefont {Stunault}},\ and\ \bibinfo {author} {\bibfnamefont {H.~M.}\ \bibnamefont {R{\o}nnow}},\ }\bibfield  {title} {\bibinfo {title} {Fractional spinon excitations in the quantum {H}eisenberg antiferromagnetic chain},\ }\href {https://doi.org/10.1038/nphys2652} {\bibfield  {journal} {\bibinfo  {journal} {Nature Physics}\ }\textbf {\bibinfo {volume} {9}},\ \bibinfo {pages} {435} (\bibinfo {year} {2013})}\BibitemShut {NoStop}%
\bibitem [{\citenamefont {Schlappa}\ \emph {et~al.}(2012)\citenamefont {Schlappa}, \citenamefont {Wohlfeld}, \citenamefont {Zhou}, \citenamefont {Mourigal}, \citenamefont {Haverkort}, \citenamefont {Strocov}, \citenamefont {Hozoi}, \citenamefont {Monney}, \citenamefont {Nishimoto}, \citenamefont {Singh}, \citenamefont {Revcolevschi}, \citenamefont {Caux}, \citenamefont {Patthey}, \citenamefont {R{\o}nnow}, \citenamefont {van~den Brink},\ and\ \citenamefont {Schmitt}}]{Schlappa2012spin}%
  \BibitemOpen
  \bibfield  {author} {\bibinfo {author} {\bibfnamefont {J.}~\bibnamefont {Schlappa}}, \bibinfo {author} {\bibfnamefont {K.}~\bibnamefont {Wohlfeld}}, \bibinfo {author} {\bibfnamefont {K.~J.}\ \bibnamefont {Zhou}}, \bibinfo {author} {\bibfnamefont {M.}~\bibnamefont {Mourigal}}, \bibinfo {author} {\bibfnamefont {M.~W.}\ \bibnamefont {Haverkort}}, \bibinfo {author} {\bibfnamefont {V.~N.}\ \bibnamefont {Strocov}}, \bibinfo {author} {\bibfnamefont {L.}~\bibnamefont {Hozoi}}, \bibinfo {author} {\bibfnamefont {C.}~\bibnamefont {Monney}}, \bibinfo {author} {\bibfnamefont {S.}~\bibnamefont {Nishimoto}}, \bibinfo {author} {\bibfnamefont {S.}~\bibnamefont {Singh}}, \bibinfo {author} {\bibfnamefont {A.}~\bibnamefont {Revcolevschi}}, \bibinfo {author} {\bibfnamefont {J.~S.}\ \bibnamefont {Caux}}, \bibinfo {author} {\bibfnamefont {L.}~\bibnamefont {Patthey}}, \bibinfo {author} {\bibfnamefont {H.~M.}\ \bibnamefont {R{\o}nnow}}, \bibinfo {author} {\bibfnamefont {J.}~\bibnamefont {van~den Brink}},\ and\ \bibinfo {author}
  {\bibfnamefont {T.}~\bibnamefont {Schmitt}},\ }\bibfield  {title} {\bibinfo {title} {Spin--orbital separation in the quasi-one-dimensional {M}ott insulator {Sr$_2$CuO$_3$}},\ }\href {https://doi.org/10.1038/nature10974} {\bibfield  {journal} {\bibinfo  {journal} {Nature}\ }\textbf {\bibinfo {volume} {485}},\ \bibinfo {pages} {82} (\bibinfo {year} {2012})}\BibitemShut {NoStop}%
\bibitem [{\citenamefont {Schlappa}\ \emph {et~al.}(2018)\citenamefont {Schlappa}, \citenamefont {Kumar}, \citenamefont {Zhou}, \citenamefont {Singh}, \citenamefont {Mourigal}, \citenamefont {Strocov}, \citenamefont {Revcolevschi}, \citenamefont {Patthey}, \citenamefont {R{\o}nnow}, \citenamefont {Johnston},\ and\ \citenamefont {Schmitt}}]{Schlappa2018}%
  \BibitemOpen
  \bibfield  {author} {\bibinfo {author} {\bibfnamefont {J.}~\bibnamefont {Schlappa}}, \bibinfo {author} {\bibfnamefont {U.}~\bibnamefont {Kumar}}, \bibinfo {author} {\bibfnamefont {K.~J.}\ \bibnamefont {Zhou}}, \bibinfo {author} {\bibfnamefont {S.}~\bibnamefont {Singh}}, \bibinfo {author} {\bibfnamefont {M.}~\bibnamefont {Mourigal}}, \bibinfo {author} {\bibfnamefont {V.~N.}\ \bibnamefont {Strocov}}, \bibinfo {author} {\bibfnamefont {A.}~\bibnamefont {Revcolevschi}}, \bibinfo {author} {\bibfnamefont {L.}~\bibnamefont {Patthey}}, \bibinfo {author} {\bibfnamefont {H.~M.}\ \bibnamefont {R{\o}nnow}}, \bibinfo {author} {\bibfnamefont {S.}~\bibnamefont {Johnston}},\ and\ \bibinfo {author} {\bibfnamefont {T.}~\bibnamefont {Schmitt}},\ }\bibfield  {title} {\bibinfo {title} {Probing multi-spinon excitations outside of the two-spinon continuum in the antiferromagnetic spin chain cuprate {Sr$_2$CuO$_3$}},\ }\href {https://doi.org/10.1038/s41467-018-07838-y} {\bibfield  {journal} {\bibinfo  {journal} {Nature
  Communications}\ }\textbf {\bibinfo {volume} {9}},\ \bibinfo {pages} {5394} (\bibinfo {year} {2018})}\BibitemShut {NoStop}%
\bibitem [{\citenamefont {Kumar}\ \emph {et~al.}(2022)\citenamefont {Kumar}, \citenamefont {Nag}, \citenamefont {Li}, \citenamefont {Robarts}, \citenamefont {Walters}, \citenamefont {Garc\'{\i}a-Fern\'andez}, \citenamefont {Saint-Martin}, \citenamefont {Revcolevschi}, \citenamefont {Schlappa}, \citenamefont {Schmitt}, \citenamefont {Johnston},\ and\ \citenamefont {Zhou}}]{Kumar2022unraveling}%
  \BibitemOpen
  \bibfield  {author} {\bibinfo {author} {\bibfnamefont {U.}~\bibnamefont {Kumar}}, \bibinfo {author} {\bibfnamefont {A.}~\bibnamefont {Nag}}, \bibinfo {author} {\bibfnamefont {J.}~\bibnamefont {Li}}, \bibinfo {author} {\bibfnamefont {H.~C.}\ \bibnamefont {Robarts}}, \bibinfo {author} {\bibfnamefont {A.~C.}\ \bibnamefont {Walters}}, \bibinfo {author} {\bibfnamefont {M.}~\bibnamefont {Garc\'{\i}a-Fern\'andez}}, \bibinfo {author} {\bibfnamefont {R.}~\bibnamefont {Saint-Martin}}, \bibinfo {author} {\bibfnamefont {A.}~\bibnamefont {Revcolevschi}}, \bibinfo {author} {\bibfnamefont {J.}~\bibnamefont {Schlappa}}, \bibinfo {author} {\bibfnamefont {T.}~\bibnamefont {Schmitt}}, \bibinfo {author} {\bibfnamefont {S.}~\bibnamefont {Johnston}},\ and\ \bibinfo {author} {\bibfnamefont {K.-J.}\ \bibnamefont {Zhou}},\ }\bibfield  {title} {\bibinfo {title} {Unraveling higher-order contributions to spin excitations probed using resonant inelastic x-ray scattering},\ }\href {https://doi.org/10.1103/PhysRevB.106.L060406}
  {\bibfield  {journal} {\bibinfo  {journal} {Phys. Rev. B}\ }\textbf {\bibinfo {volume} {106}},\ \bibinfo {pages} {L060406} (\bibinfo {year} {2022})}\BibitemShut {NoStop}%
\bibitem [{\citenamefont {Kim}\ \emph {et~al.}(2006)\citenamefont {Kim}, \citenamefont {Koh}, \citenamefont {Rotenberg}, \citenamefont {Oh}, \citenamefont {Eisaki}, \citenamefont {Motoyama}, \citenamefont {Uchida}, \citenamefont {Tohyama}, \citenamefont {Maekawa}, \citenamefont {Shen},\ and\ \citenamefont {Kim}}]{Kim2006distinct}%
  \BibitemOpen
  \bibfield  {author} {\bibinfo {author} {\bibfnamefont {B.~J.}\ \bibnamefont {Kim}}, \bibinfo {author} {\bibfnamefont {H.}~\bibnamefont {Koh}}, \bibinfo {author} {\bibfnamefont {E.}~\bibnamefont {Rotenberg}}, \bibinfo {author} {\bibfnamefont {S.~J.}\ \bibnamefont {Oh}}, \bibinfo {author} {\bibfnamefont {H.}~\bibnamefont {Eisaki}}, \bibinfo {author} {\bibfnamefont {N.}~\bibnamefont {Motoyama}}, \bibinfo {author} {\bibfnamefont {S.}~\bibnamefont {Uchida}}, \bibinfo {author} {\bibfnamefont {T.}~\bibnamefont {Tohyama}}, \bibinfo {author} {\bibfnamefont {S.}~\bibnamefont {Maekawa}}, \bibinfo {author} {\bibfnamefont {Z.~X.}\ \bibnamefont {Shen}},\ and\ \bibinfo {author} {\bibfnamefont {C.}~\bibnamefont {Kim}},\ }\bibfield  {title} {\bibinfo {title} {Distinct spinon and holon dispersions in photoemission spectral functions from one-dimensional {SrCuO$_2$}},\ }\href {https://doi.org/10.1038/nphys316} {\bibfield  {journal} {\bibinfo  {journal} {Nature Physics}\ }\textbf {\bibinfo {volume} {2}},\ \bibinfo {pages}
  {397} (\bibinfo {year} {2006})}\BibitemShut {NoStop}%
\bibitem [{\citenamefont {Kumar}\ \emph {et~al.}(2018)\citenamefont {Kumar}, \citenamefont {Nocera}, \citenamefont {Dagotto},\ and\ \citenamefont {Johnston}}]{Kumar2018multispinon}%
  \BibitemOpen
  \bibfield  {author} {\bibinfo {author} {\bibfnamefont {U.}~\bibnamefont {Kumar}}, \bibinfo {author} {\bibfnamefont {A.}~\bibnamefont {Nocera}}, \bibinfo {author} {\bibfnamefont {E.}~\bibnamefont {Dagotto}},\ and\ \bibinfo {author} {\bibfnamefont {S.}~\bibnamefont {Johnston}},\ }\bibfield  {title} {\bibinfo {title} {Multi-spinon and antiholon excitations probed by resonant inelastic x-ray scattering on doped one-dimensional antiferromagnets},\ }\href {https://doi.org/10.1088/1367-2630/aad00a} {\bibfield  {journal} {\bibinfo  {journal} {New Journal of Physics}\ }\textbf {\bibinfo {volume} {20}},\ \bibinfo {pages} {073019} (\bibinfo {year} {2018})}\BibitemShut {NoStop}%
\bibitem [{\citenamefont {Li}\ \emph {et~al.}(2021)\citenamefont {Li}, \citenamefont {Nocera}, \citenamefont {Kumar},\ and\ \citenamefont {Johnston}}]{Li2021particlehole}%
  \BibitemOpen
  \bibfield  {author} {\bibinfo {author} {\bibfnamefont {S.}~\bibnamefont {Li}}, \bibinfo {author} {\bibfnamefont {A.}~\bibnamefont {Nocera}}, \bibinfo {author} {\bibfnamefont {U.}~\bibnamefont {Kumar}},\ and\ \bibinfo {author} {\bibfnamefont {S.}~\bibnamefont {Johnston}},\ }\bibfield  {title} {\bibinfo {title} {Particle-hole asymmetry in the dynamical spin and charge responses of corner-shared {1D} cuprates},\ }\href {https://doi.org/10.1038/s42005-021-00718-w} {\bibfield  {journal} {\bibinfo  {journal} {Communications Physics}\ }\textbf {\bibinfo {volume} {4}},\ \bibinfo {pages} {217} (\bibinfo {year} {2021})}\BibitemShut {NoStop}%
\bibitem [{\citenamefont {Ament}\ \emph {et~al.}(2009)\citenamefont {Ament}, \citenamefont {Ghiringhelli}, \citenamefont {Sala}, \citenamefont {Braicovich},\ and\ \citenamefont {van~den Brink}}]{AmentPRL2009}%
  \BibitemOpen
  \bibfield  {author} {\bibinfo {author} {\bibfnamefont {L.~J.~P.}\ \bibnamefont {Ament}}, \bibinfo {author} {\bibfnamefont {G.}~\bibnamefont {Ghiringhelli}}, \bibinfo {author} {\bibfnamefont {M.~M.}\ \bibnamefont {Sala}}, \bibinfo {author} {\bibfnamefont {L.}~\bibnamefont {Braicovich}},\ and\ \bibinfo {author} {\bibfnamefont {J.}~\bibnamefont {van~den Brink}},\ }\bibfield  {title} {\bibinfo {title} {Theoretical demonstration of how the dispersion of magnetic excitations in cuprate compounds can be determined using resonant inelastic x-ray scattering},\ }\href {https://doi.org/10.1103/PhysRevLett.103.117003} {\bibfield  {journal} {\bibinfo  {journal} {Phys. Rev. Lett.}\ }\textbf {\bibinfo {volume} {103}},\ \bibinfo {pages} {117003} (\bibinfo {year} {2009})}\BibitemShut {NoStop}%
\bibitem [{\citenamefont {Bhaseen}\ \emph {et~al.}(2005)\citenamefont {Bhaseen}, \citenamefont {Essler},\ and\ \citenamefont {Grage}}]{Bhaseen_Essler_Grage_2005}%
  \BibitemOpen
  \bibfield  {author} {\bibinfo {author} {\bibfnamefont {M.~J.}\ \bibnamefont {Bhaseen}}, \bibinfo {author} {\bibfnamefont {F.~H.~L.}\ \bibnamefont {Essler}},\ and\ \bibinfo {author} {\bibfnamefont {A.}~\bibnamefont {Grage}},\ }\bibfield  {title} {\bibinfo {title} {Itinerancy effects on spin correlations in one-dimensional {M}ott insulators},\ }\href {https://doi.org/10.1103/PhysRevB.71.020405} {\bibfield  {journal} {\bibinfo  {journal} {Physical Review B}\ }\textbf {\bibinfo {volume} {71}},\ \bibinfo {pages} {020405} (\bibinfo {year} {2005})}\BibitemShut {NoStop}%
\bibitem [{\citenamefont {Klauser}\ \emph {et~al.}(2011)\citenamefont {Klauser}, \citenamefont {Mossel}, \citenamefont {Caux},\ and\ \citenamefont {Van Den~Brink}}]{Klauser_Mossel_Caux_Van_DenBrink_2011}%
  \BibitemOpen
  \bibfield  {author} {\bibinfo {author} {\bibfnamefont {A.}~\bibnamefont {Klauser}}, \bibinfo {author} {\bibfnamefont {J.}~\bibnamefont {Mossel}}, \bibinfo {author} {\bibfnamefont {J.-S.}\ \bibnamefont {Caux}},\ and\ \bibinfo {author} {\bibfnamefont {J.}~\bibnamefont {Van Den~Brink}},\ }\bibfield  {title} {\bibinfo {title} {Spin-{E}xchange {D}ynamical {S}tructure {F}actor of the {$S = 1/2$} {H}eisenberg chain},\ }\href {https://doi.org/10.1103/PhysRevLett.106.157205} {\bibfield  {journal} {\bibinfo  {journal} {Physical Review Letters}\ }\textbf {\bibinfo {volume} {106}},\ \bibinfo {pages} {157205} (\bibinfo {year} {2011})}\BibitemShut {NoStop}%
\bibitem [{\citenamefont {Jia}\ \emph {et~al.}(2016)\citenamefont {Jia}, \citenamefont {Wohlfeld}, \citenamefont {Wang}, \citenamefont {Moritz},\ and\ \citenamefont {Devereaux}}]{Jia2016_RIXSElement}%
  \BibitemOpen
  \bibfield  {author} {\bibinfo {author} {\bibfnamefont {C.}~\bibnamefont {Jia}}, \bibinfo {author} {\bibfnamefont {K.}~\bibnamefont {Wohlfeld}}, \bibinfo {author} {\bibfnamefont {Y.}~\bibnamefont {Wang}}, \bibinfo {author} {\bibfnamefont {B.}~\bibnamefont {Moritz}},\ and\ \bibinfo {author} {\bibfnamefont {T.~P.}\ \bibnamefont {Devereaux}},\ }\bibfield  {title} {\bibinfo {title} {Using {RIXS} to uncover elementary charge and spin excitations},\ }\href {https://doi.org/10.1103/PhysRevX.6.021020} {\bibfield  {journal} {\bibinfo  {journal} {Phys. Rev. X}\ }\textbf {\bibinfo {volume} {6}},\ \bibinfo {pages} {021020} (\bibinfo {year} {2016})}\BibitemShut {NoStop}%
\bibitem [{\citenamefont {Forte}\ \emph {et~al.}(2011)\citenamefont {Forte}, \citenamefont {Cuoco}, \citenamefont {Noce},\ and\ \citenamefont {van~den Brink}}]{forte2011doping}%
  \BibitemOpen
  \bibfield  {author} {\bibinfo {author} {\bibfnamefont {F.}~\bibnamefont {Forte}}, \bibinfo {author} {\bibfnamefont {M.}~\bibnamefont {Cuoco}}, \bibinfo {author} {\bibfnamefont {C.}~\bibnamefont {Noce}},\ and\ \bibinfo {author} {\bibfnamefont {J.}~\bibnamefont {van~den Brink}},\ }\bibfield  {title} {\bibinfo {title} {Doping dependence of magnetic excitations of one-dimensional cuprates as probed by resonant inelastic x-ray scattering},\ }\href {https://doi.org/10.1103/PhysRevB.83.245133} {\bibfield  {journal} {\bibinfo  {journal} {Phys. Rev. B}\ }\textbf {\bibinfo {volume} {83}},\ \bibinfo {pages} {245133} (\bibinfo {year} {2011})}\BibitemShut {NoStop}%
\bibitem [{\citenamefont {Scheie}\ \emph {et~al.}(2021)\citenamefont {Scheie}, \citenamefont {Laurell}, \citenamefont {Samarakoon}, \citenamefont {Lake}, \citenamefont {Nagler}, \citenamefont {Granroth}, \citenamefont {Okamoto}, \citenamefont {Alvarez},\ and\ \citenamefont {Tennant}}]{Scheie2021Witnessing}%
  \BibitemOpen
  \bibfield  {author} {\bibinfo {author} {\bibfnamefont {A.}~\bibnamefont {Scheie}}, \bibinfo {author} {\bibfnamefont {P.}~\bibnamefont {Laurell}}, \bibinfo {author} {\bibfnamefont {A.~M.}\ \bibnamefont {Samarakoon}}, \bibinfo {author} {\bibfnamefont {B.}~\bibnamefont {Lake}}, \bibinfo {author} {\bibfnamefont {S.~E.}\ \bibnamefont {Nagler}}, \bibinfo {author} {\bibfnamefont {G.~E.}\ \bibnamefont {Granroth}}, \bibinfo {author} {\bibfnamefont {S.}~\bibnamefont {Okamoto}}, \bibinfo {author} {\bibfnamefont {G.}~\bibnamefont {Alvarez}},\ and\ \bibinfo {author} {\bibfnamefont {D.~A.}\ \bibnamefont {Tennant}},\ }\bibfield  {title} {\bibinfo {title} {Witnessing entanglement in quantum magnets using neutron scattering},\ }\href {https://doi.org/10.1103/PhysRevB.103.224434} {\bibfield  {journal} {\bibinfo  {journal} {Phys. Rev. B}\ }\textbf {\bibinfo {volume} {103}},\ \bibinfo {pages} {224434} (\bibinfo {year} {2021})}\BibitemShut {NoStop}%
\bibitem [{\citenamefont {Laurell}\ \emph {et~al.}(2024)\citenamefont {Laurell}, \citenamefont {Scheie}, \citenamefont {Dagotto},\ and\ \citenamefont {Tennant}}]{laurell2024witnessing}%
  \BibitemOpen
  \bibfield  {author} {\bibinfo {author} {\bibfnamefont {P.}~\bibnamefont {Laurell}}, \bibinfo {author} {\bibfnamefont {A.}~\bibnamefont {Scheie}}, \bibinfo {author} {\bibfnamefont {E.}~\bibnamefont {Dagotto}},\ and\ \bibinfo {author} {\bibfnamefont {D.~A.}\ \bibnamefont {Tennant}},\ }\bibfield  {title} {\bibinfo {title} {Witnessing entanglement and quantum correlations in condensed matter: a review},\ }\href {https://arxiv.org/abs/2405.10899} {\bibfield  {journal} {\bibinfo  {journal} {arXiv:2405.10899}\ } (\bibinfo {year} {2024})}\BibitemShut {NoStop}%
\bibitem [{\citenamefont {Ren}\ \emph {et~al.}(2024)\citenamefont {Ren}, \citenamefont {Shen}, \citenamefont {TenHuisen}, \citenamefont {Sears}, \citenamefont {He}, \citenamefont {Upton}, \citenamefont {Casa}, \citenamefont {Becker}, \citenamefont {Mitrano}, \citenamefont {Dean},\ and\ \citenamefont {Konik}}]{ren2024witnessing}%
  \BibitemOpen
  \bibfield  {author} {\bibinfo {author} {\bibfnamefont {T.}~\bibnamefont {Ren}}, \bibinfo {author} {\bibfnamefont {Y.}~\bibnamefont {Shen}}, \bibinfo {author} {\bibfnamefont {S.~F.~R.}\ \bibnamefont {TenHuisen}}, \bibinfo {author} {\bibfnamefont {J.}~\bibnamefont {Sears}}, \bibinfo {author} {\bibfnamefont {W.}~\bibnamefont {He}}, \bibinfo {author} {\bibfnamefont {M.~H.}\ \bibnamefont {Upton}}, \bibinfo {author} {\bibfnamefont {D.}~\bibnamefont {Casa}}, \bibinfo {author} {\bibfnamefont {P.}~\bibnamefont {Becker}}, \bibinfo {author} {\bibfnamefont {M.}~\bibnamefont {Mitrano}}, \bibinfo {author} {\bibfnamefont {M.~P.~M.}\ \bibnamefont {Dean}},\ and\ \bibinfo {author} {\bibfnamefont {R.~M.}\ \bibnamefont {Konik}},\ }\bibfield  {title} {\bibinfo {title} {Witnessing quantum entanglement using resonant inelastic x-ray scattering},\ }\href {https://arxiv.org/abs/2404.05850} {\bibfield  {journal} {\bibinfo  {journal} {arXiv:2404.05850}\ } (\bibinfo {year} {2024})}\BibitemShut {NoStop}%
\bibitem [{\citenamefont {Weber}\ \emph {et~al.}(2015)\citenamefont {Weber}, \citenamefont {Assaad},\ and\ \citenamefont {Hohenadler}}]{Weber2015phonon}%
  \BibitemOpen
  \bibfield  {author} {\bibinfo {author} {\bibfnamefont {M.}~\bibnamefont {Weber}}, \bibinfo {author} {\bibfnamefont {F.~F.}\ \bibnamefont {Assaad}},\ and\ \bibinfo {author} {\bibfnamefont {M.}~\bibnamefont {Hohenadler}},\ }\bibfield  {title} {\bibinfo {title} {Phonon spectral function of the one-dimensional {H}olstein-{H}ubbard model},\ }\href {https://doi.org/10.1103/PhysRevB.91.235150} {\bibfield  {journal} {\bibinfo  {journal} {Phys. Rev. B}\ }\textbf {\bibinfo {volume} {91}},\ \bibinfo {pages} {235150} (\bibinfo {year} {2015})}\BibitemShut {NoStop}%
\bibitem [{\citenamefont {Johnston}\ \emph {et~al.}(2010)\citenamefont {Johnston}, \citenamefont {Vernay}, \citenamefont {Moritz}, \citenamefont {Shen}, \citenamefont {Nagaosa}, \citenamefont {Zaanen},\ and\ \citenamefont {Devereaux}}]{Johnston2010systematic}%
  \BibitemOpen
  \bibfield  {author} {\bibinfo {author} {\bibfnamefont {S.}~\bibnamefont {Johnston}}, \bibinfo {author} {\bibfnamefont {F.}~\bibnamefont {Vernay}}, \bibinfo {author} {\bibfnamefont {B.}~\bibnamefont {Moritz}}, \bibinfo {author} {\bibfnamefont {Z.-X.}\ \bibnamefont {Shen}}, \bibinfo {author} {\bibfnamefont {N.}~\bibnamefont {Nagaosa}}, \bibinfo {author} {\bibfnamefont {J.}~\bibnamefont {Zaanen}},\ and\ \bibinfo {author} {\bibfnamefont {T.~P.}\ \bibnamefont {Devereaux}},\ }\bibfield  {title} {\bibinfo {title} {Systematic study of electron-phonon coupling to oxygen modes across the cuprates},\ }\href {https://doi.org/10.1103/PhysRevB.82.064513} {\bibfield  {journal} {\bibinfo  {journal} {Phys. Rev. B}\ }\textbf {\bibinfo {volume} {82}},\ \bibinfo {pages} {064513} (\bibinfo {year} {2010})}\BibitemShut {NoStop}%
\bibitem [{\citenamefont {Tseng}\ \emph {et~al.}(2022)\citenamefont {Tseng}, \citenamefont {Thomas}, \citenamefont {Zhang}, \citenamefont {Paris}, \citenamefont {Puphal}, \citenamefont {Bag}, \citenamefont {Deng}, \citenamefont {Asmara}, \citenamefont {Strocov}, \citenamefont {Singh}, \citenamefont {Pomjakushina}, \citenamefont {Kumar}, \citenamefont {Nocera}, \citenamefont {Rønnow}, \citenamefont {Johnston},\ and\ \citenamefont {Schmitt}}]{Tseng22}%
  \BibitemOpen
  \bibfield  {author} {\bibinfo {author} {\bibfnamefont {Y.}~\bibnamefont {Tseng}}, \bibinfo {author} {\bibfnamefont {J.}~\bibnamefont {Thomas}}, \bibinfo {author} {\bibfnamefont {W.}~\bibnamefont {Zhang}}, \bibinfo {author} {\bibfnamefont {E.}~\bibnamefont {Paris}}, \bibinfo {author} {\bibfnamefont {P.}~\bibnamefont {Puphal}}, \bibinfo {author} {\bibfnamefont {R.}~\bibnamefont {Bag}}, \bibinfo {author} {\bibfnamefont {G.}~\bibnamefont {Deng}}, \bibinfo {author} {\bibfnamefont {T.~C.}\ \bibnamefont {Asmara}}, \bibinfo {author} {\bibfnamefont {V.~N.}\ \bibnamefont {Strocov}}, \bibinfo {author} {\bibfnamefont {S.}~\bibnamefont {Singh}}, \bibinfo {author} {\bibfnamefont {E.}~\bibnamefont {Pomjakushina}}, \bibinfo {author} {\bibfnamefont {U.}~\bibnamefont {Kumar}}, \bibinfo {author} {\bibfnamefont {A.}~\bibnamefont {Nocera}}, \bibinfo {author} {\bibfnamefont {H.~M.}\ \bibnamefont {Rønnow}}, \bibinfo {author} {\bibfnamefont {S.}~\bibnamefont {Johnston}},\ and\ \bibinfo {author} {\bibfnamefont {T.}~\bibnamefont
  {Schmitt}},\ }\bibfield  {title} {\bibinfo {title} {Crossover of high-energy spin fluctuations from collective triplons to localized magnetic excitations in {${\mathrm{{S}r_{14-x}{C}a_x{C}u_{24}{O}_{41}}}$} ladders},\ }\href {https://doi.org/https://doi.org/10.1038/s41535-022-00502-1} {\bibfield  {journal} {\bibinfo  {journal} {npj Quantum Materials}\ }\textbf {\bibinfo {volume} {7}},\ \bibinfo {pages} {92} (\bibinfo {year} {2022})}\BibitemShut {NoStop}%
\bibitem [{\citenamefont {Chen}\ \emph {et~al.}(2021)\citenamefont {Chen}, \citenamefont {Wang}, \citenamefont {Rebec}, \citenamefont {Jia}, \citenamefont {Hashimoto}, \citenamefont {Lu}, \citenamefont {Moritz}, \citenamefont {Moore}, \citenamefont {Devereaux},\ and\ \citenamefont {Shen}}]{ChenScience2021}%
  \BibitemOpen
  \bibfield  {author} {\bibinfo {author} {\bibfnamefont {Z.}~\bibnamefont {Chen}}, \bibinfo {author} {\bibfnamefont {Y.}~\bibnamefont {Wang}}, \bibinfo {author} {\bibfnamefont {S.~N.}\ \bibnamefont {Rebec}}, \bibinfo {author} {\bibfnamefont {T.}~\bibnamefont {Jia}}, \bibinfo {author} {\bibfnamefont {M.}~\bibnamefont {Hashimoto}}, \bibinfo {author} {\bibfnamefont {D.}~\bibnamefont {Lu}}, \bibinfo {author} {\bibfnamefont {B.}~\bibnamefont {Moritz}}, \bibinfo {author} {\bibfnamefont {R.~G.}\ \bibnamefont {Moore}}, \bibinfo {author} {\bibfnamefont {T.~P.}\ \bibnamefont {Devereaux}},\ and\ \bibinfo {author} {\bibfnamefont {Z.-X.}\ \bibnamefont {Shen}},\ }\bibfield  {title} {\bibinfo {title} {Anomalously strong near-neighbor attraction in doped {1D} cuprate chains},\ }\href {https://doi.org/10.1126/science.abf5174} {\bibfield  {journal} {\bibinfo  {journal} {Science}\ }\textbf {\bibinfo {volume} {373}},\ \bibinfo {pages} {1235} (\bibinfo {year} {2021})}\BibitemShut {NoStop}%
\bibitem [{\citenamefont {Wang}\ \emph {et~al.}(2021)\citenamefont {Wang}, \citenamefont {Chen}, \citenamefont {Shi}, \citenamefont {Moritz}, \citenamefont {Shen},\ and\ \citenamefont {Devereaux}}]{Wang2021phonon}%
  \BibitemOpen
  \bibfield  {author} {\bibinfo {author} {\bibfnamefont {Y.}~\bibnamefont {Wang}}, \bibinfo {author} {\bibfnamefont {Z.}~\bibnamefont {Chen}}, \bibinfo {author} {\bibfnamefont {T.}~\bibnamefont {Shi}}, \bibinfo {author} {\bibfnamefont {B.}~\bibnamefont {Moritz}}, \bibinfo {author} {\bibfnamefont {Z.-X.}\ \bibnamefont {Shen}},\ and\ \bibinfo {author} {\bibfnamefont {T.~P.}\ \bibnamefont {Devereaux}},\ }\bibfield  {title} {\bibinfo {title} {Phonon-mediated long-range attractive interaction in one-dimensional cuprates},\ }\href {https://doi.org/10.1103/PhysRevLett.127.197003} {\bibfield  {journal} {\bibinfo  {journal} {Phys. Rev. Lett.}\ }\textbf {\bibinfo {volume} {127}},\ \bibinfo {pages} {197003} (\bibinfo {year} {2021})}\BibitemShut {NoStop}%
\bibitem [{\citenamefont {Chen}\ \emph {et~al.}(2019)\citenamefont {Chen}, \citenamefont {Wang}, \citenamefont {Jia}, \citenamefont {Moritz}, \citenamefont {Shvaika}, \citenamefont {Freericks},\ and\ \citenamefont {Devereaux}}]{chen2019theory}%
  \BibitemOpen
  \bibfield  {author} {\bibinfo {author} {\bibfnamefont {Y.}~\bibnamefont {Chen}}, \bibinfo {author} {\bibfnamefont {Y.}~\bibnamefont {Wang}}, \bibinfo {author} {\bibfnamefont {C.}~\bibnamefont {Jia}}, \bibinfo {author} {\bibfnamefont {B.}~\bibnamefont {Moritz}}, \bibinfo {author} {\bibfnamefont {A.~M.}\ \bibnamefont {Shvaika}}, \bibinfo {author} {\bibfnamefont {J.~K.}\ \bibnamefont {Freericks}},\ and\ \bibinfo {author} {\bibfnamefont {T.~P.}\ \bibnamefont {Devereaux}},\ }\bibfield  {title} {\bibinfo {title} {Theory for time-resolved resonant inelastic x-ray scattering},\ }\href {https://doi.org/10.1103/PhysRevB.99.104306} {\bibfield  {journal} {\bibinfo  {journal} {Phys. Rev. B}\ }\textbf {\bibinfo {volume} {99}},\ \bibinfo {pages} {104306} (\bibinfo {year} {2019})}\BibitemShut {NoStop}%
\bibitem [{\citenamefont {Zawadzki}\ \emph {et~al.}(2023)\citenamefont {Zawadzki}, \citenamefont {Nocera},\ and\ \citenamefont {Feiguin}}]{zawadzki2023time}%
  \BibitemOpen
  \bibfield  {author} {\bibinfo {author} {\bibfnamefont {K.}~\bibnamefont {Zawadzki}}, \bibinfo {author} {\bibfnamefont {A.}~\bibnamefont {Nocera}},\ and\ \bibinfo {author} {\bibfnamefont {A.~E.}\ \bibnamefont {Feiguin}},\ }\bibfield  {title} {\bibinfo {title} {{A time-dependent momentum-resolved scattering approach to core-level spectroscopies}},\ }\href {https://doi.org/10.21468/SciPostPhys.15.4.166} {\bibfield  {journal} {\bibinfo  {journal} {SciPost Phys.}\ }\textbf {\bibinfo {volume} {15}},\ \bibinfo {pages} {166} (\bibinfo {year} {2023})}\BibitemShut {NoStop}%
\bibitem [{\citenamefont {Jeckelmann}(2002)}]{PhysRevB.66.045114}%
  \BibitemOpen
  \bibfield  {author} {\bibinfo {author} {\bibfnamefont {E.}~\bibnamefont {Jeckelmann}},\ }\bibfield  {title} {\bibinfo {title} {Dynamical density-matrix renormalization-group method},\ }\href {https://doi.org/10.1103/PhysRevB.66.045114} {\bibfield  {journal} {\bibinfo  {journal} {Phys. Rev. B}\ }\textbf {\bibinfo {volume} {66}},\ \bibinfo {pages} {045114} (\bibinfo {year} {2002})}\BibitemShut {NoStop}%
\end{thebibliography}%
